\documentclass{emulateapj}
\usepackage{epsfig}
\newcommand{\beq}{\begin{equation}}
\newcommand{\eeq}{\end{equation}}
\def\be{\begin{equation}}
\def\ee{\end{equation}}
\def\bea{\begin{eqnarray}}
\def\eea{\end{eqnarray}}



\def \logTd6 {\hbox{log$( T/6 \kev)$} }

\def\myputfigure#1#2#3#4#5%
{\vskip#5pt\makebox[0pt]{\hskip#2in
\includegraphics[width=#3\textwidth]{#1}}\vskip#4pt\hfill}

\def \etal      {et al.\ }

\def \kev       {{\rm\ keV}}
\def \msol      {{\rm\ M}_\odot}
\def \hmsol     {h^{-1}{\rm\ M}_\odot}
\def \hMpc      {h^{-1}{\rm\ Mpc}}
\def \hkpc      {h^{-1}{\rm\ kpc}}

\newcommand{\Om}{\mbox{$\Omega_m$}}

\input epsf
\bibpunct{(}{)}{;}{a}{}{,}

\begin{document}

\lefthead{SHEAR SELECTED CLUSTERS}
\righthead{HENNAWI \& SPERGEL}

\title{Shear Selected Cluster Cosmology: Tomography and Optimal Filtering} 

\author{Joseph F. Hennawi\altaffilmark{1,2,3}, 
        David N. Spergel\altaffilmark{2}
} 

\altaffiltext{1}{Department of Astronomy, University of California
  Berkeley, Berkeley, CA 94720}

\altaffiltext{2}{Princeton University Observatory, Princeton, NJ 08544}

\altaffiltext{3}{Hubble Fellow}

\begin{abstract}
  We study the potential of weak lensing surveys to detect clusters of
  galaxies, using a fast Particle Mesh cosmological N-body simulation
  algorithm specifically tailored to investigate the statistics of
  these \emph{shear selected} clusters. In particular, we explore the
  degree to which the radial positions of galaxy clusters can be
  determined \emph{tomographically}, by using photometric redshifts of
  \emph{background} source galaxies.  We quantify the errors in the
  tomographic redshifts, $\Delta z \equiv z_{\rm tomography}-z_{\rm
    real}$, and study their dependence on mass, redshift, detection
  significance, and filtering scheme. For clusters detected with
  signal to noise ratio $S/N\gtrsim 4.5$, the fraction of clusters
  with tomographic redshift errors $|\Delta z| \leq 0.2$ is $45\%$ and
  the root mean square deviation from the real redshifts is
  $\langle(\Delta z)^2\rangle^{1/2}=0.41$, with smaller errors for
  higher $S/N$ ratios.  A Tomographic Matched Filtering (TMF) scheme,
  which combines tomography and matched filtering, is introduced which
  optimally detects clusters of galaxies in weak lensing surveys.  The
  TMF exploits the extra information provided by photometric redshifts
  of background source galaxies, neglected in previous studies, to
  optimally weight the sources. The efficacy and reliability of the
  TMF is investigated using a large ensemble of mock observations from
  our simulations and detailed comparisons are made to other filters.
  Using photometric redshift information with the TMF enhances the
  number of clusters detected with $S/N\gtrsim 4.5$ by as much as $76
  \%$, and it increases the dynamic range of weak lensing searches for
  clusters, detecting more high redshift clusters and extending the
  mass sensitivity down to the scale of large groups. Furthermore, we
  find that coarse redshift binning of source galaxies with as few as
  three bins is sufficient to realize the gains of the TMF. Thus, the
  tomographic filtering techniques presented here can be applied to
  current ground based weak lensing data in as few as three bands,
  since two colors and magnitude prior are sufficient to bin source
  galaxies this coarsely. Cosmological applications of shear selected
  cluster samples are also discussed.

\end{abstract}

%------------------------------------------------------------------------------
% User-supplied List of keywords.

%------------------------------------------------------------------------------

\section{Introduction}
\label{sec:intro}

Mapping the distribution of dark matter in the universe is now one of
the primary goals of observational cosmology.  The traditional
approach has been to survey the sky for a population of luminous
objects, assuming a direct relationship between luminosity and dark
matter. While large galaxy surveys such as the 2dF Galaxy Redshift
Survey \citep{Coll01} and the Sloan Digital Sky Survey \citep{York00}
are perhaps the first examples that come to mind, future wide field
X-ray \citep{Romer01} and Sunyaev-Z'eldovich (SZ) surveys
\citep{Carl02,Koso03} will usher in an era where maps of the universe
will be available at multiple frequencies.  Generically, all objects
that emit radiation will be biased tracers of the underlying dark
matter distribution. Hence, the interpretation of this data is limited
by our ability to quantify the relationship between dark and luminous
matter.

Weak gravitational lensing, or the coherent distortion of images of
faint background galaxies by the foreground matter distribution,
provides a unique opportunity to map the dark matter directly without
making any assumptions about how baryons trace dark matter (see e.g.,
\citet{BaSc01} or \citet{vWM03} for a review).  Recently, weak lensing
by large scale structure, or ``cosmic shear'', has been independently
detected by several groups (see Table 1 of \citet{vWM03} for a
compilation of references).  The primary focus of these observations
has been to measure the two point statistical properties of the shear
field, which probes the matter power spectrum in projection.  However,
as first suggested by \citet{Kaiser95} and \citet{Sch96}, weak lensing
surveys can also be used to detect individual mass concentrations,
allowing one to construct a \emph{shear selected} sample of dark
halos. Typically these will be galaxy cluster size $M \gtrsim 
10^{13.5} \ \msol$ objects, as the finite number of background sources
and their intrinsic ellipticity place a lower limit on the size of a
halo that can be detected.  To date, there are only two such cases of
spectroscopically confirmed cluster detections from `blank field'
weak lensing observations \citep{Witt01,Witt03}. Although, recently
\citet{Schirm03,Schirm04} have used weak lensing to confirm several
color-selected optical cluster candidates. 

A shear selected sample of galaxy clusters would be of great
astrophysical and cosmological interest. Probable biases in optical,
X-ray, and SZ selected cluster samples with respect to richness,
baryon fraction, morphology, and dynamical state, would be revealed
\citep{Hughes04}.  Furthermore, if there does exist a population of
high mass to light clusters that has hitherto gone undetected, this
would cast serious doubt upon the ``fair sample'' hypotheses
\citep{White93}, invoked to determine the matter density parameter,
$\Omega_{\rm m}$, from comparisons of cluster baryon fraction to total
cluster mass measurements (see e.g., \citet{Allen03} or \citet{VV04} for 
recent examples).

Perhaps a more contentious issue is the possible existence of a
population of truly dark clusters. Recently, several groups have purportedly
detected dark mass concentrations from weak lensing \citep[though see
 Erben \etal 2003]{Fis99,Erben00,UF00,Mir02,Dahle03}.  If these
objects are actually `dark' virialized mass concentrations as
opposed to non-virialized objects \citep{Wein02} or projections of large
scale structure \citep{MWL01,WvWM02,Paddy03,HTY03}, this would
pose serious challenges for our current structure formation paradigm
and theories of galaxy formation. The absence of galaxies would
require a complex form of feedback that suppresses galaxy formation in
clusters.  If such a dark cluster were X-ray or SZ faint, one would have
to invoke some exotic mechanism to segregate dark matter from
baryons on Mpc scales.

It is well known that a large sample of clusters of galaxies out to
moderate redshift $z \sim 0.5$ can be used to impose stringent
constraints on cosmological parameters.  Several authors have recently
suggested using the number count distribution of galaxy clusters, from
X-ray and SZ observations \citep{HMH01,WB03}, deep optical surveys
\citep{New02}, and weak lensing \citep{BPB02,Wein03}, to probe the
nature of the dark energy believed to be causing the acceleration of
the expansion of the universe.  Besides probing the dark energy, these
cluster samples would constrain the matter density and amplitude of
mass fluctuations \citep[see e.g.,][]{BF98,HHM01,Pier03}. In addition,
the clustering of such cluster samples would provide another means to
measure the matter power spectrum \citep{TED98,Schue01} which would
provide complimentary parameter constraints \citep{MM03b} and might
possibly allow for a measurement of the angular diameter
distance-redshift relation \citep{Cooray01} or baryon wiggles
\citep{HH03}.

Yet, a limitation of the `baryon selection' in optical, X-ray, and 
SZ cluster samples, is that their selection function depends on
observables (e.g., richness, velocity dispersion, flux, temperature, SZ
decrement) that serve as a proxy for mass. Attempts to model these
mass-observable relationships depend on the uncertain astrophysics of
galaxy formation and the state of baryons in clusters. While these can
be determined empirically, scatter in the correlation between any two
such observables is significant \citep[see e.g.,][]{Koch03}; furthermore,
these mass-observable relations surely evolve with redshift.  The mass
function of galaxy clusters is steep and exponentially sensitive to
changes in limiting mass; moreover, errors in the mass-observable
relations mimic cosmological parameter changes
\citep{VL99,LSW02,MM03a,Hu03}.  Inclusion of uncertainties in the
mass-observable relations significantly degrades constraints on
cosmological parameters obtained from surveys that rely on a baryonic
proxy for mass \citep{LSW02,MM03a}. Although consistency checks
provide additional leverage \citep{Hu03}, expensive follow up mass
measurements are in general required to empirically calibrate the
relationships between light and mass (Majumdar \& Mohr 2003a; though
see Lima \& Hu 2004).

In this regard, the great advantage of a shear selected sample of
galaxy clusters is that the selection function can be predicted
\emph{ab initio}, given a model for structure formation.  Because on
the scales of interest for weak lensing, only gravity is involved and
the mass is dominated by dark matter, precision cosmological measurements
will in principle be limited by instrumental systematics rather than
unknown astrophysics.  An accurate determination of the statistics of
shear selected clusters from weak lensing requires N-body simulations
of cosmological structure formation \citep[though see][for simple
  analytical treatments]{KS99,BPB02,Wein03}, which brings us to the subject
of this work.

In this paper we consider the detection of shear selected clusters
using fast Particle Mesh (PM) N-body simulations of cosmological
structure formation specifically tailored to investigate the
statistics of weak lensing by clusters.  Similar numerical studies
carried out by other groups \citep{Reb99,WvWM02,Paddy03,HTY03}, have
focused on the completeness and efficiency of weak lensing cluster
searches.  Devising an optimal filtering scheme to detect clusters has
received little attention and furthermore, the additional information
provided by photometric redshifts of background source galaxies has
been neglected. In this paper we introduce the fast PM simulation
algorithm, deduce the optimal cluster detection strategy, and
investigate cluster tomography. 

%In a future paper \citep{Me04} we use
%the simulations and filtering techniques discussed here to explore the
%potential of shear selected cluster samples to probe the equation of
%state parameter of the dark energy.

A fundamental obstacle for weak lensing studies of the matter
distribution is the lack of radial information. All of the matter
along the line-of-sight to a distant source contributes to the lensing
and so the distortion reflects a two dimensional projection of the
dark matter. This limitation can be overcome by using
photometric redshifts of the background source population, which will
be distributed in redshift, thus allowing for a \emph{tomographic}
reconstruction of the three dimensional foreground matter
distribution.

A variant of this technique has already been applied to real weak 
lensing data. \citet{Witt01,Witt03} tomographically
determined the redshifts of two \emph{shear selected} clusters at
$z_{\rm d}=0.30$ and $z_{\rm d}=0.70$ respectively, which were both
confirmed by follow up spectroscopy to be at $z=0.28$ and $z=0.68$.
Recently, \citet{Tay04} conducted a weak lensing analysis of the
previously known Abell supercluster A901/2 at $z=0.16$, and
tomographically confirmed its redshift from weak lensing alone. They
even constructed the first 3-D map of the dark matter potential of
this cluster using the inversion method introduced by
\citet{Tay01}.  \citet{HK02} and \citet{BT03} applied the
\citet{Tay01} tomographic inversion method to analytical models of 
clusters put in `by hand', with mixed success.  However, the
reliability of these tomographic reconstruction techniques have yet to
be characterized on realistic distributions of matter from N-body
simulations; although it is clear that the fidelity of the
reconstructions will be severely degraded by line of sight projections
of large scale structure.

In this work we introduce a complementary tomographic technique which is
effectively two dimensional. It is qualitatively similar to that used
by \citet{Witt01,Witt03} and \citet{Tay04} and also to maximum
likelihood methods developed to study cluster mass profiles
\citep{GS98,SKE00,KS01}. The efficacy and reliability of the method is
investigated by applying it to our large ensemble of N-body
simulations.  Additionally, a Tomographic Matched Filtering (TMF)
scheme, which combines tomography and matched filtering, is introduced
which takes advantage of the additional radial information provided by
photometric redshifts of source galaxies. The TMF is similar in spirit
to the matched filtering algorithms used to find clusters in optical
surveys \citep{Postman96,Kepner99,WK02,Koch03}.  Applying it to our ensemble of
simulations, we find that it is superior to filtering techniques
used previously \citep{Reb99,WvWM02,Paddy03,HTY03},
detecting more clusters per square degree and probing lower masses and
higher redshifts.

The outline of this paper is as follows. In \S 2 we describe our
implementation of a PM code to simulate weak lensing observations.
The formalism behind our maximum likelihood tomographic technique and
the adaptive matched filtering method is introduced in \S 3. In \S 4
we apply the TMF and other filters to a large ensemble of simulations
and determine the `optimal' filter for cluster detection. The completeness
of this optimal filter is discussed in \S 5. We assess
the reliability of tomographic redshifts in \S 6 and conclude in \S 7.

%There are several reasons why numerical simulations must be employed
%for such a study, which we briefly elucidate. First, on subdegree
%scales weak lensing observations probes highly non-linear scales
%(Jain \& Seljak 1997) so that numerical simulations are required for
%a full description (c.f. Jain, Seljak, \& White 2000; White \& Hu
%2000).  Second, alignments and projection effects can be quite severe
%(Metzler et al. 1999; White et al 2002; Padmanabhan et al. 2003) and
%must be properly simulated. Hence, naive calculations using the
%Press-Schecter formalism (Kruse \& Schneider 1999; Weinberg \&
%Kamionkowski 2003) although qualitatively informative, fail to
%represent the relevant physics.  statistics of clusters from weak
%lensing must be investigated using N-body simulations of cosmological
%structure formation, which brings us to the subject of this work.  an
%investigation of the the Because gravitational lensing This important
%the need to

%Clearly a fundamental limitation of weak lensing surveys in constructing 
%maps of the dark matter is the lack of radial inormation. 

\section{Simulating Weak Lensing}
\label{sec:numdetail}
\subsection{The  PM Code}

To evolve the dark matter distribution into the nonlinear regime, we
use a particle-mesh (PM) N-body code. This code, written by Changbom
Park, is described and tested in \citet{Park90} and \citet{Park91},
and has been used in studies of peculiar velocities \citep{BeNa00} and
biased galaxy formation \citep{NBW00,BeNa01}. The simulation uses a
staggered mesh to compute forces on particles
\citep{Me86,Cent88,Park90} , and those employed here
use $256^3$ particles and a $512^3$ force mesh.  

The initial conditions are generated by displacing particles from a
regular grid using the Z'eldovich approximation \citep[see
  e.g.,][]{Efs85}.  The force on each particle is calculated by finite
differencing the potential field, which is calculated from the density
field in Fourier space using the kernel $-1/k^2$ and fast Fourier
transforms (FFT). The gridded density field is computed from the
particles using the cloud-in-cell (CIC) charge assignment scheme
\citep{HoEa81}. The simulations are started at $1+z=50$ and are
evolved using equal steps in the expansion factor $a$ with a
symplectic leapfrog integrator described in \citet{Quinn97}.  The time
step is taken to be less than the Courant condition.  

%Tests on the
%resolution of our simulations will be presented in a future paper
%\citep{Me04}.

%The symplectic integrator allows us to take larger time steps than are
%typical with integrators using adaptive steps, and 

%(for a
%particle with $v\sim 1000$ \ \kms \ at $z=0$).  
%Tests on the resolution of our
%simulations will be presented in a companion paper \citep{Me04}.

Before we proceed we should justify our use of the PM algorithm for
the N-body simulations.  It is well known that the PM algorithm is
`memory limited' in the sense that higher spatial resolution comes
at the cost of storing an exceedingly large force mesh in random
access memory. The main drawback of PM simulations is thus limited
dynamic range, whereas the primary advantage of the algorithm is
speed. We compensate for the lack of dynamic range by adopting the
``tiling'' algorithm introduced by \citet{WH00}, whereby the light
cone is tiled with a telescoping sequence of N-body simulation cubes
of increasing resolution.  The speed of the PM algorithm is essential for
characterizing the statistics of shear selected clusters for the
following reasons.

First, massive clusters of galaxies are rare events. In order to
quantify their statistical properties it is essential to properly
sample the primordial Gaussian distribution of random phases of the
large scale structure along the light cone.  Similar numerical studies
using N-body simulations \citep{Reb99,WvWM02,Paddy03,HTY03} recycle
the output of one simulation by reprojecting across the same
simulation cube numerous times, hence avoiding the computational
challenge of simulating the entire $\sim {\rm Gpc}^3$ volume. However,
the frequency of one very rare event could be grossly misrepresented by
such a procedure, and there is a danger that the variance and tails of
the mass and redshift distributions of clusters will be
incorrect. This misrepresentation is exacerbated by the fact that the
primary leverage of cluster counts as a cosmological probe often comes
from precisely those objects on the exponential tail of the mass
function. To avoid this problem we tile nearly the entire light cone
volume with unique simulations, clearly requiring a fast
algorithm. 

%The degree to which recycling output misrepresents the
%statistics of rare clusters will be explored in our companion paper
%\citep{Me04}.

Second, a primary goal of this simulation program will be to explore
how the distribution of shear selected clusters depends on cosmological
parameters \citep{Me04}, which requires simulating many cosmological
models that span a large region of parameter space.

Finally, high resolution simulations are not required to accurately
represent weak lensing because of the small scale shot noise limit of
the observations \citep{WH00}. This can be understood heuristically
from simple scaling arguments \citep{Mir91,Bland91}.  For a galaxy
cluster with a singular isothermal profile, the effective lensing
signal, $\gamma\sqrt{N}$, scales logarithmically with $\theta$, the
size of the smoothing aperture.  This logarithmic scaling ensures that
the dominant contribution to cluster weak lensing comes from scales of
order the smoothing aperture, rather than much smaller scales which
are not resolved. Since the noise in the observations we simulate will
require averaging over apertures $\gtrsim 1^\prime$, resolving small
scale power in the simulations is not essential.
%sincethese scales are swamped by shot noise in the observations.

All simulations described in this paper are carried out using the
currently favored cold dark matter model with a cosmological constant
($\Lambda$CDM). The parameters we simulated are a total (dark matter
+ baryons) matter density parameter $\Omega_{\rm m}=0.295$, a density
of baryons $\Omega_{\rm b}=0.045$, a cosmological constant
$\Omega_{\Lambda}=0.705$, and a dimensionless Hubble constant
$h=0.69$, and normalization parameter $\sigma_8=0.84$, which are very
close to the \emph{WMAP} best fit model of \citet{Sperg03}.  The
power spectrum is given by a scale invariant spectrum of adiabatic
perturbations ($n=1$), and for the transfer function we use the
fitting function of \citet{EH98}.

\subsection{Tiling the Line of Sight}

In what follows we describe our implementation of the \citet{WH00}
algorithm to tile the line of sight with PM simulation cubes.
\citep[For a comprehensive discussion of numerical issues in weak
  lensing simulations, see e.g.,][]{JSW00,VW03,WV03}.

The distortion of a source galaxy at redshift $z$ in the direction
$\hat{\bf n}$ on the sky is determined by the shear field
$\gamma(\hat{\bf n},z)$ of the matter between the observer
and redshift $z$. In the weak lensing approximation, the shear is
completely specified by the convergence between the observer and
comoving distance $D_{\rm z} \equiv D\left(z\right)$, given by
\beq
\kappa(\hat{\bf n},z)=\frac{3}{2}\frac{\Om}{H_0^{-2}}
\int_0^{D_{\rm z}} dD D\left(1-\frac{D}{D_z}\right)
\frac{\delta(D\hat{\bf n},D)}{a}, \label{eqn:kappa}
\eeq
and
\beq 
\frac{dD}{dz}=\frac{1}{H(z)},
\eeq
which is valid for small angles in the Limber approximation 
\citep{JSW00,VW03}.  Here $\delta$ is the density contrast field,
$a=1\slash(1+z)$ is the scale factor normalized to unity
today, and $H(z)$ is the redshift dependent Hubble
constant. The shear can then be obtained via the Fourier relations
\begin{eqnarray}
  \widetilde{\gamma_1}=\frac{l_1^2-l_2^2}{l_1^2+l_2^2}\widetilde{\kappa} 
\nonumber \,,\\
  \widetilde{\gamma_2}=\frac{2l_1 l_2}{l_1^2+l_2^2}\widetilde{\kappa} 
\,, \label{eqn:gamma}
\end{eqnarray}
where $\widetilde{\kappa}$ is the two-dimensional Fourier Transform
(FT) of the convergence field, and ${\bf l}=(l_1,l_2)$ is
the Fourier variable conjugate to position on the sky.

Following \citet{WH00}, we tile the line of sight with a sequence of
telescoping simulation cubes of successively higher resolution.  The
integral in eqn.~(\ref{eqn:kappa}) is directly integrated across the
tiles to the observer at $z=0$ using the overdensity field $\delta$
from the simulations. Specifically, for each tile along the light
cone, the matter distribution is evolved from $1+z=50$ to the redshift
$z_{\rm rear}$, corresponding to the next segment of the integral in
eqn.~ (\ref{eqn:kappa}) and the rear of the current simulation
cube. In addition, we begin calculating a new convergence integral,
$\kappa(\hat{\bf n},z_{\rm rear})$ from $z_{\rm rear}$ to the observer
at each tile, so that the final output of our simulation is a sequence
of convergence planes densely spaced in redshift.

By evaluating the full convergence integral from the tiling sequence,
the geometry of the rays and the evolution of the potential are
accurately represented.  The lines of sight originate on a square grid
at $z_{\rm rear}$ and converge on an observer at $z=0$.  The planar
lattice has the same dimensionality as the force grid $N_{\rm
  grid}=512$. For staggered mesh PM algorithms, the effective
resolution of the simulation is set by the particle grid $N_{\rm
  part}=256$; however, as we will be interested in the densest regions
occupied by clusters of galaxies, there will be no danger of
undersampling the particle distribution and the use of the finer grid
is justified. We work in the small angle approximation, so that the
sky can be treated as flat and hence all photons propagate along the
z-axis of our simulation cube for all lines of sight. The convergence
of the geodesics as the light cone narrows is implemented by
multiplying by the appropriate geometric factor for each grid point,
and by the tiling. Also as mentioned previously, the use of a unique
simulation for each tile ensures the statistical independence of the
fluctuations and accurately represents the statistics of rare events.

%We thus avoid the unwanted loss of angular resolution and
%discreteness effects inherent in the technique commonly used in the
%literature, of projecting the density distribution onto a finite
%number of discrete lens planes.  

The tiling sequence is uniquely specified by an opening angle and a
maximum and minimum redshift. The redshift $z_{\rm max}=4$ is the
highest source redshift for which we evaluate the convergence. The
source redshift distributions we consider peak at $z \sim 1.0$, thus
there will be a negligible number of source galaxies at redshift
higher than $z_{\rm max}$. Below $z_{\rm min}=0.2$, we tile the
remainder of the light cone with the same simulation, recycling the
output of the last cube. This is necessary because we cannot shrink
the box size beyond the nonlinear scale, otherwise the PM code will
not evolve the density correctly because of the absence of mode
coupling to waves larger than the box.  Furthermore, below $z_{\rm
  min}=0.2$, it becomes computationally impractical to continue to
shrink the box and tile the light cone with successively smaller
unique simulations. The time step for the last leg of the simulation
with $z<z_{\rm min}$ is adjusted so as to resolve the next segment of
the convergence integral in eqn.~(\ref{eqn:kappa}).

To maintain statistical independence of the fluctuations during the
short sequence where we recycle output, we project across a different
axis of the simulation each time and the convergence grid is displaced
by a random offset relative to the density grid. Rays that leave the
box are remapped by periodicity.  The sizes of the simulation cubes in
our tiling sequence are chosen to exactly reproduce the comoving
volume of the light cone for $z_{\rm min}=0.2 < z < z_{\rm max}=4$,
ensuring that the number of cluster size halos in the field of view is
accurately represented.  Accurately reproducing the volume of the
light cone is not essential for studying two point statistics, but is
crucial for predicting cluster statistics and may become increasingly
relevant for higher order statistics which depend more sensitively on
rare events.  Below $z_{\rm min}=0.2$, we misrepresent the volume and
geometry of the light cone. The extra volume implies a slight excess
of clusters with $z_{\rm min}\leq 0.2$ and the slab geometry for this
last stretch results in a deficit of small scale power in the
convergence field, since the slabs probe larger scales than the
converging light cone. Both of these discrepancies have a negligible
effect on the tomography and filtering which we study in this paper. 
%,however they will be resolved in future studies that use this
%algorithm.

Our simulation fields are $4^\circ$ on a side and the angular pixel
size of the convergence grids is $4^\circ/512=0.47^{\prime}$. The
simulations use 46 tiles of which 32 are unique PM simulations (the
last 14 tiles are recycled from the same simulation).  Relevant
parameters for the simulations in our tiling scheme are listed in
Table \ref{table:tile}.  The right panels of Figure \ref{fig:tiling}
show the dimensionless 3-d mass power spectrum, $\Delta^2_{\rm
  mass}(k,z_{\rm rear})$, for four tiles, obtained by averaging
individual power spectra from 38 independent simulations. They are
compared to analytical fitting formulae for the nonlinear power
spectrum using the prescription of \citet{Smith03}.

From Table \ref{table:tile} is is apparent that we are forced to
simulate small simulation cubes at low redshift, which reflects the
need to achieve subarcminute angular resolution while reproducing the
volume of the light cone above $z_{\rm min}=0.2$.  However, there is a
danger that small cubes misrepresent the evolution of structure if the
fundamental mode of the simulation goes non-linear, because of mode
coupling to smaller scales that occurs during nonlinear
evolution.  However, note that the nonlinear scale $r_{\rm nl}=2\pi/k_{\rm
  nl}$ at $z=0$ defined by \beq \int_0^{k_{\rm nl}}\Delta^2_{\rm
  mass}(k)d\ln k = 1, \eeq is $r_{\rm nl}=20.3 \ \ \hMpc$ for the
cosmological model used in this paper, where $\Delta^2_{\rm mass}$ is
the dimensionless linear power spectrum.  Accordingly, because even
our smallest cubes $L_{\rm box}=40.0$ for $z<z_{\rm min}$ are larger
than the nonlinear scale, finite box effects should not be an issue. 
% Knebe paper changed removed this part.
%Knebe (2002) has investigated the effect of simulating cubes smaller 
%than the 
%nonlinear scale on the mass function. He concluded that it is safe to 
%simulate small boxes for structures with characteristic size smaller than the 
%box being simulated, and further even the box to box scatter of the most 
%massive halos is roughly the same as the scatter among the most massive halos 
%in much larger simulations. 

\begin{table}
  \begin{center}
    \caption{Tiling Solution\label{table:tile}}
    \begin{tabular}{cccc}
%      \tablevspace{3pt}
      \hline
      \hline
   $z_{\rm rear}$ \quad  & \quad $L_{\rm box}$ \quad            & \quad $L_{\rm grid}$    \quad        & \quad $M_{\rm part}$  \\
      & \quad $\left(\hMpc\right)$ \quad & \quad $\left(\hkpc\right)$ \quad  & \quad $\left(10^8 \  \hmsol\right)$    \\ 
      \hline
      4.00 & \quad 340.7 & \quad   665 & \quad 1933.1 \\ 
      3.37 & \quad 317.7 & \quad   621 & \quad 1567.6 \\ 
      2.88 & \quad 296.3 & \quad   579 & \quad 1271.2 \\ 
      2.49 & \quad 276.3 & \quad   540 & \quad 1030.9 \\ 
      2.18 & \quad 257.6 & \quad   503 & \quad 836.0  \\ 
      1.92 & \quad 240.3 & \quad   469 & \quad 677.9  \\ 
      1.70 & \quad 224.0 & \quad   438 & \quad 549.7  \\ 
      1.52 & \quad 208.9 & \quad   408 & \quad 445.8  \\ 
      1.37 & \quad 194.8 & \quad   381 & \quad 361.5  \\ 
      1.23 & \quad 181.7 & \quad   355 & \quad 293.2  \\ 
      1.11 & \quad 169.4 & \quad   331 & \quad 237.7  \\ 
      1.01 & \quad 158.0 & \quad   309 & \quad 192.8  \\ 
      0.92 & \quad 147.3 & \quad   288 & \quad 156.3  \\ 
      0.84 & \quad 137.4 & \quad   268 & \quad 126.8  \\ 
      0.77 & \quad 128.1 & \quad   250 & \quad 102.8  \\ 
      0.71 & \quad 119.5 & \quad   233 & \quad  83.4  \\ 
      0.65 & \quad 111.4 & \quad   218 & \quad  67.6  \\ 
      0.60 & \quad 103.9 & \quad   203 & \quad  54.8  \\ 
      0.55 & \quad 96.9 & \quad   189 & \quad  44.5  \\ 
      0.51 & \quad 90.4 & \quad   176 & \quad  36.1  \\ 
      0.47 & \quad 84.3 & \quad   165 & \quad  29.2  \\ 
      0.43 & \quad 78.6 & \quad   153 & \quad  23.7  \\ 
      0.40 & \quad 73.3 & \quad   143 & \quad  19.2  \\ 
      0.37 & \quad 68.3 & \quad   133 & \quad  15.6  \\ 
      0.34 & \quad 63.7 & \quad   124 & \quad  12.6  \\ 
      0.32 & \quad 59.4 & \quad   116 & \quad  10.3  \\ 
      0.29 & \quad 55.4 & \quad   108 & \quad   8.3  \\ 
      0.27 & \quad 51.7 & \quad   101 & \quad   6.7  \\ 
      0.25 & \quad 48.2 & \quad    94 & \quad   5.5  \\ 
      0.23 & \quad 44.9 & \quad    88 & \quad   4.4  \\ 
      0.22 & \quad 41.9 & \quad    82 & \quad   3.6  \\ 
      0.20 & \quad 40.0 & \quad    78 & \quad   3.1  \\ 
      0.19 & \quad 40.0 & \quad    78 & \quad   3.1  \\ 
      0.17 & \quad 40.0 & \quad    78 & \quad   3.1  \\ 
      0.16 & \quad 40.0 & \quad    78 & \quad   3.1  \\ 
      0.14 & \quad 40.0 & \quad    78 & \quad   3.1  \\ 
      0.13 & \quad 40.0 & \quad    78 & \quad   3.1  \\ 
      0.12 & \quad 40.0 & \quad    78 & \quad   3.1  \\ 
      0.10 & \quad 40.0 & \quad    78 & \quad   3.1  \\ 
      0.09 & \quad 40.0 & \quad    78 & \quad   3.1  \\ 
      0.07 & \quad 40.0 & \quad    78 & \quad   3.1  \\ 
      0.06 & \quad 40.0 & \quad    78 & \quad   3.1  \\ 
      0.05 & \quad 40.0 & \quad    78 & \quad   3.1  \\ 
      0.03 & \quad 40.0 & \quad    78 & \quad   3.1  \\ 
      0.02 & \quad 40.0 & \quad    78 & \quad   3.1  \\ 
      0.01 & \quad 40.0 & \quad    78 & \quad   3.1  \\ 
      \hline
    \end{tabular}
  \end{center}
  \footnotesize NOTES.--- Parameters for the 46 simulation cubes in
  our tiling solution for the $\Lambda$CDM model. The sequence
  comprises output from 32 unique PM simulations (the last 14 tiles
  recycle the same simulation).  The column column $z_{\rm rear}$
  corresponds to the redshift at which the particle distributions of
  each tile are output. Is is also the source redshift the densely
  spaced convergence planes $\kappa(\hat{\bf n},z_{\rm rear})$. The
  size of the simulation cube used for that tile (in comoving $\hMpc$), the
  grid spacing (in comoving $\hkpc$), and the particle mass (in $\hmsol$) are
  also given.
\vskip -0.5cm
\end{table}

\subsection{Group Catalogs}

For each tile used to evaluate the convergence, the particle
distributions are dumped and a halo catalog is produced by running a
``friends--of--friends'' (FOF) group finder\footnote{We used the
  University of Washington NASA HPCC ESS group's publicly available
  FOF code at \url{http://www-hpcc.astro.washington.edu}} \citep[see
  e.g.,][]{Davis85} with the canonical linking length (in units of the
mean interparticle separation) $b=0.2$. The FOF algorithm groups the
particles into equivalence classes by linking together all pairs
separated by less than $b$.  We impose a minimum halo mass of $10^{12.7}
\ \hmsol$, where we follow \citet{Jenkins01} and define the mass of a
halo as the mass of all the particles in the FOF group (with
$b=0.2$). Note that this minimum mass corresponds to different numbers
of particles for different tiles, as our larger tiles have poorer mass
resolution.  A cluster is defined as a halo with $M > 10^{13.5}
\ \hmsol$.  It is clear from Table \ref{table:tile} that although our
mass resolution varies from $M_{\rm part} \sim 3 \times 10^{8} - 2
\times 10^{11} \ \hmsol$ from the smallest tile to the largest, it is
always sufficient to resolve cluster size halos.  As we will see in \S
3, weak lensing selected clusters span the redshift range $0.2
\lesssim z \lesssim 1.0$, so that we also spatially resolve the virial
radii of cluster halos in this range.

Although there exist several methods to define the center of a FOF
group, such as the location of the density (potential) maximum
(minimum), these require an $\sim N_{\rm group}^2$ calculation for each
group, which is not computationally feasible for a large ensemble of
simulations. Instead we opt for a simpler, faster definition of the
group center which scales as $N_{\rm group}$.  A sphere of radius
$r_{\rm sphere}$ is placed at the location of a random particle in the
group, and the center of mass of the particles interior to the sphere
is calculated. Moving the sphere center to this new location, the
procedure is iterated until the sphere center converges to the center
of mass of the particles inside of it. We set $r_{\rm sphere}=0.4 \ 
\hMpc$ of order the size of a cluster virial radius, which gives halo
centers that match the peaks in our lensing maps accurately. This
moving sphere method is more robust than using the center of mass of
all the particles in the FOF group because the latter quantity often
does not coincide with the density maximum for disturbed or elongated
structures.

\begin{figure}[t]
\centering
\centerline{\epsfig{file=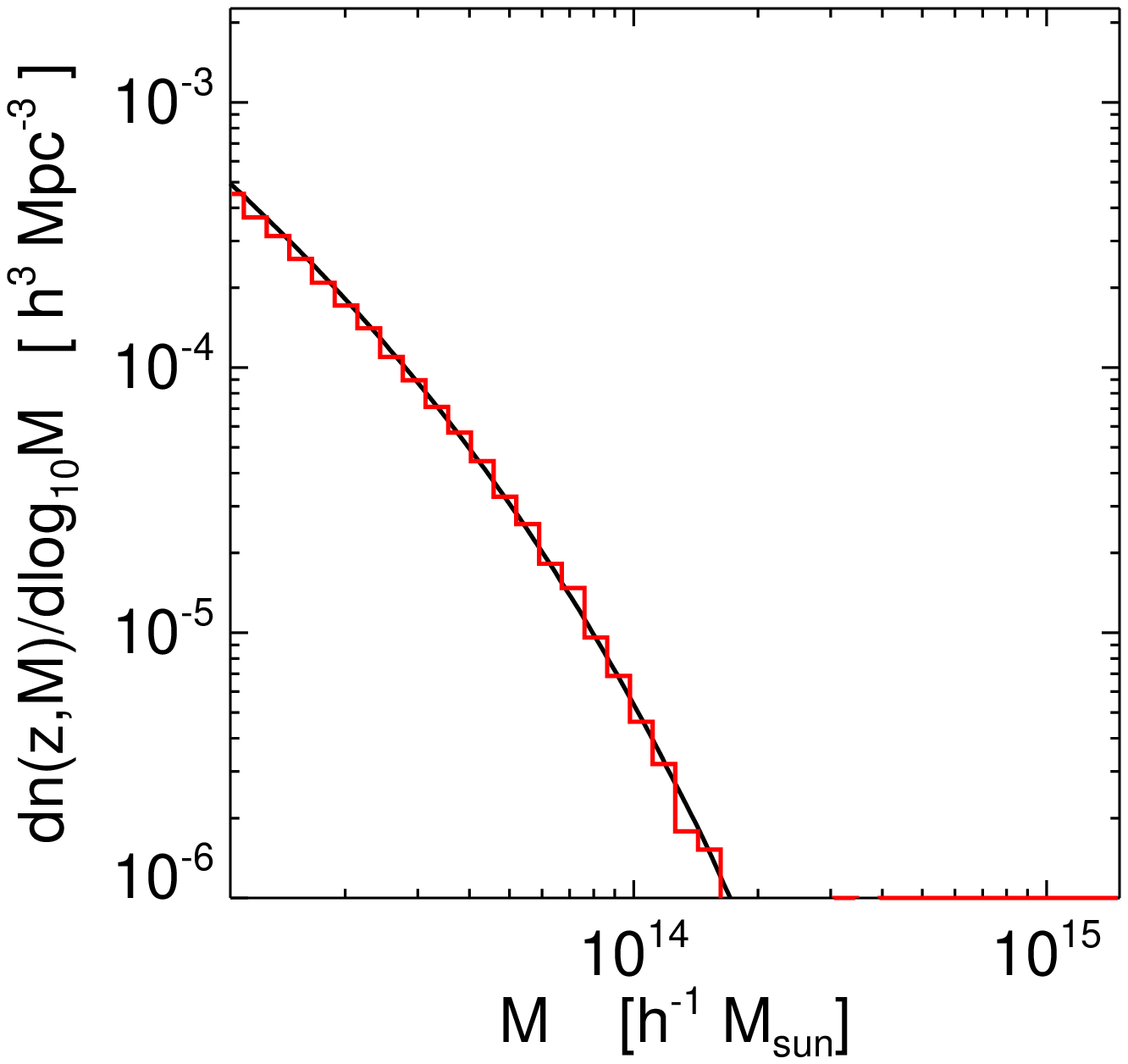,bb=0 30 410 460,width=0.25\textwidth}
\epsfig{file=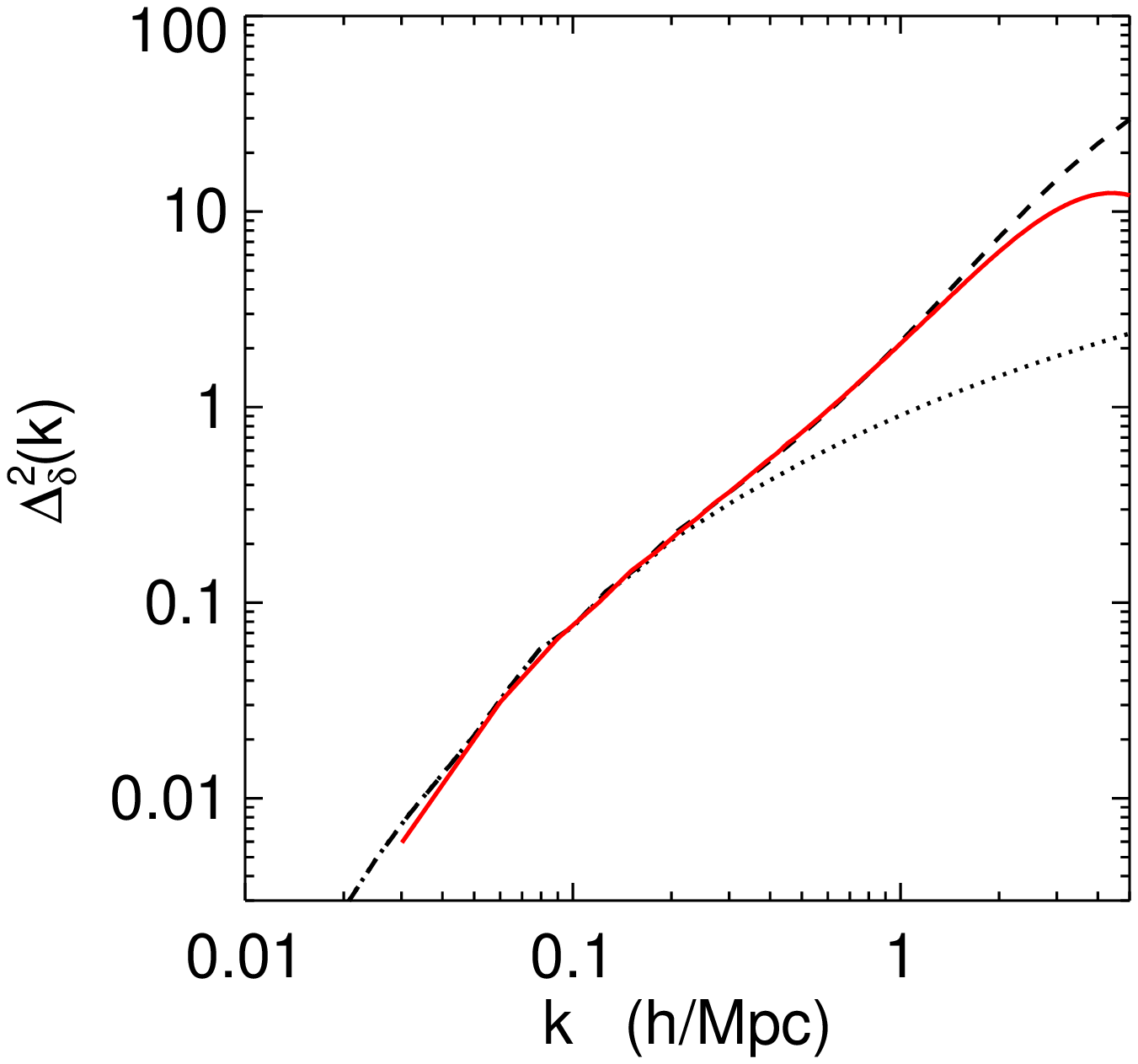,bb=0 30 410 460,width=0.25\textwidth}}
\centerline{\epsfig{file=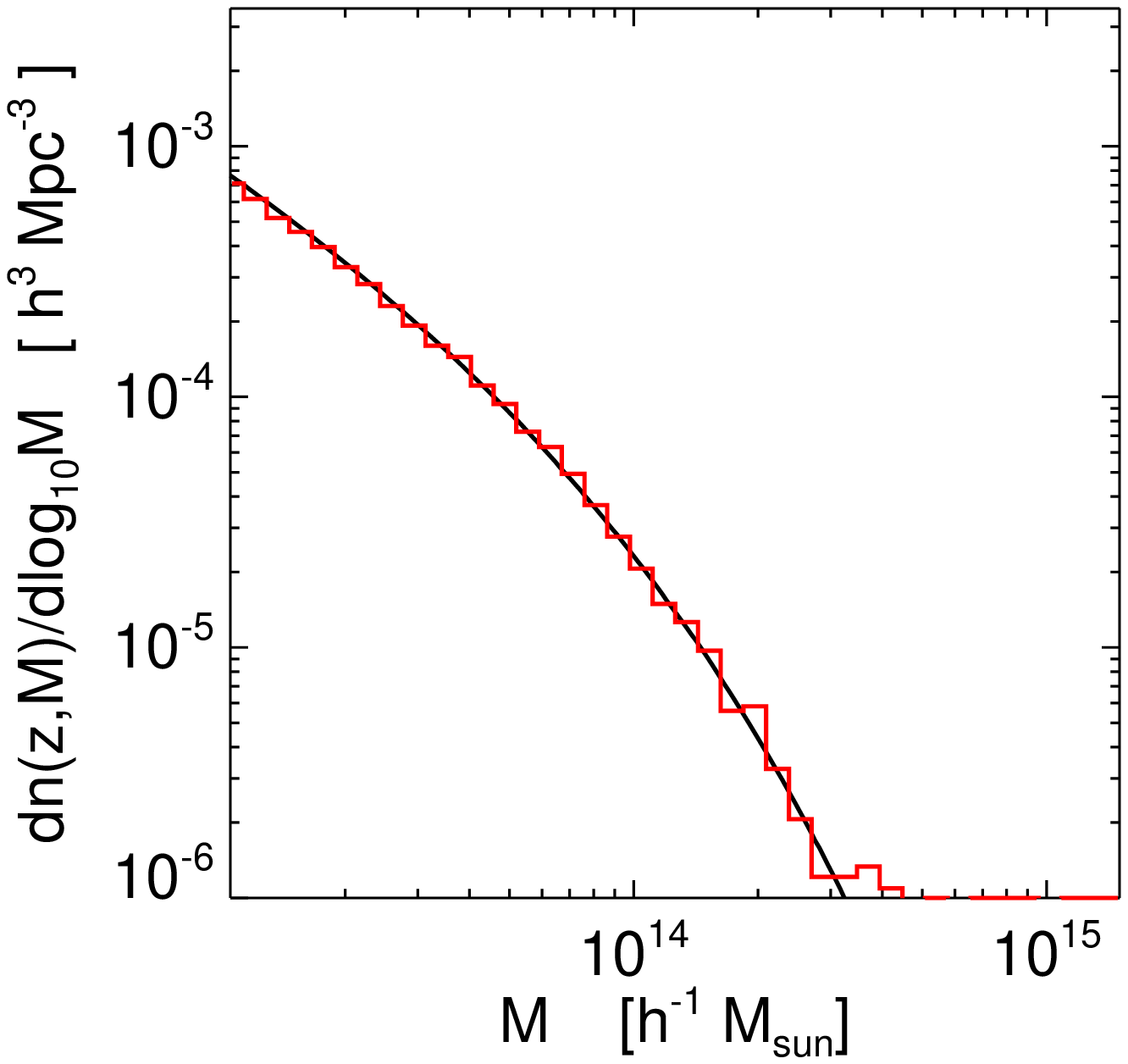,bb=0 30 410 400,width=0.25\textwidth}
\epsfig{file=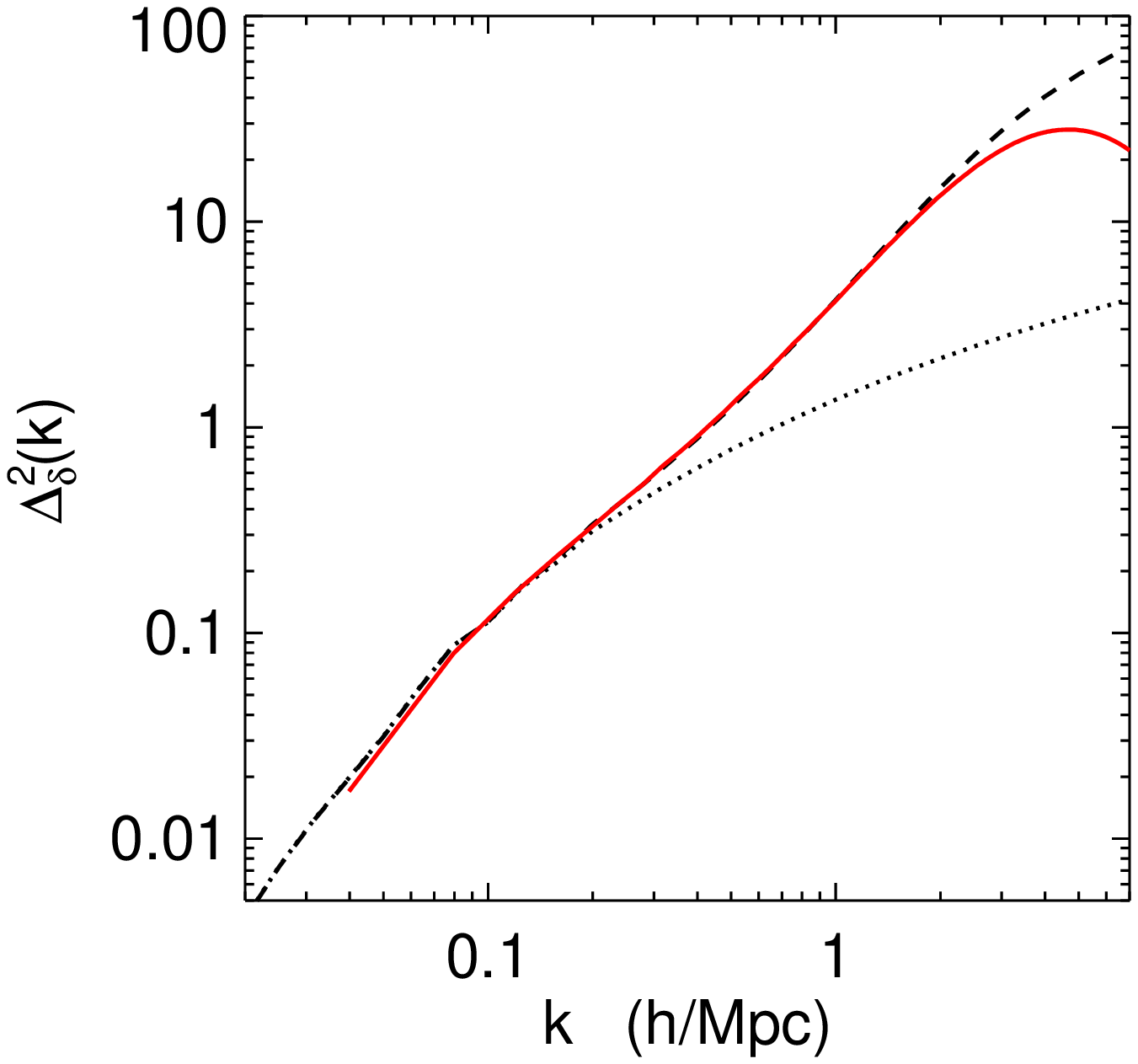,bb=0 30 410 400,width=0.25\textwidth}}
\centerline{\epsfig{file=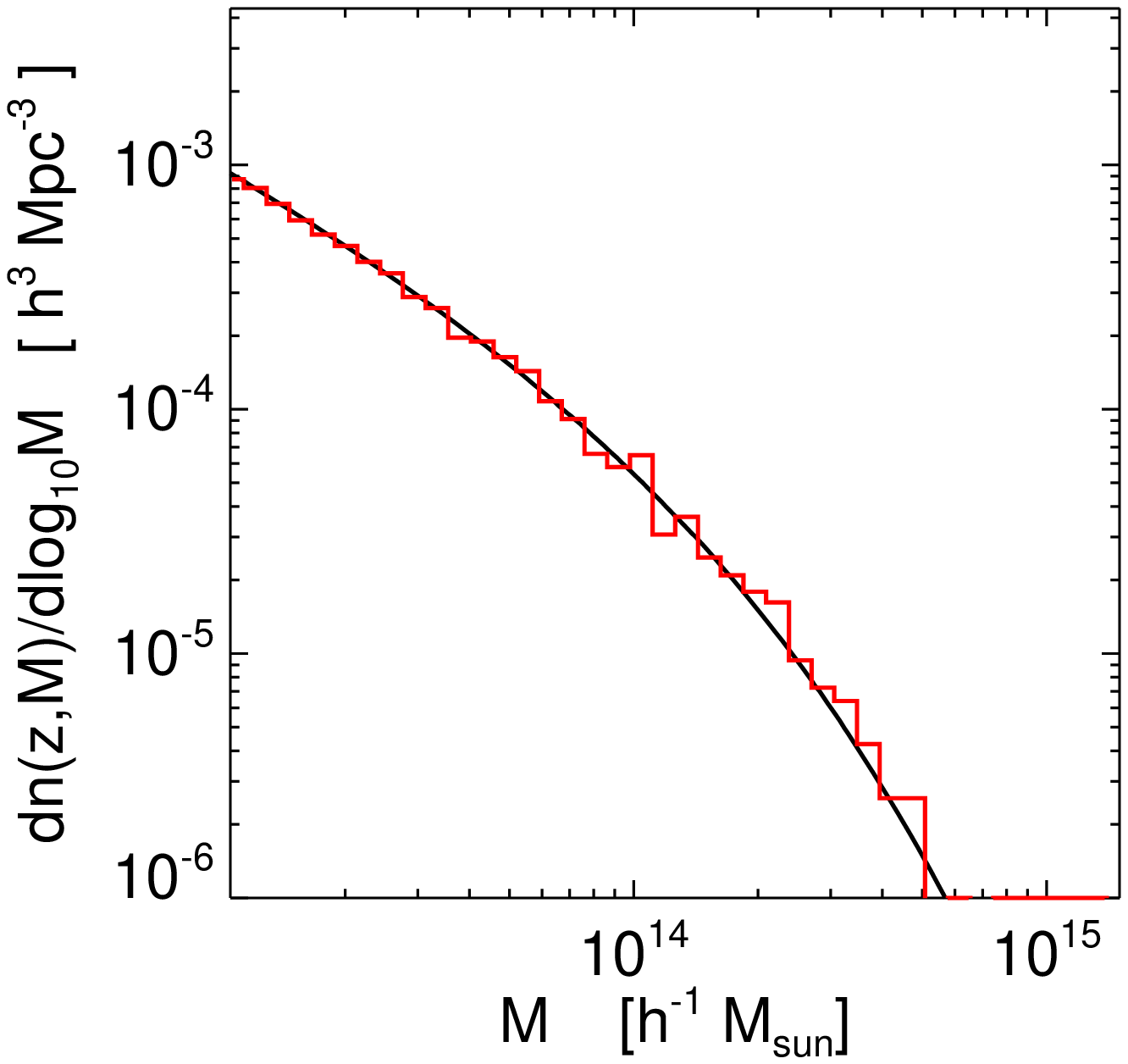,bb=0 30 410 400,width=0.25\textwidth}
\epsfig{file=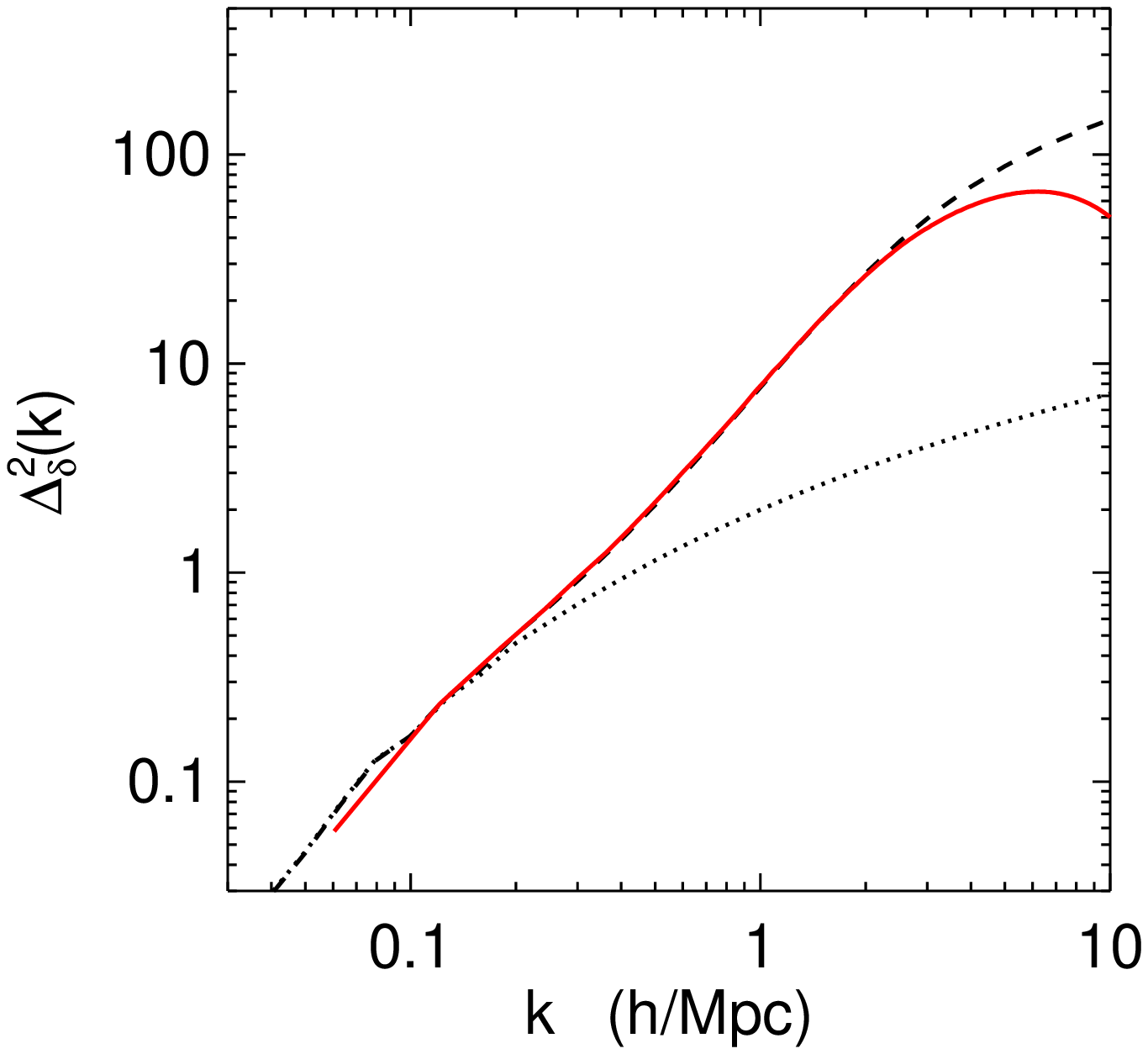,bb=0 30 410 400,width=0.25\textwidth}}
\centerline{\epsfig{file=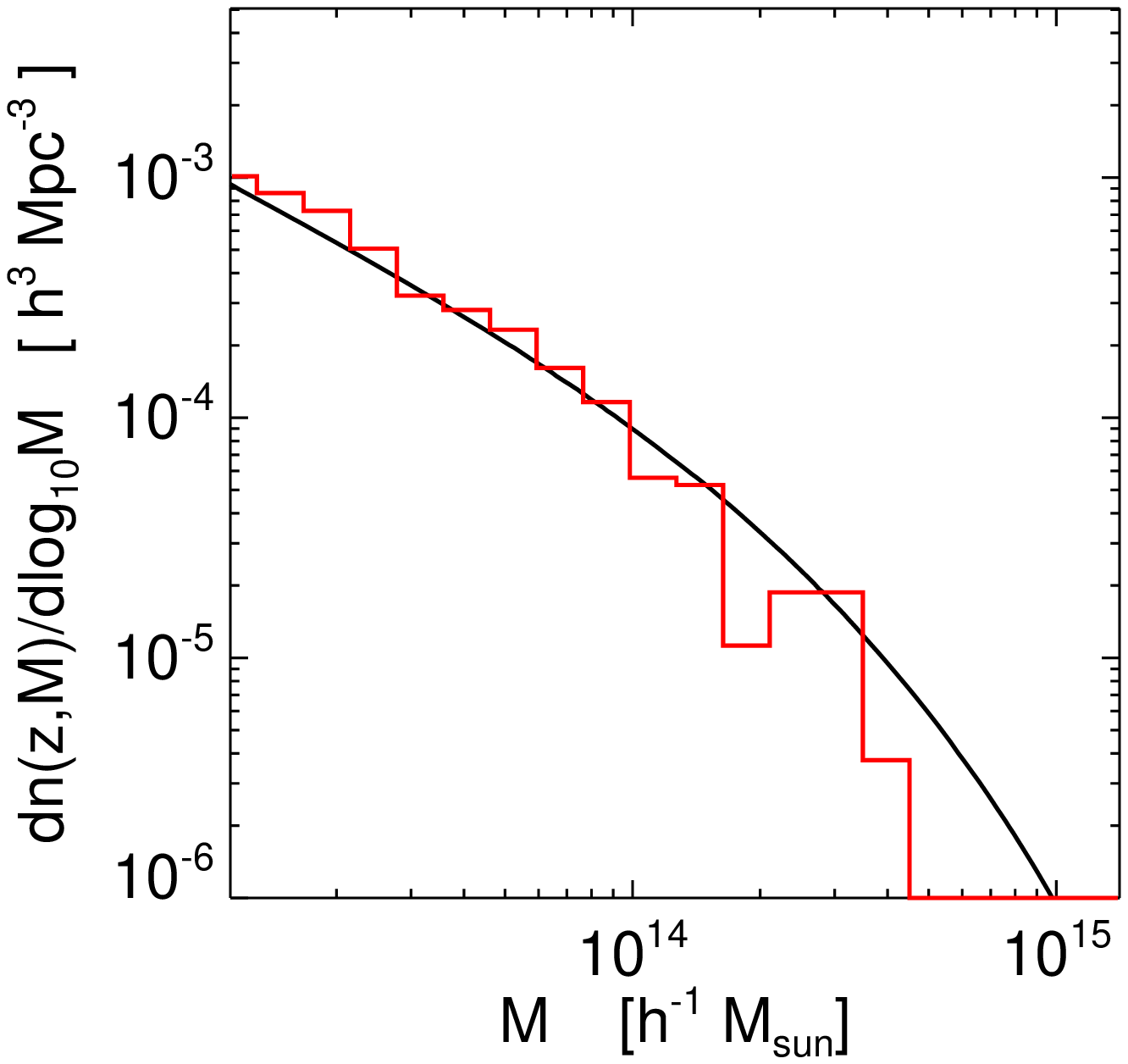,bb=0 30 410 400,width=0.25\textwidth}
\epsfig{file=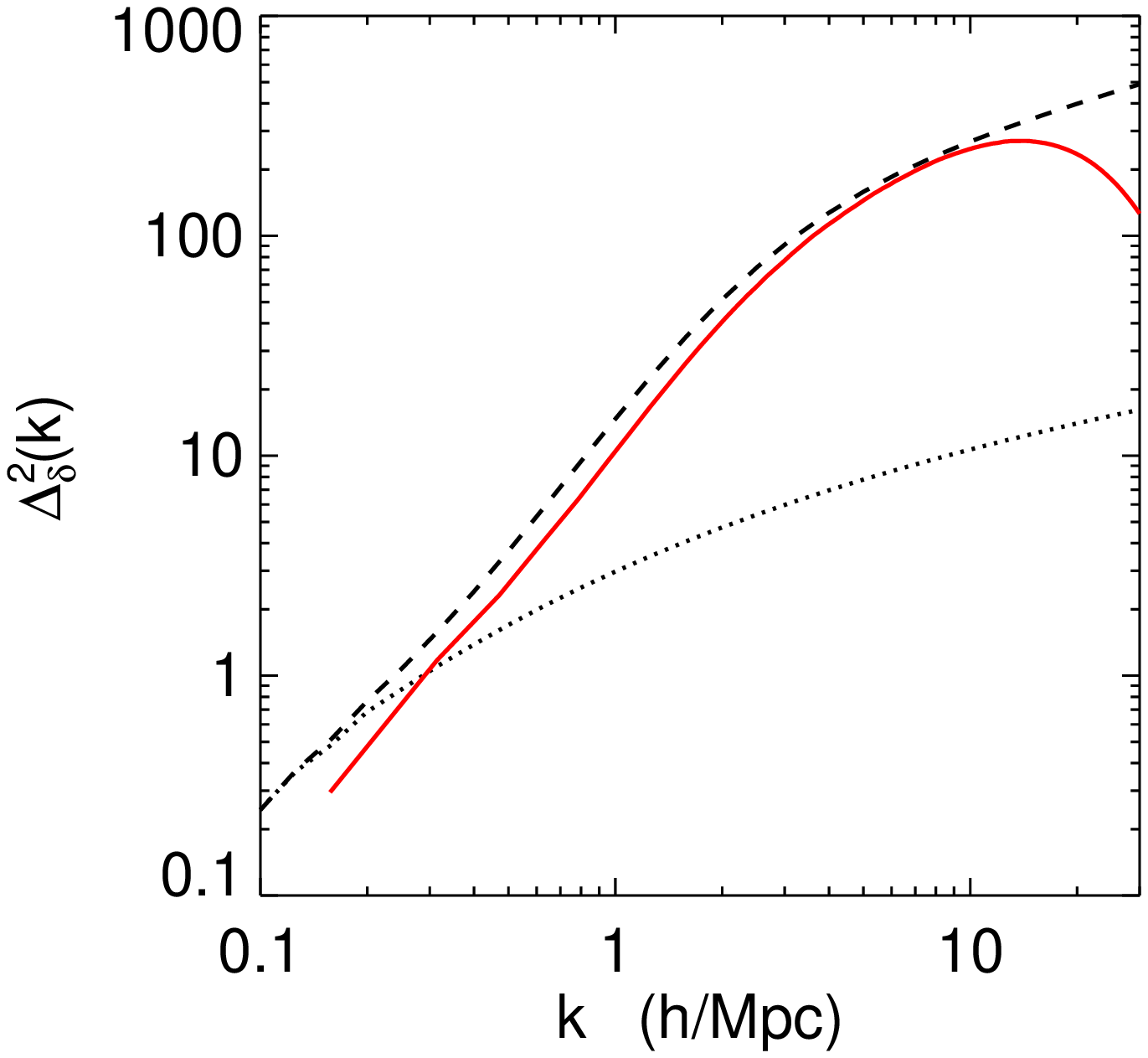,bb=0 30 410 400,width=0.25\textwidth}}
\caption{Mass functions and 3-d power spectra for four tiles. From top
  to bottom the redshift of the tiles are $z_{\rm
    rear}=1.52,1.01,0.60,0.20$ (see Table \ref{table:tile} for simulation
  parameters).  The left panels show the mass function (histogram)
  averaged over 38 different simulations of the same tile compared to
  the fitting formula (solid) of \citet{Jenkins01}.  The right panels
  show the 3-d power spectra of the density field for each tile. Solid
  lines are the power spectra averaged over 38 simulations. Dashed
  curves show the nonlinear power spectra from the analytical fitting
  formulae of \citet{Smith03} and dotted curves are the linear theory
  power spectra.
\label{fig:tiling}
}
\end{figure}

\begin{figure}[t]
\centering
\centerline{\epsfig{file=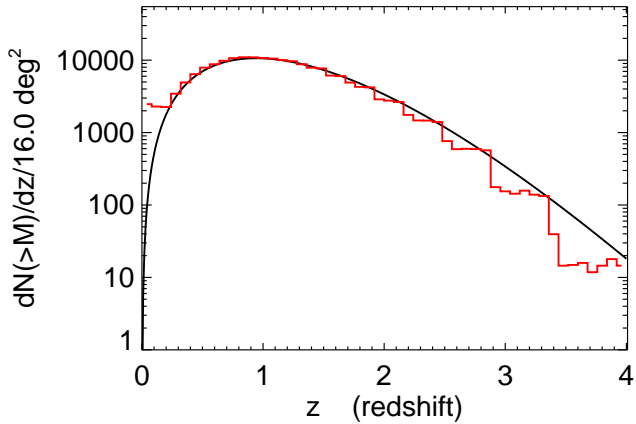,width=0.5\textwidth}}
\caption{
  Total number of clusters above $M=10^{13.5} \ \hmsol $ in a $16 \ 
  {\rm deg}^{-2}$ patch of sky as a function of redshift. The
  histogram shows the redshift distribution averaged over 38
  simulations. The solid line is the redshift distribution obtained
  from the fitting formula of \citet{Jenkins01}.}
\label{fig:dndz}
\end{figure}

The left panel of Figure \ref{fig:tiling} compares the differential
mass function for several tiles, again averaged over 38 simulations,
with the universal mass function fitting formula of
\citet{Jenkins01}. The close agreement between the mass function in
the tiles and the fits indicate that our simulations robustly
reproduce the mass function down to $10^{13}\ \hmsol$.  Figure
\ref{fig:dndz} shows the cumulative number of halos in the light cone,
$dN(> M=10^{13} \ \hmsol)/dz$ as a function of redshift, averaged over
the 38 $16 \ {\rm deg}^2$ fields, again compared to fitting
formula. Because we have been careful to accurately reproduce the
volume of the light cone, the total number of clusters as a function
of redshift is reproduced for $0.2 \lesssim z \lesssim
2.5$. Below $z_{\rm min}=0.2$, we overproduce clusters because of the
excess volume of our simulations where our tiling becomes
inefficient. Above $z \sim 2.5$, there is a tendency to underproduce
collapsed objects because the resolution of our larger tiles
approaches the size of the virial radius $r_{\rm vir} \simeq 1.0
\ \hMpc$ of clusters. As the efficiency for lensing is only
appreciable for $0.2 \lesssim z \lesssim 1.0$, the overproduction of
clusters at $z \lesssim 0.2$ and under production for $z \gtrsim 2.5$
will not effect our conclusions on clusters from weak lensing.

\subsection{Mock Observations}

The observables in weak lensing observations are the ellipticities of
background source galaxies which have been distorted by the foreground
matter distribution. Even for the ideal case of no instrumental
noise, observations are limited by the intrinsic ellipticities of the
sources and the finite number of Poisson distributed sources on the
sky. Here we simulate this ideal case by drawing a number density $n$
of source galaxies from the source redshift distribution 
\beq
p_z(z)=\frac{1}{2z_0^3}z^2 e^{-z/z_{0}}, \label{eqn:p_z} 
\eeq
and placing them at random positions on our $4^{\circ} \times
4^{\circ}$ simulated field. This redshift distribution peaks at
$2z_0$, has mean redshift $\langle z \rangle=3z_0$, and has been used
in previous studies of cosmic shear \citep{Witt00}. A fraction
$f_z$ of source galaxies will have photometric redshifts, where the
errors are drawn from a Gaussian distribution with dispersion
$\sigma_z=\delta_z (1+z)$ and added to the source
redshifts. We take $n=40 \ {\rm arcmin}^{-2}$, $z_0=0.50$
corresponding to a peak redshift of $z=1.0$, and $\delta_z=0.12$.
Henceforth in this work we set $f_z=1$ and assume all source galaxies
have photometric redshifts in order to determine the efficacy of our
tomographic and adaptive matched filtering techniques for the best
case. 

%In a future paper, \citet{Me04}, we explore the
%effects of different number densities, redshift distributions, and
%photometric redshift parameters on the statistics of shear selected
%clusters.

%consistent with expectations for future deep wide field weak lensing observations (Tony Tyson, private communication). 

The intrinsic ellipticities, $\epsilon_{\rm int}$, of the source
galaxies are drawn from a Gaussian distribution with a single
component rms dispersion $\gamma_{\rm rms}=0.165$ \citep{BJ02}.  Note
that there has been some confusion in the literature over the
intrinsic ellipticity, and larger values have been employed in other
studies. The value used here is achievable if the optimal weighting
discussed in \citep{BJ02} is applied to the sources for
the asymptotic case of no instrumental noise (Gary Bernstein, private
communication).

The simulations output convergence planes $\kappa(\hat{\bf n},z_{\rm
  rear})$ for $0\leq z_{\rm rear}\leq 4$ densely spaced in redshift as
listed in Table \ref{table:tile}, and the shear fields
$\gamma(\hat{\bf n},z_{\rm rear})$ are obtained from the convergence
via FFT using eqns. (\ref{eqn:gamma}). A source galaxy at redshift $z$
is sheared by the nearest shear plane with $z_{\rm rear} < z$, via the
weak lensing relation \beq \epsilon=\epsilon_{\rm int} +
\gamma, \label{eqn:epsilon} \eeq which is valid for $\kappa \ll 1$,
$|\gamma| \ll 1$.  The shear is interpolated from the grid onto the
source galaxy position using bicubic spline interpolation.  The end
result is a mock catalog of source galaxy ellipticities with
photometric redshifts, which we use to explore the statistics of
clusters below.

In what follows we will want to compare mock observations with the
noise properties described above with \emph{noiseless} weak lensing
observations in order to assess the intrinsic limitations of
tomography and weak lensing searches for clusters.  We define
noiseless observations to be an infinite number of source galaxies
with zero intrinsic ellipticity and photo-z error (i.e. $\gamma_{\rm
  rms}=0.0$ and $\delta_z=0$) drawn from the source redshift
distribution in eqn.~(\ref{eqn:p_z}), and placed on the simulation
grid rather than at random positions.  Placing the source galaxies on
the grid removes the Poisson clustering of the source galaxies, so
that the angular resolution is then limited by the resolution of the
simulation rather than shot noise. Because we cannot simulate an
infinite number of sources, we draw 50 galaxies from eqn.
(\ref{eqn:p_z}) for each of the $512^2$ grid points, which we find is
sufficient to converge to the limiting case of no noise. The left
panel of Figure \ref{fig:demo} shows a Kaiser-Squires reconstruction
\citep{KS93} of the mean convergence field for noiseless data.

\subsection{Matching Peaks with Clusters}

Given a mock catalog of source galaxy ellipticities, we construct
smooth maps by convolving the data with some filters. Candidate mass
selected clusters will correspond to the peaks in these smoothed maps.
We elaborate on the filtering techniques in the next section, but
focus here on the details of matching the peaks in smoothed maps with
clusters from the simulation tiles. We locate peaks in the maps using
a simple algorithm whereby a peak is identified if it is higher than
all of its neighboring pixels.  While more complex algorithms exist to
find peaks in pixelised data, because our maps have already been
smoothed, searching for local maxima is sufficient.  Given a list of
peaks from the smoothed map and the cluster catalog constructed by
applying FOF to the simulation tiles, we can correlate the peaks with
the cluster catalog and locate all halos along the line of sight of a
given peak.  A peak in a map is considered a cluster detection if
there is a cluster in the light cone within an aperture $\theta_{\rm
  match}=3^{\prime}$ centered on the peak. Although it might seem more
appropriate to set this aperture size to the field of view of some
`follow up' instrument, $3^{\prime}$ is well matched to the angle
subtended by the virial radius of a typical cluster at $z=0.4$, a
typical redshift for a shear selected cluster. Further our interest
here is in the primary object responsible for the lensing signal and
not objects at larger angular separation that happen to be in the
vicinity of the peak.  In the event that there are multiple clusters
within the aperture centered on the peak, we take the most massive
cluster to be the `match' to that peak.  Depending on the method of
follow up, the object most likely to be detected will be some function
of mass and redshift, though we neglect this subtlety here. Finally,
to avoid the problem of multiple peaks in the map corresponding to the
same cluster, we keep only the highest peak in an aperture
$\theta_{\rm iso}=1^{\prime}$ centered on each peak, so that less
significant peaks near higher peaks are discarded.

%Though it should be noted that the efficiency defined in eqn.~(\label{eqn:eff}) will  increase if a larger matching aperture is used. Finally, to avoid the problem of multiple peaks in the map corresponding to the same cluster, we keep only the highest peak in an aperture $\theta_{\rm iso}=1^{\prime}$ centered on each peak, so that less significant peaks near higher peaks are discarded.

\section{Tomography and Matched Filtering}

Previous studies \citep{Reb99,WvWM02,Paddy03,HTY03}, have neglected
the extra information provided by photometric redshifts of source
galaxies when searching for clusters in weak lensing data.  Without
redshift information, the source galaxy ellipticities provide a noisy
measure of the \emph{mean shear} 
\beq {\bar \gamma}(\hat{\bf n})
\equiv \int dz p_z(z)\gamma(\hat{\bf n},z), \label{eqn:gam_bar} 
\eeq
which is the shear out to a given redshift averaged over the source
redshift distribution $p_{z}(z)$. Knowledge of the source redshift
enables one to measure the shear more accurately, and in this section
we present a tomographic matched filtering scheme which fully
incorporates this extra information.

The tomographic matched filtering (TMF) technique is similar in spirit
to matched filtering algorithms used to find clusters in optical
surveys \citep{Postman96,Kepner99,WK02,Koch03} and is also
qualitatively similar to the maximum likelihood techniques developed
to study cluster mass profiles from weak lensing
\citep{GS98,SKE00,KS01}. In what follows we present a well-defined
procedure for identifying clusters of galaxies in weak lensing surveys
that makes optimal use of both the shape and photometric redshift
information of the background source galaxies.  It can be applied to
weak lensing data for which only the overall source redshift
distribution is known, for which photometric redshifts of the source
galaxies are available, as well as combinations of the two scenarios
(i.e. a fraction of galaxies with photometric redshifts and the rest
with only shapes).  Furthermore, we will see in \S 4 that a few source
redshift bins perform nearly as well as full photometric redshift
information, so simple color cuts and apparent magnitude priors can be
used to extract the extra tomographic information.
 
The matched filter identifies clusters in weak lensing data by finding
the peaks in a cluster likelihood map generated by convolving the
source galaxy ellipticities with a filter that models the distortion
caused by the foreground mass distribution. The peaks in the
likelihood map will correspond to the locations where the match
between the cluster model and the data is maximized, hence giving the
two dimensional location of the cluster.  In addition, the algorithm
produces a tomographic estimate of the cluster redshift using the
lensing signal, similar to the tomographic technique used by
\citet{Witt01,Witt03} and \citet{Tay04}.  

%Parameters of the cluster
%profile, such as a scale angle and amplitude, may also be extracted in
%principle \citep{GS98,SKE00,KS01}.
%( see also Geiger \& Schneider 1998;Schneider et al. 2001;
%King \& Schneider; King \& Schneider, \& Erben 2002).

\subsection{Formalism}

        The goal of the matched filter is to match the data to a model 
that describes the  distortion of background galaxies by the foreground mass 
distribution, here a galaxy cluster.  For the sake of generality, we 
parameterize the cluster lens by three quantities, a cluster redshift 
$z_{\rm d}$, an angular scale $\theta_{\rm s}$, and an amplitude $A$. 

For a cluster at redshift $z_{\rm d}$ the convergence for a source
galaxy at position $\vec{\theta}_j$ and redshift $z_j$ can be written
\beq  
\kappa(\vec{\theta}_j,z_j) = Z(z_j;z_{\rm d})\kappa_{\rm \infty}(\vec{\theta}_j) 
\eeq 
where 
\beq
Z(z;z_{\rm d}) \equiv \frac{D_{\rm ds}}{D_{\rm s}}H(z-z_{\rm d}),
\label{Z} \label{eqn:bigZ} 
\eeq 
and 
\beq
\kappa_{\rm \infty}(\vec{\theta}_j) =\frac{\Sigma(\vec{\theta}_j)}
{c^2D_{\rm d}\slash 4\pi G}. 
\eeq  
Here $\Sigma_{\rm crit}(z_{\rm d},z_{\rm s})=\frac{c^2}{4\pi G}\frac{D_{\rm
    s}}{D_{\rm d}D_{\rm ds}}$ is the critical density for lensing with
$D_{\rm d}$, $D_{\rm ds}$, $D_{\rm s}$ the deflector,
deflector-source, and source distances, respectively. The Heaviside
step function accounts for the fact that sources in the foreground of
the deflector are not lensed, and $\kappa_{\infty}$ is the convergence
for a hypothetical source at infinite distance \citep[see e.g. ][for similar 
treatments]{SS97,KS01}. The same relationship holds for
the shear, $\gamma(z)=Z(z;z_{\rm d})\gamma_{\infty}$. For a flat
universe, eqn.~(\ref{eqn:bigZ}) simplifies to 

\beq Z(z;z_{\rm d}) =
\max\left(1-\frac{D_{\rm d}}{D_{\rm
    z}},0\right).  \label{eqn:bigZ_final} 
\eeq

In the weak lensing regime $\kappa \ll 1$, $|\gamma| \ll
1$, the observed complex ellipticity of a source galaxy is related to
the shear field by eqn.~(\ref{eqn:epsilon}).  For a spherically
symmetric mass distribution centered about the position
$\vec{\theta_0}$, the shear will be entirely tangential \beq
\gamma(\theta)=-\gamma^{T}(\vec{\theta};\vec{\theta_0})e^{2i\phi}
\eeq where $\phi$ is the azimuthal angle about the position
$\vec{\theta_0}$. We employ the standard definition of the tangential
shear
\bea
\gamma^{T}(\vec{\theta};\vec{\theta_0}) &&\equiv -
[\gamma_1\cos(2\phi)+\gamma_2\sin(2\phi)]\nonumber \\ 
&& =-\Re[\gamma(\vec{\theta}+\vec{\theta_0})e^{-2i\phi}]
\eea
and analogously for the tangential component of $\epsilon$, and hence
the tangential ellipticity is related to the tangential shear by
$\epsilon^{T} = \epsilon_{\rm int}^{T}+\gamma^{T}$.  For our model
cluster lens we take the mass distribution to be spherically
symmetric, and henceforth work primarily with the tangential shear and
tangential ellipticity.

%We parameterize the cluster lens with the three parameters  $(z_{\rm d},\theta_s,A)$, 
%where $z_d$ is the cluster redshift, $\theta_s$ is a scale angle of the cluster profile, 
%and $A$ is the surface density in units of the critical surface density for source galaxies at
%infinity. 

For a cluster centered at position $\vec{\theta_0}$ with the parameter vector
$(z_{\rm d},\theta_s,A)$ the convergence can be written 
\bea
\kappa_{\rm model}(\vec{\theta}_j,z_j;z_{\rm d},\theta_{\rm s},A) &&= \frac{\Sigma_{\rm model}}{\Sigma_{\rm crit}} \\
 && = Z(z_j;z_{\rm d}) A \ K(x_j), 
\nonumber
\eea 
where $x_j\equiv|\vec{\theta_j}-\vec{\theta_0}|/\theta_{\rm s}$, $K(x)$ is the cluster 
convergence profile in units of $\theta_s$, and the amplitude $A$ is given by
\beq
A\equiv \frac{\Sigma_0 D_{\rm
    d}}{c^2/4\pi G},   
\eeq
where $\Sigma_0$ is the surface density normalization. 

The tangential shear for this mass distribution is 
\beq 
\gamma_{\rm model}(\vec{\theta}_j,z_j;z_{\rm d},\theta_s,A)=Z(z_j;z_{\rm d}) A
\ G(x_j) \label{eqn:model} 
\eeq 
where $G$ is related to $K$ by
\beq
G(x) = \frac{2}{x^2}\int_0^{x}K(y)y dy-K(x), \label{eqn:GfromK} 
\eeq 
from the relation
$\gamma=\bar{\kappa}-\kappa$.  We omit the tangential superscript on
$\gamma_{\rm model}$ and $\epsilon_{\rm model}$ for notational
simplicity.

The probability of measuring a source galaxy with tangential
ellipticity $\epsilon^T_j$ at position $\vec{\theta}_j$ with source
redshift $z_j$, given the lens model in eqn.~(\ref{eqn:model}) is 
\beq
P_j=\frac{1}{\sqrt{2\pi\sigma_j^2}}\exp\left[-\frac{1}{2}\frac{(\epsilon^T_j-\gamma_{\rm
      model})^2}{\sigma_j^2}\right] 
\eeq 

The measurement error (assumed to be Gaussian) $\sigma_j^2$ in principle
includes contributions from the intrinsic ellipticities of the source galaxies
as well as an error term due to the photometric redshift error of the
source galaxy at $z_j$, but in practice the former will dominate because
of the the slow variation of the function $Z\left(z;z_d\right)$ in
eqn.~(\ref{eqn:bigZ_final}) with source redshift.

If the only knowledge of the source redshift comes from the aggregate
source redshift distribution, one must work with the first moment of
$Z(z)$ \citep{SS97,KS01},
$\langle Z\rangle=\int dz
\ p_{z}(z)Z(z;z_{\rm d})$ where $p_{z}$ is the
source redshift distribution in eqn.~(\ref{eqn:p_z}). In this case,
$\langle Z\rangle$ must be substituted for $Z$ in
eqn.~(\ref{eqn:model}).
  
The likelihood of finding a cluster at position $\vec{\theta_0}$ with
parameters $(z_{\rm d},\theta_s,A)$, is the product over
all the data of the individual probabilities 
\beq
\mathcal{L}(\vec{\theta_0},z_{\rm
  d},\theta_s,A)=\prod_{j}P_j 
\eeq 
and the log likelihood is
\beq
\ln\mathcal{L}=-\frac{1}{2}\sum_{j}\ln(2\pi\sigma_j^{2})-\frac{1}{2}\sum_{j}
\frac{(\epsilon^T_j-\gamma_{\rm model})^2}{\sigma_j^2} \label{eqn:lhood}. 
\eeq 

Expanding the expression in eqn.~(\ref{eqn:lhood}) and dropping 
terms that do not depend on the parameter vector gives
%\beq
%\ln\mathcal{L}=-\frac{1}{2}\sum_{j}\frac{\left([\epsilon^T_j]^2-
%2\epsilon^T_j\gamma_{\rm
%    model}+\gamma_{\rm model}^2\right)}{\sigma_j^2}
%\label{lhood_ex}
%\eeq 
%Dropping the first term of this equation, which is a constant, the log 
%likelihood becomes 
\beq
\ln\mathcal{L}=\frac{1}{2}\sum_{j}\frac{(2\epsilon^T_j\gamma_{\rm
    model}-\gamma^2_{\rm
    model})}{\sigma_j^2}. \label{eqn:lhood_expand} 
\eeq
This expression has a linear dependence on $A$, which can be exploited
to lower the dimensionality of the fit. Setting the derivative of
eqn.~(\ref{eqn:lhood_expand}) with respect to $A$ to zero gives 
\beq
A(z_{\rm d},\theta_s)=\frac{\sum_j\frac{1}{\sigma_j^2}\epsilon^T_j Z
  G}{2\sum_j\frac{1}{\sigma_j^2} Z^2 G^2} \label{eqn:A} 
\eeq
Substituting this back into
eqn.~(\ref{eqn:lhood_expand}) finally gives 
\beq
\ln\mathcal{L}(\vec{\theta_0},z_{\rm d},\theta_s)=\frac{1}{2}\sum_{j}
\frac{\epsilon^T_j\gamma_{\rm model}}{\sigma_j^2}.
\label{eqn:lhood_conv}
\eeq 

Thus, the log likelihood is  a convolution of the data with a
`matched filter', with each source galaxy inverse weighted by its
respective error.  The technique is `adaptive' in that it searches
for parameters $(z_{\rm d},\theta_s,A)$ that maximizes the
contrast between the cluster and the background noise, making full use
of the additional information provided by photometric redshifts of
source galaxies, which is reflected in the redshift `weights' used
in eqn.~(\ref{eqn:model}).

%In the event that the scale of the matched filter, $\theta_s$, is kept
%fixed, the convolution in eqn.~(\ref{eqn:lhood_conv}) can be done in
%Fourier space using FFT methods. This will dramatically speed up the
%filtering, as convolving the source galaxies with the matched filter
%in real space is a costly procedure for number densities of typical
%for ground based weak lensing data of $n \gtrsim 200,000
%\ \ \mathrm{deg}^{-2}$.

Ideally, one would perform the convolution in
eqn.~(\ref{eqn:lhood_conv}) with the fast algorithms introduced and
applied to weak lensing by \citet{Paddy03}, which do not require
spatial binning. Here for simplicity, we opt to use FFT methods.
Doing the convolutions with FFT's requires binning the source galaxies
both spatially and in redshift. In the event that the scale of the
matched filter, $\theta_{\rm s}$, is kept fixed when maximizing the
likelihood, then the convolution in eqn.~(\ref{eqn:lhood_conv}) can be
done with a single FFT for each redshift bin, as we see below.

For the spatial binning we follow \citet{SS96} and write the complex
ellipticity at any gridpoint $k$ as 
\beq \epsilon (\vec{\theta_k}) =
\frac{\sum_{j} W_j \epsilon_j}{\sum_{j}W_{j}} \label{eqn:bin} 
\eeq
with 
\beq
W_j=\exp\left(-\frac{|\vec{\theta_j}-\vec{\theta_k}|^2}{2\Delta
  \theta^2}\right) 
\eeq 
where $\Delta \theta$ is a smoothing length.

In choosing the source galaxy redshift bins, we take $p_{z}(z)$ into
account and require each bin to have an equal fraction of the
probability. Because the function $Z\left(z;z_d\right)$ in
eqn.~(\ref{eqn:bigZ_final}) varies slowly with redshift for $z>z_{\rm
  d}$, this binning can be relatively coarse.  We consider two
different binnings: a coarse binnning with only 3 bins, as might be
achieved by using as little as two colors and an apparent magnitude
prior, and a finer set of 5 bins which would require multi color
data. We don't simulate the effects of errors in assigning galaxies to
their respective bins. The spacing of the bins is chosen to contain
equal probability as given by the source redshift distribution in
eqn.~(\ref{eqn:p_z}). The redshift ranges of the bins and the central
redshift are listed in Table \ref{table:bins}.  We will henceforth
refer to the TMF with three coarse bins as the TMF3 and that with five
finer bins as the TMF5.

\begin{table}[t!]
\vskip -0.2cm
\begin{center}
  \caption{Source Redshift Binning \label{table:bins}}
  \begin{tabular}{lccccc}
     %      \tablevspace{3pt}
    \hline
    \hline
Type \  & \ $z_{\rm low}$ \ &\ -- \ & \ $z_{\rm high}$ \  & \ $z_{\rm center}$ \ \\
\hline
         &    0 &\ -- \ & 1.02     & \ 0.70 \ \\
Coarse   & 1.02 &\ -- \ &  1.72    & \ 1.34 \ \\
         & 1.72 &\ -- \ & $\infty$ & \ 2.28 \ \\
\hline
         &  0   &\ -- \ & 0.77     & \ 0.55 \ \\
         & 0.77 &\ -- \ & 1.14     & \ 0.96 \ \\
Fine     & 1.14 &\ -- \ & 1.55     & \ 1.34 \ \\
         & 1.55 &\ -- \ & 2.14     & \ 1.81 \ \\
         & 2.14 &\ -- \ & $\infty$ & \ 2.66 \ \\
\hline
  \end{tabular}
\end{center}
\footnotesize NOTES.--- Source redshift binnings used in this
paper. The ``Coarse'' binning uses three redshift bins, while the
``Fine'' binning uses five. Bins are chosen to contain equal
probability from the source redshift distribution in
eqn. \ref{eqn:p_z}. Lower, upper, and central redshift are denoted by
$z_{\rm low}$, $z_{\rm high}$, and $z_{\rm center}$, respectively.
\end{table}

Combining eqns. (\ref{eqn:A}) and (\ref{eqn:bin})
with the log likelihood in eqn.~(\ref{eqn:lhood_conv}) gives 
\beq
\ln\mathcal{L}(\vec{\theta_0},z_d,\theta_s)=\frac{1}{2\sigma^2}
\frac{\left(\sum_{i}^{n_z+1}Z(z_i;z_d)\sum_{k}^{n_g}\epsilon^T_{ik}
G(x_k)\right)^2}{\sum_{i}^{n_z+1}Z^2(z_i;z_d)\sum_{k}^{n_g}G^2(x_k)} 
\label{eqn:lhood_bin}
\eeq 

Here $\epsilon^T_{ik}$ is the tangential ellipticity about
$\vec{\theta_0}$ at gridpoint $k$ for source galaxies in the $i$th
redshift bin.  The number of grid points and source redshift bins are
denoted by $n_g$ and $n_z$ respectively. The sum over $i$ extends to
$n_z+1$ to indicate that the last bin will be those galaxies that do
not have photometric redshifts, for which the mean value of $\langle
Z\rangle(z_d)$ must be used.  For simplicity we have taken all the
ellipticity errors to be the same, $\sigma\equiv \sigma^2_j$.  This
can be easily generalized to the general case of different errors for
each source galaxy, which would amount to modifying
eqn.~(\ref{eqn:bin}) to take the different errors into account.  Note
that the sum over $k$ in the numerator of this last expression for the
likelihood is just a convolution of the tangential shear with the
`matched filter' $G(x_k)$ which can be done using an FFT.

If we define
\beq
M_i(\vec{\theta_0})\equiv\frac{\sum_{k}^{n_g}\epsilon^T_{ik}G(x_k)}
{\left[\sum_{k}^{n_g}G^2(x_k)\right]^{1/2}},
\eeq
then the likelihood can finally be written 
\beq
\ln\mathcal{L}(\vec{\theta_0},z_d,\theta_s)=\frac{1}{2\sigma^2}
\frac{\left(\sum_{i}^{n_z+1}Z(z_i;z_d)M_i(\vec{\theta_0})\right)^2}
{\sum_{i}^{n_z+1}Z^2(z_i;z_d)} \label{eqn:lhood_final},
\eeq
where $M_i(\vec{\theta_0})$ is the convolution of the ellipticity of
the source galaxies in the $i$th redshift bin with the normalized
kernel $G(x_k)/[\sum_{k}G^2(x_k)]^{1/2}$.  If there is no photometric
redshift information, $n_z=0$, the redshift weight factors of $\langle
Z\rangle^2(z_d)$ in the numerator and denominator of
eqn.~(\ref{eqn:lhood_final}) will cancel, and the log likelihood just
reduces to the square of a single map $M(\vec{\theta_0})$, which is a
convolution of the tangential shear with the normalized kernel.  It is
only this limiting case which has previously been considered by studies
that neglect photometric redshift information
\citep{Reb99,WvWM02,Paddy03,HTY03}.

The expression for the likelihood in eqn.~(\ref{eqn:lhood_final}) can be 
maximized at each point on the sky, allowing one to create a cluster 
likelihood map. Clusters will be located at the peaks in this map and the 
maximum likelihood $z_d$ will be the tomographic estimate for the lens 
redshift.

\begin{figure*}[t]
  \centering \centerline{\epsfig{file=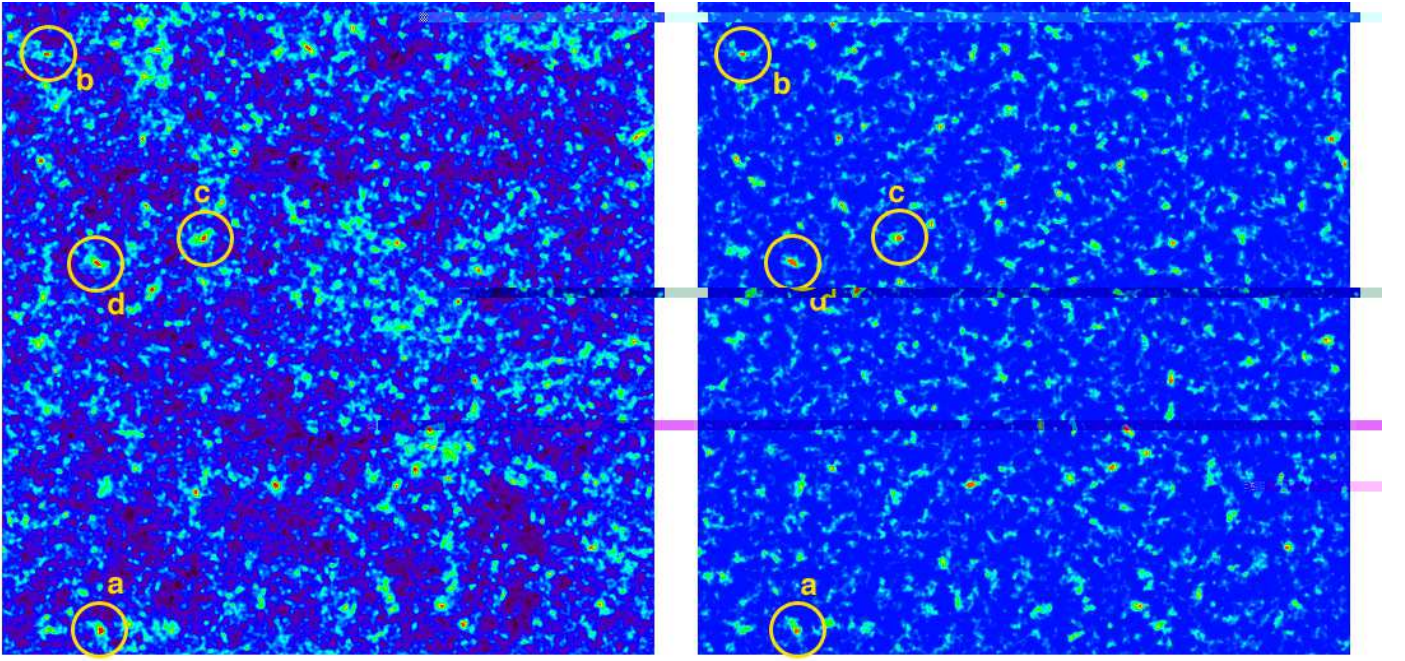,bb= 30 10 1095
      524,width=1.0\textwidth}}
  \caption{\emph{Left panel:} Kaiser-Squires reconstruction of the
    mean convergence field, ${\bar \kappa}$, for noiseless mock
    data. \emph{Right panel:} Cluster likelihood map constructed with
    the TMF applied to noisy data. Source galaxies were binned into
    three coarse redshift bins (TMF3). Cluster statistics and the
    tomographic redshift probability distributions for the four
    clusters labeled \textbf{a}-\textbf{d} are shown in Figure
    \ref{fig:pdf}.}.
  \label{fig:demo}
\vskip -0.1cm
\end{figure*}

An example of a likelihood map constructed from binned source galaxies
is depicted in Figure \ref{fig:demo}. The left panel shows a
Kaiser-Squires reconstruction of the \emph{mean} convergence field
(recall eqn.~\ref{eqn:gam_bar}) for the case of no noise. The right
panel shows the cluster likelihood map constructed by applying the TMF
to mock data which includes noise. Three coarse source redshift bins
have been used to make this map (TMF3).  For the function $G(x)$ we
have used the `optimal filter' determined in \S 4 below
(eqn.~(\ref{eqn:G}) with a scale angle,$\theta_s=0.50^{\prime}$, and
gaussian truncation radius, $\theta_{\rm out}=5.5^{\prime}$), which
maximizes the number of clusters detected.  This filter approximates a
projected Navarro-Frenk-White \citet{NFW97} (NFW) profile, but is
truncated with a Gaussian to prevent the likelihood sum from being
diluted by distant uncorrelated source galaxies.  Peaks in the
likelihood map are candidate shear selected clusters and four of the
more significant peaks in this map are circled and labeled {\bf a-d}.

\begin{figure*}[t]
\begin{picture}(400,230)(0,-10)
\centering
\centerline{\epsfig{file=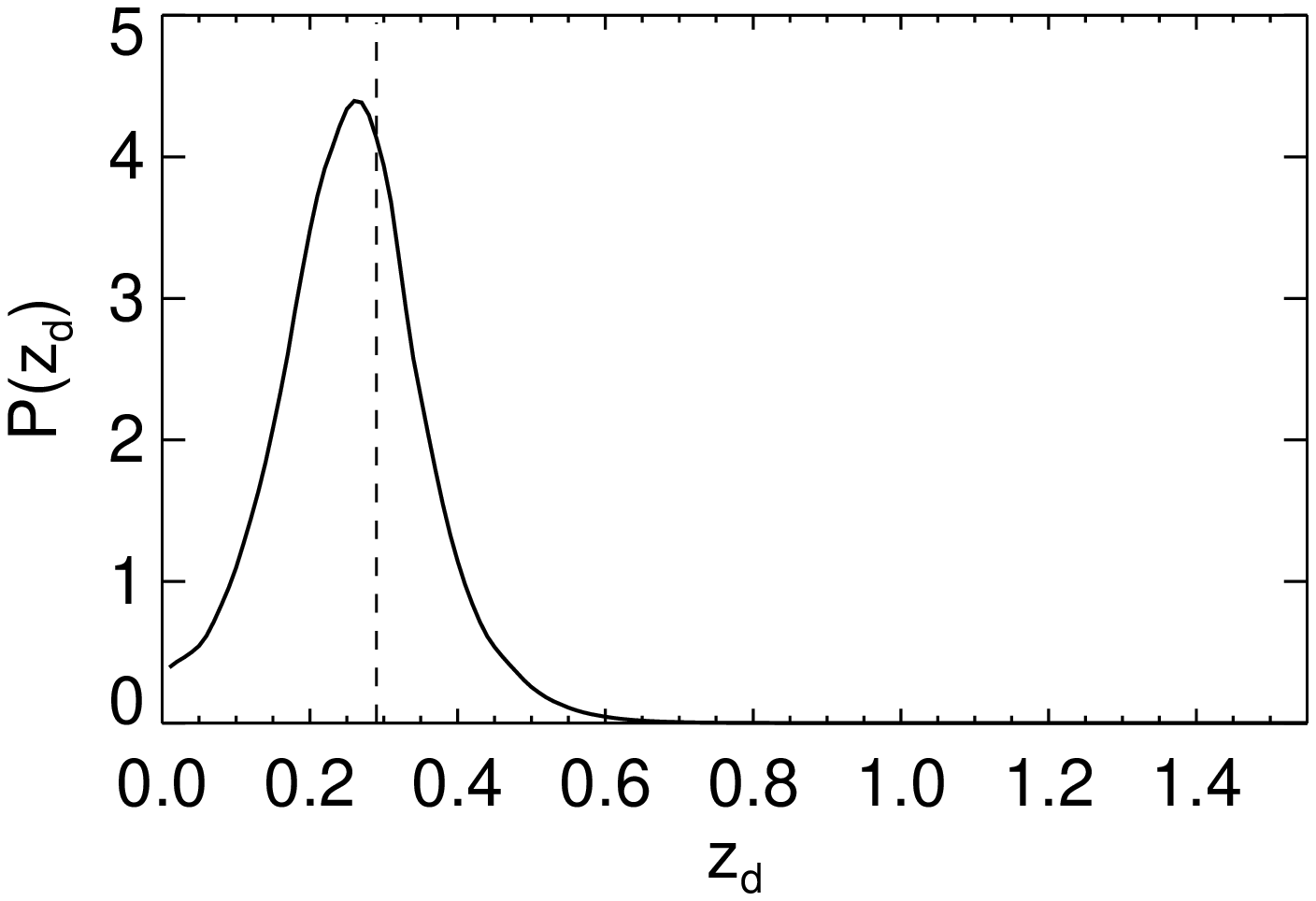,bb= 10 0 445 400,width=0.5\textwidth}\epsfig{file=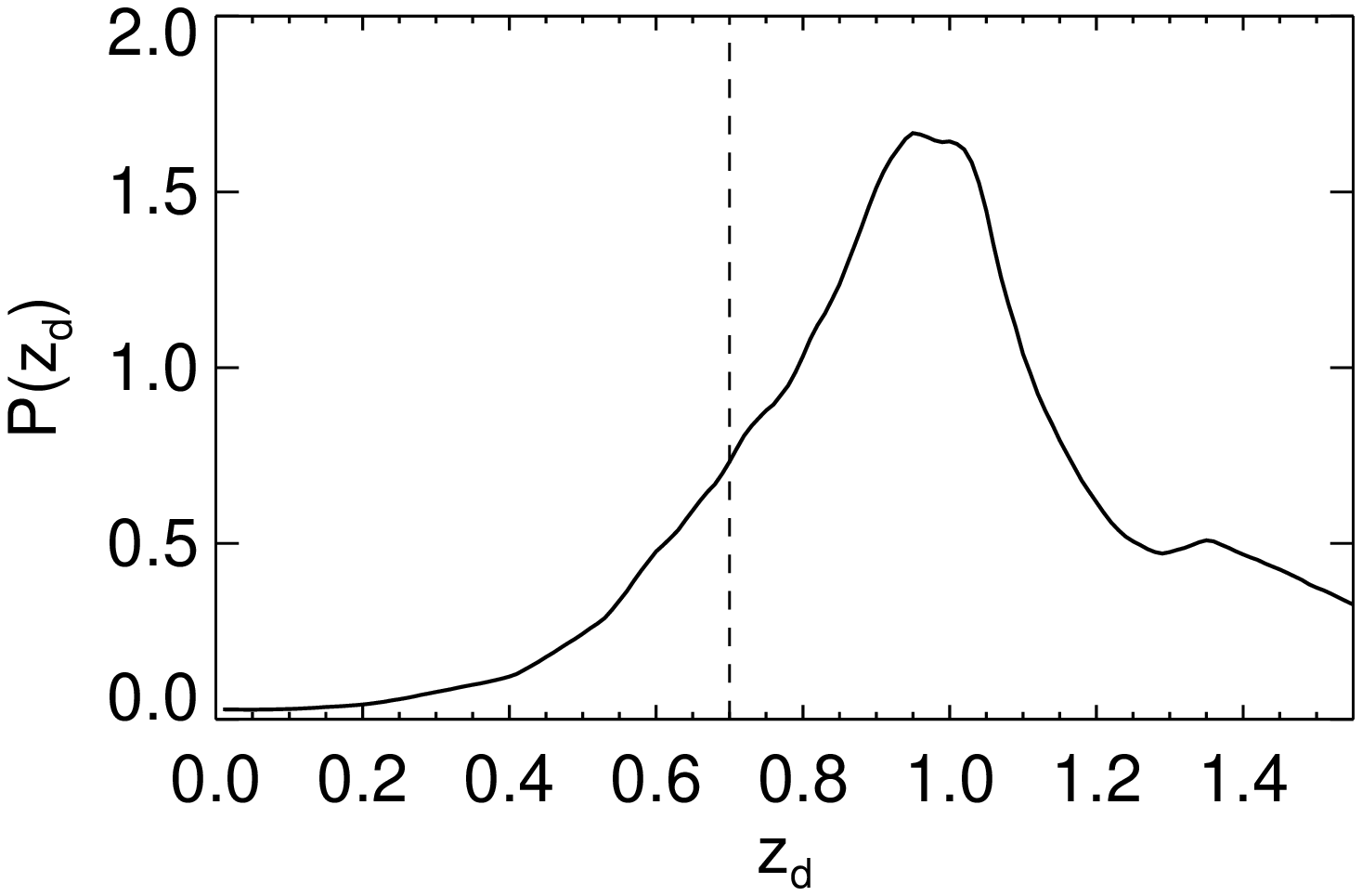,bb= 10 0 445 400,width=0.5\textwidth}}
\put(-465,190){\Huge \bf a}
%\put(-410,170){\vector(-1,0){15.0}}
%\put(-405,165){\Large Ground}
%\put(-380,110){\Large Space}
%\put(-385,115){\vector(-1,0){15.0}}
\put(-370,180){\Large $z_d=0.26$}
\put(-370,160){\Large $z=0.29$}
\put(-370,140){\Large $M_{14}=4.1$}
\put(-210,190){\Huge \bf b}
\put(-210,170){\Large $z_d=0.95$}
\put(-210,150){\Large $z=0.70$}
\put(-210,130){\Large $M_{14}=5.0$}
\end{picture}
\begin{picture}(400,245)(0,-85)
\centerline{\epsfig{file=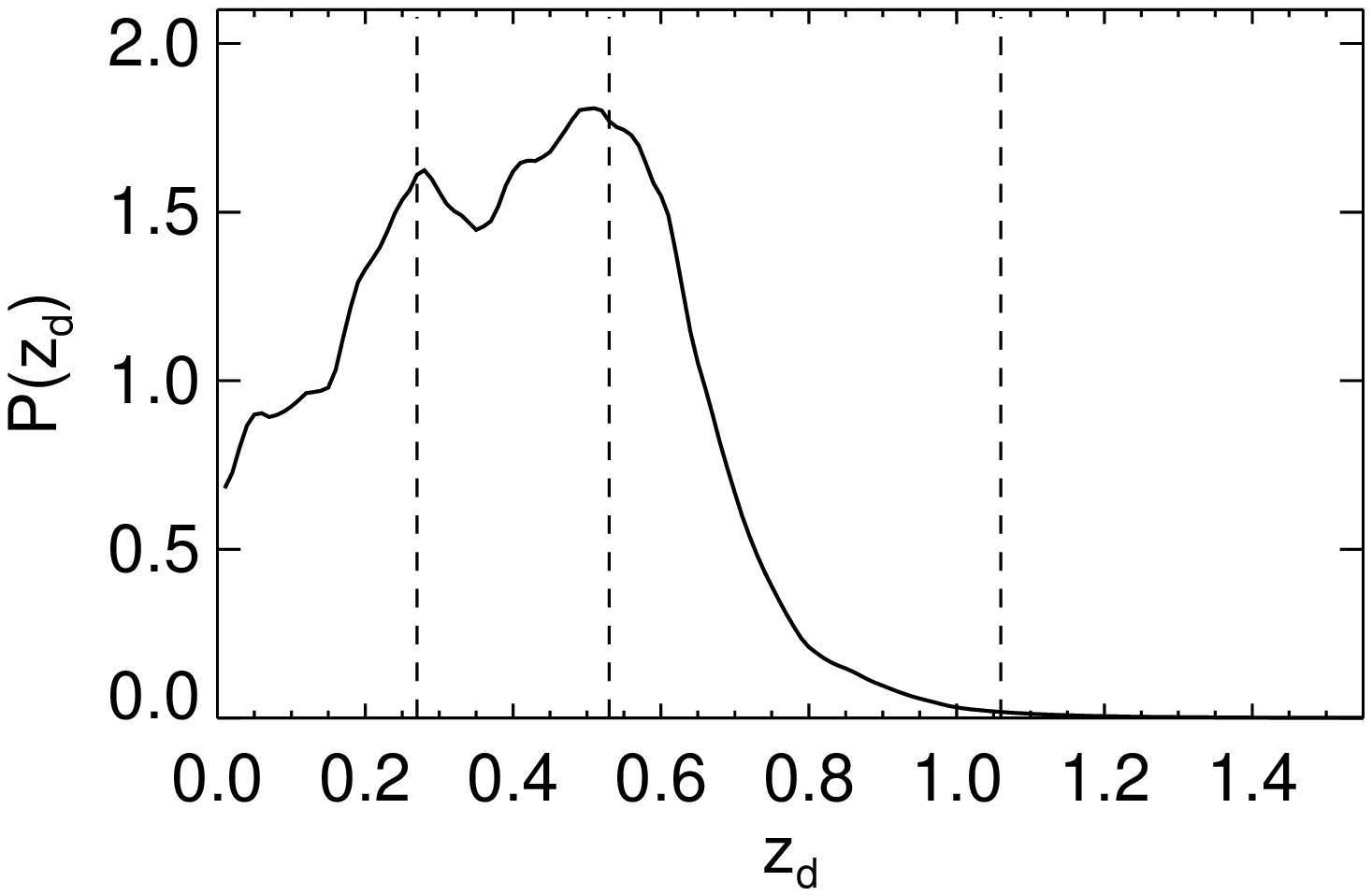,bb= 10 0 445 280,width=0.5\textwidth}\epsfig{file=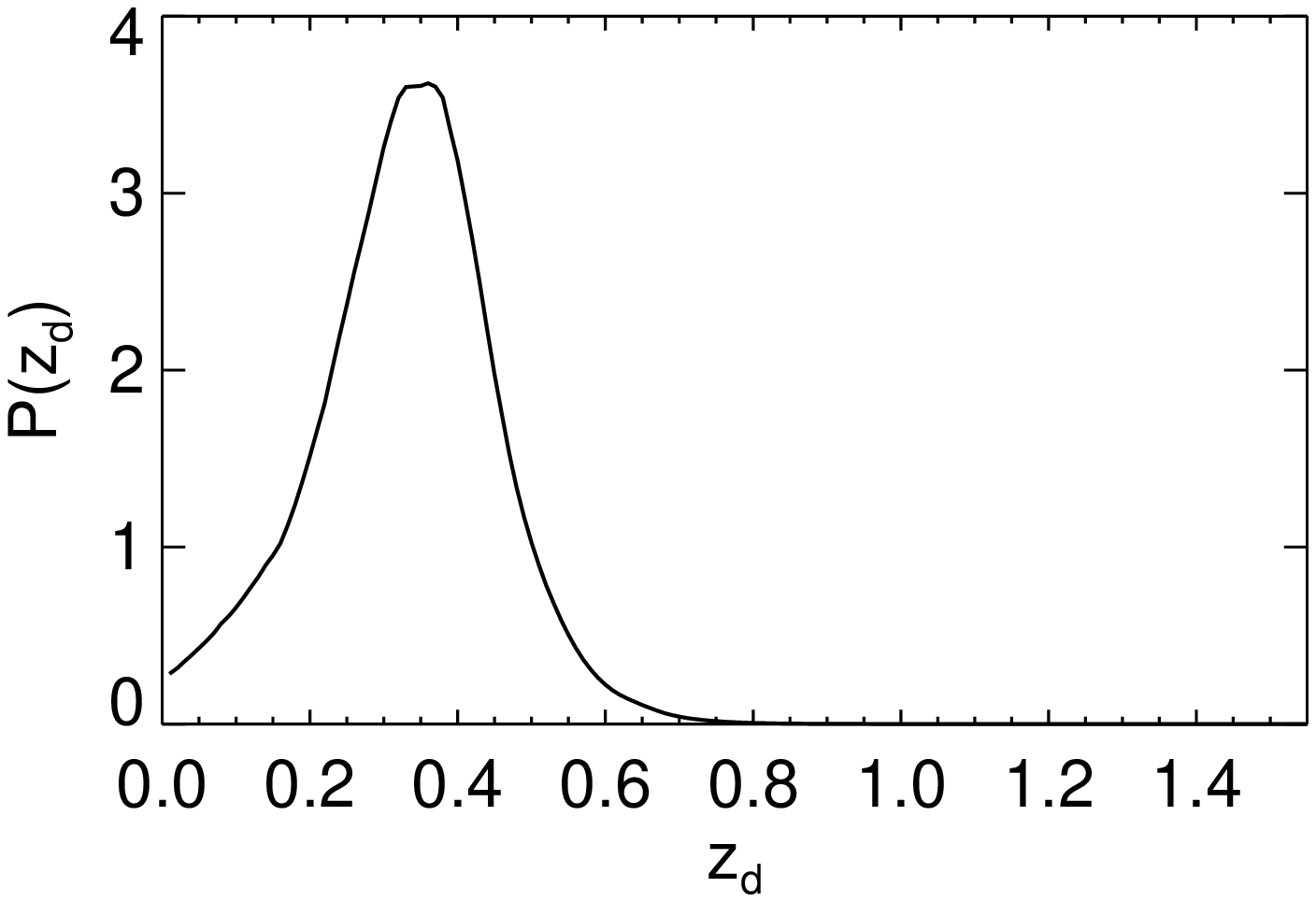,bb= 10 0 445 280,width=0.5\textwidth}}
\put(-465,190){\Huge \bf c}
\put(-330,200){\normalsize $z_d=0.51$}
\put(-330,185){\normalsize $z=0.27$}
\put(-330,170){\normalsize $M_{14}=1.9$}
\put(-330,155){\normalsize $z=0.53$}
\put(-330.5,140){\normalsize $M_{14}=0.47$}
\put(-330,125){\normalsize  $z=1.06$}
\put(-330.5,110){\normalsize $M_{14}=0.47$}
\put(-210,190){\Huge \bf d}
\put(-130,170){\Large $z_d=0.36$}
\put(-130,150){\Large \bf PROJECTION}
\end{picture}
\vskip -4.75cm
\caption{Cluster statistics and the tomographic redshift probability
  distributions for the four clusters labeled \textbf{a}-\textbf{d} in
  Figure \ref{fig:demo}. The mass, $M_{14}=M\slash 10^{14} \ \hmsol$,
  and redshift, $z$, of the cluster(s) responsible for each peak are
  labeled. Vertical dashed lines indicate the true redshifts of the
  halos along the line of sight. The location of the peak of the probability  
  is the \emph{tomographic} redshift, $z_{d}$, which is also labelled.}
\label{fig:pdf}
\end{figure*}

The probability distributions of the tomographic redshift for each of
these four peaks is shown in Figure \ref{fig:pdf}. To create the
likelihood map in Figure \ref{fig:demo} the source galaxies were
binned spatially and in redshift as in eqn.~(\ref{eqn:lhood_bin});
however, to compute the probability distributions shown in Figures
\ref{fig:pdf}, the convolution in eqn.~(\ref{eqn:lhood_conv}) is
performed exactly in real space with no binning.  The mass and
redshift of the cluster(s) responsible for the peaks are labeled, and
the vertical dashed lines indicate the true lens redshift(s). 

%The dotted
%lines show the probability distributions if the source distribution is
%increased to $n=250 \ {\rm arcmin}^{-2}$, as could be achieved with
%deep observations from a space based platform (Massey \etal
%2003) \footnote{For simplicity we have kept the instrinsic ellipticity
%  fixed in this comparison, whereas in reality the instrinsic
%  ellipticity distribution of sources depends on surface brightness,
%  and will hence differ for deeper observations (Bernstein \& Jarvis
%  2000; Massey \etal 2003)}

%Probability
%distributions ares shown for two different truncation radii: figure
%\ref{fig:pdf} uses $\theta_{\rm out}=2.5'$ while figure \ref{fig:pdf5}
%uses $\theta_{\rm out}=5.0'$. Probability distributions are shown for
%two different truncations to emphasize the subtle point that the
%tomographic redshift will depend on the shape of the assumed profile
%$G(\theta)$ as well the number of source galaxies included in the
%aperture. 

Although we will consider the accuracy of the tomographic redshifts in
detail in \S 5, the examples in Figure \ref{fig:pdf} illustrate some
qualitative features of the tomographic technique which we elaborate on
them briefly.

%First note that the resulting probability distributions and maximum
%likelihood tomographic redshifts depend on the size of the aperture
%used to compute the likelihood as well as on the number of background
%sources. Tests show that changing the shape of the profile $G(\theta)$
%also produces similar changes in the individual
%distributions. Furthermore, increasing the background source density
%does not necessarily result in narrower probability distributions. For
%some clusters the tomography \emph{saturates} such that increasing the
%source density no longer improves the determination of the lens
%redshift. This saturation is due to the width of the lensing
%efficiency, which is a broad bell shaped curve that peaks at
%half-distance to the background source population. This broad kernel
%combined with line of sight projections of large scale structure
%bastardize the radial information.  In particular, because the
%function used to weight the background galaxies
%(eqn.~\ref{eqn:bigZ_final}) in the likelihood is a steeper function of
%source redshift for higher lens redshifts, increasing the source
%density can provide additional information for higher redshift
%lenses.

Peak {\bf a} corresponds to a fairly massive cluster at $z=0.29$ and
the tomography performs well, giving a tomographic redshift of
$z=0.26$. For the higher redshift cluster coincident with peak {\bf
  b}, at $z=0.70$, the tomography gives $z_{d}=0.95$, which is
incorrect, but the tomography succeeds to indicate that the
deflector is at high redshift, which is still helpful as the source
galaxies can be weighted for a high redshift deflector, increasing the
contrast between the cluster and the noise.

Projections of several halos at different redshifts are quite common,
which is illustrated by peak {\bf c} which is a projection of a
$2\times 10^{14} \ \hmsol$ cluster at $z=0.27$ and two large group
size halos of $5\times 10^{13} \ \hmsol$ at $z=0.53$ and $z=1.06$,
respectively. Notice in Figure \ref{fig:pdf}, that the tomography
gives a deflector probability distribution with two likelihood maxima
at redshifts corresponding to the two redshifts of the lower redshift
halos. The fact that the peaks in the probability line up with the
deflector redshifts is somewhat of a coincidence, since we are fitting
a model to the shear with only a single deflector.  

Finally, as we will elaborate on further in the next section, mass
selected cluster samples are plagued by projection effects due to
large scale structure and small halos along the line of
sight. Although the peak labeled {\bf d} is the highest peak in the
likelihood map, there is no corresponding halo in the light cone above
$10^{13.5} \ \hmsol$ within an aperture of radius $3^{\prime}$ (which
is our peak-halo matching criteria). However, there is no indication
that this peak is a projection from the tomography, which gives a
probability distribution similar to the others indicating the
tomographic redshift is $z_d=0.36$. It is perhaps not surprising that
$z_d \approx 0.4$, when one considers that the lensing efficiency
peaks at this redshift, and large scale structure along the line of
sight weighted by this lensing kernel is likely to yield a tomographic
redshift at its peak.

Although the tomographic redshift probability distribution radially
resolved two projected clusters for peak {\bf c}, it led us to believe
that peak {\bf d} was a shear selected cluster just like any other, at
$z_{d}=0.36$, when in reality it was a projection of large scale
structure.  A detailed investigation of the degree to which cluster
tomography can distinguish projections from real clusters is beyond
the scope of this work. If it were possible to make a such a
distinction based on the tomography, one could imagine flagging
certain these peaks, so that a shear selected cluster sample could be
cleaned of such projections, which will complicate attempts to
determine cosmological parameters from the redshift distribution of
shear selected clusters.  Of course, the projection hypothesis can
also be tested by searching for overdensities of early type galaxies
(i.e. cluster galaxies associated with the lenses) as a function of
photometric redshift
%We defer a detailed investigation of the degree to
%which cluster tomography can distinguish projections to a future
%work.  

\citep{Dahle03,Schirm03,Schirm04,Tay04}. 


\section{The Optimal Filter}

\subsection{Discussion of Filter Functions}

\label{sec:filter}

In this section we will consider several filter functions $G(x)$ to convolve
the tangential shear data with for the TMF introduced in \S 3. But 
first, a few general remarks about searching for clusters in weak lensing
data are in order.

Like the $\zeta$ statistics \citep{Fahl94} and the aperture mass
measures $M_{\rm ap}$ \citep{Sch96}, The TMF has the virtue that it
operates directly on the shear, which is the observable quantity in
weak lensing observations. Some studies of shear selected cluster
detection search for clusters by first performing a Kaiser-Squires
density reconstruction, and then smoothing the reconstructed
convergence field with a filter and searching for peaks
\citep{Reb99,WvWM02,HTY03}. However, working directly with the shear
is crucial, when one recalls that the surface mass density at any
point depends on the galaxy ellipticities at all distances \citep[see
  e.g.,][]{BaSc01}. The kernel that the ellipticity data is convolved
with in a density reconstruction is not local, decaying as
$\theta^{-2}$. Because of edge effects from the field boundaries and
all the masked bright stars, diffraction spikes, and bleed trails (see
Figure 14 of \citet{vWM03} for a nice example), the non-local
Kaiser-Squires reconstruction will propagate ringing from this `window
function' into other regions of the mass map, significantly amplifying
noise. It should thus be avoided for the purposes of finding clusters.

Besides avoiding the deleterious effect of the noisy window function
or mask, convolving the shear with a compact kernel seems sensible
because clusters of galaxies have characteristic size on the
sky. However, weak lensing statistics using small apertures are
complicated by the non-local relationship between shear and mass. Any
filter convolved with the shear within some aperture will have
fluctuations imprinted on it from scales \emph{outside} that
aperture. The simplest case is the mass sheet degeneracy, whereby the
convergence field reconstructed from a source galaxies in a finite
field can only be determined up to an additive constant or sheet of
mass \citep{BaSc01}.  While the mass sheet can be intuitively
identified with fluctuations in the $\ell \sim 0$ mode, all large
scale structure fluctuations on scales larger than the aperture
$\theta$, $\ell \lesssim 1/\theta$, result in a variance of any
statistic we apply to the shear over a finite
aperture. \citet{Hoek01,Hoek03} and \citet{Dod03} have studied the
effect of the variance due to large scale structure on measurements of
cluster masses and density profiles. For the purposes of identifying
clusters, variance in our aperture statistic will increase the number
of false detections, lowering the efficiency of cluster finding.

The aperture mass measures \citep{Sch96,SvWJK98}
constitute a class of filters which mitigate the aforementioned problem.  
Convolution of the $\kappa$ map with a kernel $U$ can be shown to be 
equivalent to convolving the tangential shear map with a related kernel $Q$,
provided that the kernel $U$ is compensated 
\beq 
\int_{0}^{1} dx x U(x)=0.  \label{eqn:comp}
\eeq 
In this work we consider the most widely used pair of
kernels  $U$ and $Q$ which are 
\bea 
U(x) &=& \frac{9}{\pi\vartheta^2}(1-x^2)\left(\frac{1}{3}-x^2\right) \label{eqn:M_ap}\nonumber\\
Q(x) &=& \frac{6}{\pi\vartheta^2}x^2\left(1-x^2\right), 
\eea 
where $\vartheta$ is the size of the aperture and $x=\theta/\vartheta$.

The aperture mass has the appealing property that it 
provides a lower limit on the mass contained within the
aperture $\theta$, because of its equivalence to a convolution of
$\kappa$. Furthermore, it is relatively unaffected by modes larger 
than the filter scale. As shown by \citet{SvWJK98}, its variance can be 
written 
\beq
\langle M^2_{\rm ap}\rangle = \int d\ln\ell~\Delta_\kappa^2(\ell)F(\ell\vartheta) 
\label{eqn:ap_var}
\eeq
where $F(\ell\vartheta)$ is a notch filter which peaks at 
$\ell \sim \vartheta$ and suppresses fluctuations from scales 
smaller and larger. The fact that the aperture mass is insensitive to the 
\emph{mass sheet degeneracy} is manifest by the fact that $F(\ell=0)=0$. 
Because modes larger and smaller than the aperture scale (matched to the size
of a cluster) don't cause fluctuations in the aperture mass, 
we might expect this filter to more efficiently locate clusters in weak 
lensing data. We return to this point below.

A potential problem with using the aperture mass is that because it is 
compensated (eqn.~\ref{eqn:comp}) it does not provide a very good fit
to the tangential shear profile of a galaxy cluster. Hence, we don't expect
it to work as well with the tomographic technique introduced in the previous
section. 

\citet{Paddy03} advocate using a matched filter with convergence
profile 
\beq 
K_{\rm NFW}(x) = \frac{1}{(1+x)^2}. \label{eqn:uros},
\eeq
where $x=\theta\slash\theta_s$, for several different values of the 
scale angle $\theta_{\rm s}$. This profile approximates an NFW density 
profile \citep{NFW97} in projection, and has been used in studies of 
optical clusters (White \& Kochanek 2002; White et al. 2002).  
The tangential shear corresponding to this convergence profile is 
\beq G_{\rm NFW}(x) =\frac{2\ln(1+x)}{x^2}-\frac{2}{x(1+x)}- \frac{1}{(1+x)^2},
\label{eqn:G_NFW}
\eeq 
where we have applied eqn.~(\ref{eqn:GfromK}). 

The soft core in this profile is a desirable feature, since convolving
shear data with a divergent profile gives a large weight to a single
noisy galaxy resulting in an ultraviolet divergence, as is well known
in the literature on weak lensing mass reconstructions
\citep[see][]{KS93}.  However, these NFW convergence and shear
profiles decay asymptotically as a power law $\sim \theta^{-2}$, and
are not compact. As we will see, this has a deleterious effect on the
number of clusters detected and on the efficiency of the cluster
search, because of a dramatic increase in confusion from large scale
structure.

The foregoing discussion motivates the truncated projected NFW profile
\beq
  G(x) =  G_{\rm NFW}(x)\exp\left(-\frac{\theta^2}{2\theta_{\rm out}^2}\right), 
\label{eqn:G}
\eeq 
where we have truncated the filter in eqn.~(\ref{eqn:G_NFW}) by
multiplying with a Gaussian of truncation angle $\theta_{\rm out}$.

Note that the model shear profile in eqn.~(\ref{eqn:G}) now has two
free scale parameters $\theta_{\rm s}$ and $\theta_{\rm out}$, whose
values we must determine.  In principle, we could leave them as
parameters, and maximize the likelihood in eqn.~(\ref{eqn:lhood_conv})
with respect to redshift and both angular scale factors.  Tests on the
simulations indicate that for fixed $\theta_{\rm out}$, maximizing the
likelihood with respect to scale angle $\theta_{\rm s}$ and $z_d$
performs slightly worse than the case where $\theta_{\rm s}$ is fixed
and the likelihood is maximized only with respect to $z_{\rm d}$.
Furthermore, varying the scale angle $\theta_s$ would require
performing a whole sequence of convolutions with different size
filters, which is more expensive computationally.  For these reasons
we choose to fix $\theta_{\rm s}=0.50^{\prime}$, which is roughly
matched to both the pixel size in our simulated maps and the angle
$\theta_{\rm s}$ subtended by the scale radius $r_{\rm s}$ of a
$10^{14} \ \hmsol$ NFW cluster at $ z \simeq 0.4$. Varying this scale
angle from $\theta_{\rm s} = 0.10-0.70$ produces negligible changes in
our results.

Maximizing the likelihood with respect to the truncation angle
$\theta_{\rm out}$ also performs poorly, because of a tendency to
overfit noise and large scale structure on scales much larger than the
size of a cluster. Also, the likelihood for small and large
$\theta_{\rm out}$ will be over a different number of source galaxies
and hence degrees of freedom.  Thus their distributions
will have different means and variances, complicating a comparison of
the resulting likelihoods. For these reasons, we take the truncation
radius $\theta_{\rm out}$ to be a `prior' that we vary by hand,
until the number of clusters detected for a given noise model is
maximized.  

%Finally, note that the optimal truncation radius will depend on the
%amount of noise, with noisier data requiring smoothing over a larger
%aperture.

%There will be a tradeoff as $\theta_{\rm out}$ is varied between averaging over more source galaxies, and hence increasing the signal on the one hand, but increasing noise from tangential aliasing of large scale structure and merging multiple clusters together on the other.  Finally, the truncation radius will depend on the amount of noise, with noisier data requiring smoothing over a larger aperture.

In addition to the aperture mass (eqn.~(\ref{eqn:M_ap})), the projected NFW 
profile (eqn.~(\ref{eqn:G_NFW})), and the truncated NFW (eqn.~{\ref{eqn:G})),
we also try simply convolving the tangential shear with a Gaussian filter of
size $\theta_{\rm out}$.  

In general, the degree of smoothing required to 
detect clusters from weak lensing data will depend on the level of noise for 
the particular set of observations, as higher noise will require more 
smoothing. Here we focus on the noise model described in \S 2. Adapting a 
filtering scheme to different levels of noise will amount to varying the 
overall extent of the filter.

\subsection{Comparison of Filters}

Searching for shear selected clusters entails convolving the ellipticity data 
with a filter and searching for peaks in a map. In general some signal to 
noise ratio threshold, $\nu$,  will be chosen such that peaks above this 
threshold are considered candidate clusters and peaks below it are discarded.  
Given a threshold $\nu$, a fraction of all peaks larger than $\nu$ will 
be identified as galaxy clusters and the remainder will correspond to false 
detections. 

We define the efficiency as a function of this threshold as
\beq
e(\nu)\equiv \frac{n_{\rm clusters}(> \nu)}{n_{\rm peaks}(>\nu)}. \label{eqn:eff}
\eeq
It is clear that raising the threshold $\nu$, will increase the efficiency, 
but at the expense of fewer detections $n_{\rm clusters}(>\nu)$,
 since there will be fewer total peaks greater than $\nu$.  Rather than fix 
$\nu$ at say 5$\sigma$, where $\sigma$ represents the 
variance of the noise, we consider the number of clusters detected as this 
threshold is varied. There are several reasons why one may wish to work at 
low thresholds and hence low efficiency, but we defer a discussion of this 
point to \S 5.  Note that the definition in eqn.~(\ref{eqn:eff}) allows us to 
assign an efficiency to any peak  $\nu$ in our maps, and we may speak of an 
efficiency cutoff of say 75\%, where it is understood that the we keep all 
clusters above the threshold $\nu_{75}$ where $e(\nu_{75})=75 \%$. 

%\myputfigure{eff.eps}{3.0}{0.43}{-5}{+5} 
\begin{figure*}[t]
  \centering
  \centerline{\epsfig{file=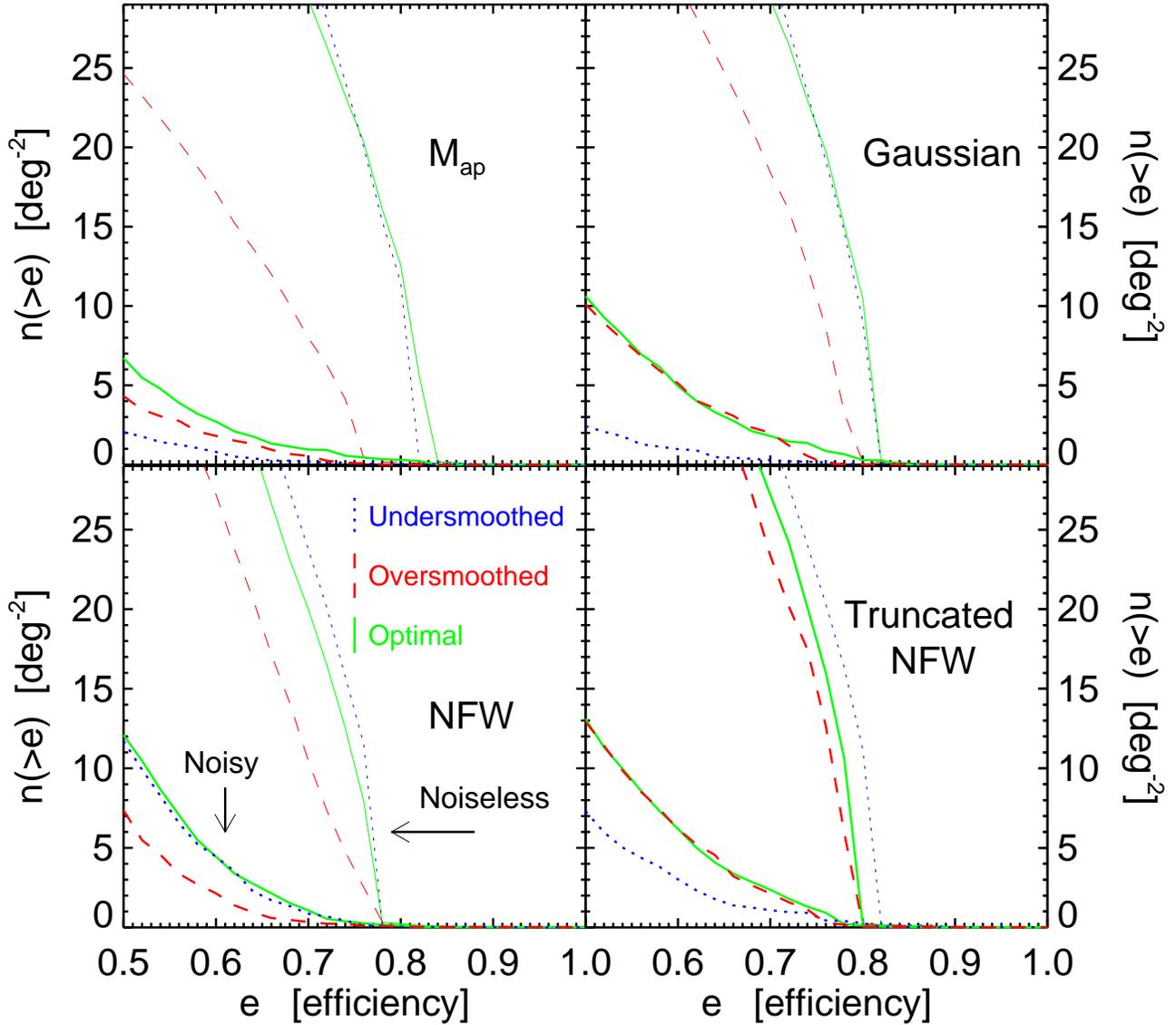,bb=20 35 490 440,width=1.0\textwidth}}
  \caption{ Number density of clusters versus efficiency for four
    filters.  Dotted (blue) curves are undersmoothed, dashed (red)
    curves are over smoothed, and solid (black) curves indicate the
    filter size that performs best. The inner set of curves are noisy
    mock data and the thin outer curves are for noiseless data. A
    summary of information on the filters and aperture scales used in
    this figure is given in Table \ref{table:filters}.}
    \label{fig:eff_noz}
%\emph{Upper left:} The aperture mass for
%    aperture sizes $\vartheta=1.5^{\prime}$ (dotted), $2.5^{\prime}$
%    (solid), and $3.5^{\prime}$ (dashed). \emph{Upper right:} Gaussian
%    filter for cutoff angles $\theta_{\rm out}=1.0^{\prime}$ (dotted),
%    $1.5^{\prime}$ (solid), and $3.0{\prime}$ (dashed). \emph{Lower
%      Left:} The set of non-compact filters advocated by PSP with
%    scale angles $\theta_{\rm s}= 0.1^{\prime}$ (solid),
%    $0.5^{\prime}$ (dotted), $1.0^{\prime}$ (dot-dashed),and
%    $5.0^{\prime}$ (dashed). \emph{Lower Right} The TMF for
%    $\theta_{\rm out}=1.0^{\prime}$ (dotted), $2.5{\prime}$ (solid),
%    and $10.0^{\prime}$ (dashed).}

\end{figure*}

In Figure \ref{fig:eff_noz} we plot the number of clusters detected per
square degree versus efficiency for the filters discussed above.  All the
aforementioned filters have an angular scale parameter, and will yield
more or less clusters at a given efficiency as this scale is varied.
For the aperture mass this scale is the size of the aperture
$\vartheta$.  The only scale in the NFW profile in eqn.~(\ref{eqn:uros})
is the scale angle $\theta_{\rm s}$. For the truncated
NFW and the Gaussian, the scale parameter is the Gaussian cutoff angle
$\theta_{\rm out}$. For each filter we plot two sets of curves
corresponding to noisy and noiseless mock weak lensing data, with the
outermost thin curves corresponding to the noiseless case. 

As the amount of smoothing is increased there is a tradeoff between
averaging over more source galaxies, and hence increasing the signal
to noise on the one hand, but increasing the level of confusion from
cosmic structure and merging distinct clusters together into one
another. For a given amount of noise, the optimal scale for a filter
corresponds to the curve furthest to the right in Figure
\ref{fig:eff_noz}, which maximizes the number of clusters detected at
each efficiency.   The dotted (blue) and dashed
(red) curves correspond to under and over smoothing, respectively, and
the solid (green) curves correspond to the `optimal' scale for the
noisy data. Table \ref{table:filters} summarizes the information for the
filters discussed in this section and plotted in Figure
\ref{fig:eff_noz}. The optimal smoothing scale is printed in bold.

\begin{table}
\begin{center}
  \caption{Filters used in Figure \ref{fig:eff_noz}\label{table:filters}}
  \begin{tabular}{lcll}
     %      \tablevspace{3pt}
    \hline
    \hline
    Filter & \ Profile Eqn & \quad Scale Angle   & \quad\quad  Angles\\
    \hline
     NFW         & (\ref{eqn:G_NFW})   &\quad\quad\quad $\theta_{\rm s}$  &\quad  0.01,\textbf{0.1},1.0\\ 

  $M_{\rm ap}$& (\ref{eqn:M_ap})&  \quad\quad\quad $\vartheta $       & \quad 2.0,\textbf{3.0}, 5.0\\ 
  Gaussian       &  ---                & \quad\quad\quad $\theta_{\rm out}$ & \quad 1.5,\textbf{2.0}, 4.5\\ 
  Truncated NFW         & (\ref{eqn:G})    & \quad\quad\quad  $\theta_{\rm out}$  & \quad 2.0, \textbf{5.5}, 10.0 \\
    \hline
  \end{tabular}
\end{center}
\footnotesize NOTES.--- Parameters for the four filters functions
$G(x)$ (see eqn. \ref{eqn:model}) used in Figure
\ref{fig:eff_noz}. The table lists the defining equation of the
profile, the angular scale parameter varied to set the level of
smoothing, and the three angles used in the figure. The three angles
correspond to undersmoothing (dotted curves in Figure
\ref{eqn:model}), the optimal smoothing, (solid curves), and
oversmoothing (dashed curves), respectively. The optimal smoothing is
listed in bold and all angles are in arcminutes.
\vskip -0.5cmt
\end{table}

\begin{figure*}[t]
  \centering
  \centerline{\epsfig{file=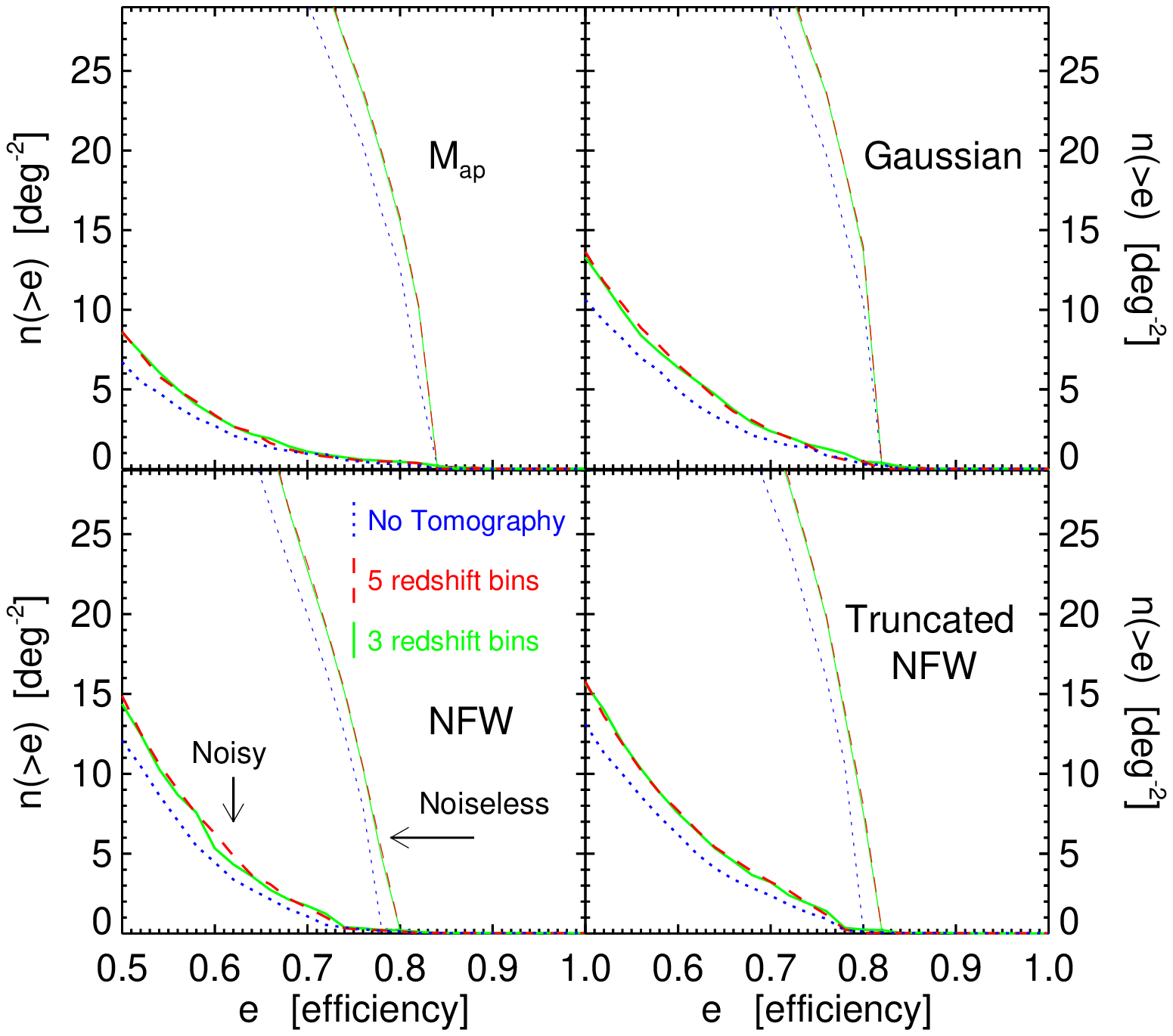,bb=20 35 490 440,width=1.0\textwidth}}
  \caption{ Number density of clusters versus efficiency for four
    filters using varying degrees of photometric redshift information.
    Dotted (blue) curves use no photometric redshift information,
    dashed (red) are for sources binned into five fine photometric
    redshift bins (TMF5), and solid (green) curves are for the sources
    binned into three coarse redshift bins (TMF3). The inner set of
    curves are noisy mock data and the thin outer curves are for
    noiseless data.  The TMF exploits the information provided by
    source photometric redshifts to increase the number of clusters
    detected at all efficiencies, even for crude redshift bins. A
    comparison of the performance of these filters is presented in
    Table \ref{table:eff_tomo}.}
    \label{fig:eff_tomo}
\end{figure*}

\begin{figure*}[t]
  \centerline{ \epsfysize=3.6truein \epsffile{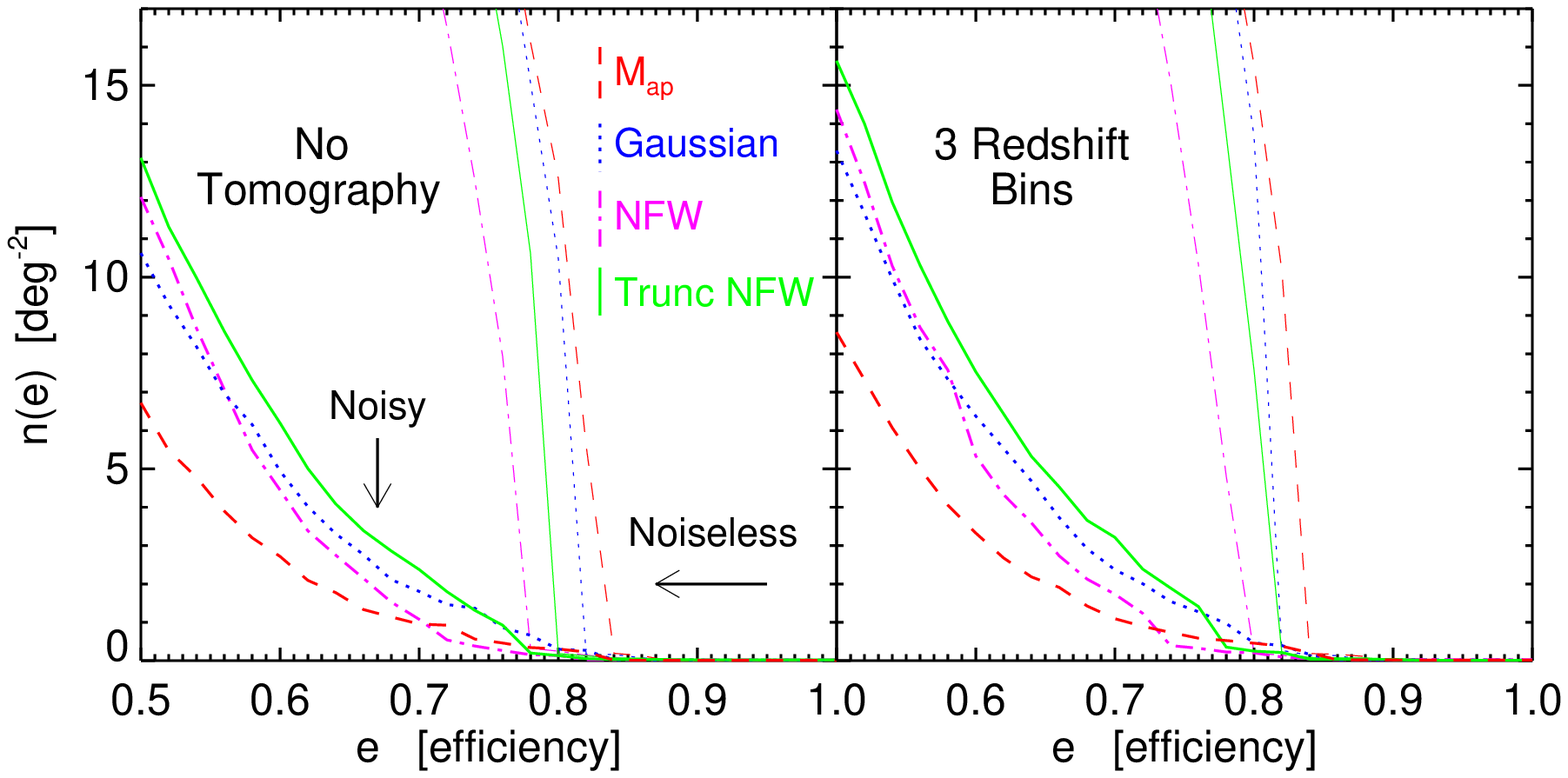} }
  \figcaption{ \footnotesize Comparison of the four filter functions
    with and without tomography. The left panel uses no photometric
    redshift information, whereas the right panel uses tomography with
    source galaxies binned into three coarse redshift bins (TMF3). The
    optimal filtering scale is used for each filter (see Table
    \ref{table:filters}). The inner set of curves are for noisy data
    and the thin outer curves are noiseless. The dot-dashed (magenta)
    curve is for the NFW profile, the dashed (red) curve is the
    aperture mass, the dotted (blue) curve is the Gaussian, and the
    solid (green) curve is the truncated NFW. The amount by which
    tomography enhances the performance of each filter is listed in 
    Table \ref{table:eff_tomo}. \label{fig:eff_comp}
  }
\end{figure*}

The benefit of using photometric redshift information is illustrated
in Figure \ref{fig:eff_tomo}, where we smoothed with the optimal
filter scale for each filter as determined from Figure
\ref{fig:eff_noz} and listed in Table \ref{table:filters}, but varied the
amount of photometric redshift information used. The dotted (blue)
curves ignore photo-z information, the solid (green) curves correspond
to the TMF with the three coarse photometric redshift bins (TMF3), and
the dashed (red) curves correspond to the TMF with the five finer
(TMF5) photometric redshift bins (see Table \ref{table:bins}).  Both
TMF's detect significantly more clusters for each filter at high and
low efficiencies.

We quantify the increase in clusters detected with the TMF in
Table \ref{table:eff_tomo}, where we list the number of clusters per
square degree with and without tomography. We compare all four filters
at a high efficiency cutoff of $e=75\%$ and a lower cutoff of
$e=60\%$. To get an idea of the signal to noise ratios corresponding
to the efficiency thresholds, we also list the signal to noise ratio at
the efficiency cutoff, $\nu_{75}$ and $\nu_{60}$, of the filtered maps
which did not use tomography.  The noise in the maps was computed by
calculating the variance of smoothed maps of \emph{unlensed} source
galaxies which had the same positions and intrinsic ellipticities as
the lensed sources.  An efficiency of $e\geq 75\%$
corresponds to peaks with a signal to noise ratio $\nu \gtrsim 4.5$,
whereas $e\geq 60\%$ corresponds to signal to noise cutoff of $\nu \gtrsim
3.5$.  The TMF increases the number of $\nu \gtrsim 4.5$ peaks by up to
$76\%$.  For lower thresholds, the increases in the number of
clusters detected is less substantial ($\sim 30\%$), because the
tomography is less effective at determining the redshift of the
cluster for lower signal to noise detections.  The TMF increases the
number of clusters detected for all filters; however, the gains are
more substantial for the truncated NFW profile which best
approximates the tangential shear profile of clusters.

Besides increasing the total number of clusters detected, the TMF
substantially increases the dynamic range of the resulting cluster
sample. The redshift and mass number count distributions detected by
the truncated NFW profile for an efficiency cutoff of $e\geq
75\%$ are shown in Figure \ref{fig:eff_dist}. The solid histogram is
for the TMF3 whereas the dotted (blue) line does not use photometric
redshifts.  At low redshifts and high mass, the number of clusters
detected are nearly the same, however the TMF detects many more high
redshift clusters and it extends the mass sensitivity down to the
scale of large groups.

As is evident from Figures \ref{fig:eff_tomo}, \ref{fig:eff_dist}, and
Table \ref{table:eff_tomo}, dividing source galaxies into three coarse
redshift bins is sufficient to reap the benefits of tomographic
information and increase the number of clusters detected. Such a
coarse redshift division can be achieved with as few as two colors and an
apparent magnitude prior, so that the TMF can be used with deep
imaging data in as few as three pass bands.

A comparison of the four filters we consider is presented in Figure
\ref{fig:eff_comp}. The filters shown are the optimal aperture mass
($\vartheta=3.0^{\prime}$) , Gaussian ($\theta_{\rm
  out}=2.0^{\prime}$), NFW ($\theta_{\rm s}=0.1^{\prime}$), and
truncated NFW ($\theta_{\rm out}=5.5^{\prime}$). The left panel
doesn't use tomography and the right panel is for the TMF3. The thin
outermost set of curves are for noiseless data.

For noisy data, the truncated NFW is most effective at finding
clusters for efficiency cuts between $60-75\%$, corresponding to
detection significance of $\nu=3.5-4.5$, detecting $\sim 20-30\% $
more clusters, especially if tomography is used. The non-truncated NFW
filter performs significantly worse than the others, even at very high
signal to noise ratios, because it is not compact enough, dropping off
as $\theta^{-2}$.
 
At the highest efficiencies $e\gtrsim 0.8$ corresponding to peaks with
$\nu_{80}\gtrsim 6$, the aperture mass detects more clusters than the other
filters. For these highest peaks, the aperture mass is less affected
by large scale structure and noise, and does the best job of sorting
out projection effects.  Whereas the truncated NFW ranks peaks by
some combination of mass and also goodness of fit to the NFW profile,
by definition, the aperture mass ranks peaks by mass, since it
provides a lower limit on the mass enclosed within the aperture. Further, 
this result is not unexpected in light of our discussion of the variance of
the aperture mass (eqn.~\ref{eqn:ap_var}) in \S \ref{sec:filter}. Notch
filtering of the shear field results in a smaller variance from both
noise and large scale structure, which for the most massive 
clusters $\sim 10^{15} \ \hmsol$ clusters, increases the efficiency 
of the cluster search.

A conspicuous feature of Figures \ref{fig:eff_noz}, \ref{fig:eff_tomo},
and \ref{fig:eff_comp}, is that even for the unachievable case of no
intrinsic ellipticity or Poisson noise, the maximum \emph{intrinsic}
efficiency of weak lensing cluster searches is only $\sim 85 \%$.
Thus $\sim 15 \%$ of even the most significant peaks detected in
noiseless weak lensing maps do not have a collapsed halo with $M
>10^{13.5} \ \hmsol $ within 3 arcmintues (which is our peak halo
matching criteria). This intrinsic inefficiency is due to projection
effects and to confusion from cosmic structures, and has been noted by
previous studies \citep{MWL01,WvWM02,Paddy03,HTY03}.  This fact sheds light
on the purported detections of `dark clumps' reported recently by
several groups \citep{Fis99,Erben00,UF00,Mir02,Dahle03}. These objects
could be entirely consistent with the $\sim 15 \%$ false detections,
though a definitive conclusion will have to wait for future wide field
weak lensing surveys.  Because of the intrinsic inefficiency that we
find here, only statistical statements can be made about a population
of dark clusters.

\begin{table*}[t]
\begin{center}
  \caption{Filter Comparison\label{table:eff_tomo}}
  \begin{tabular}{lcccccccccc}
  \hline
  \hline
  Filter \hfill\vline & $\nu_{60}$ & $n(>60\%)$ &  $n_{\rm TMF3}(>60\%)$ & $n_{\rm TMF5}(>60\%)$    & \% increase \hfill\vline & $\nu_{75}$ & $n(>75\%)$ &  $n_{\rm TMF3}(>75\%)$ & $n_{\rm TMF5}(>75\%)$    & \% increase\\
%  \vskip 0.5cm\\
  \hline 
    NFW           \hfill\vline & 4.1    & 4.45 & 5.34 & 6.27  & \quad \quad 41 \hfill\vline & 6.3  & 0.28 & 0.34 & 0.29  & 19 \\
    $M_{\rm ap}$   \hfill\vline & 3.5   & 2.72 & 3.33 & 3.38  & \quad \quad 25 \hfill\vline  & 4.5  & 0.48 & 0.63 & 0.51  & 33 \\
    Gaussian      \hfill\vline & 3.4   & 4.94 & 6.38 & 6.54  & \quad \quad 32 \hfill\vline  & 4.4  & 1.10 & 1.47 & 1.28  & 34 \\
    Truncated NFW \hfill\vline & 3.4   & 6.20 & 7.53 & 7.70  & \quad \quad 24 \hfill\vline  & 4.9  & 0.99 & 1.75 & 1.45  & 76 \\
  \hline
\end{tabular}
\end{center}
\footnotesize NOTES.--- Comparison of filters with and without
tomography. The number of clusters detected per square degree without
tomography, with coarse redshift tomography (TMF3), and fine redshift
tomography (TMF5), are denoted by $n$, $n_{\rm TMF3}$, and $n_{\rm
  TMF5}$. Two efficiency thresholds are compared, $60\%$ and $75\%$,
and the $S\slash N$ ratio threshold (computed from the maps without
tomography) corresponding to each efficiency cut, $\nu_{60}$ and
$\nu{75}$, are listed. The column labeled ``\% increase'' shows the
improvement when tomography is used of either the TMF3 or TMF5,
whichever is greater.
\vskip -0.5cm
\end{table*}

\begin{figure*}[t]
\centering 
\centerline{\epsfig{file=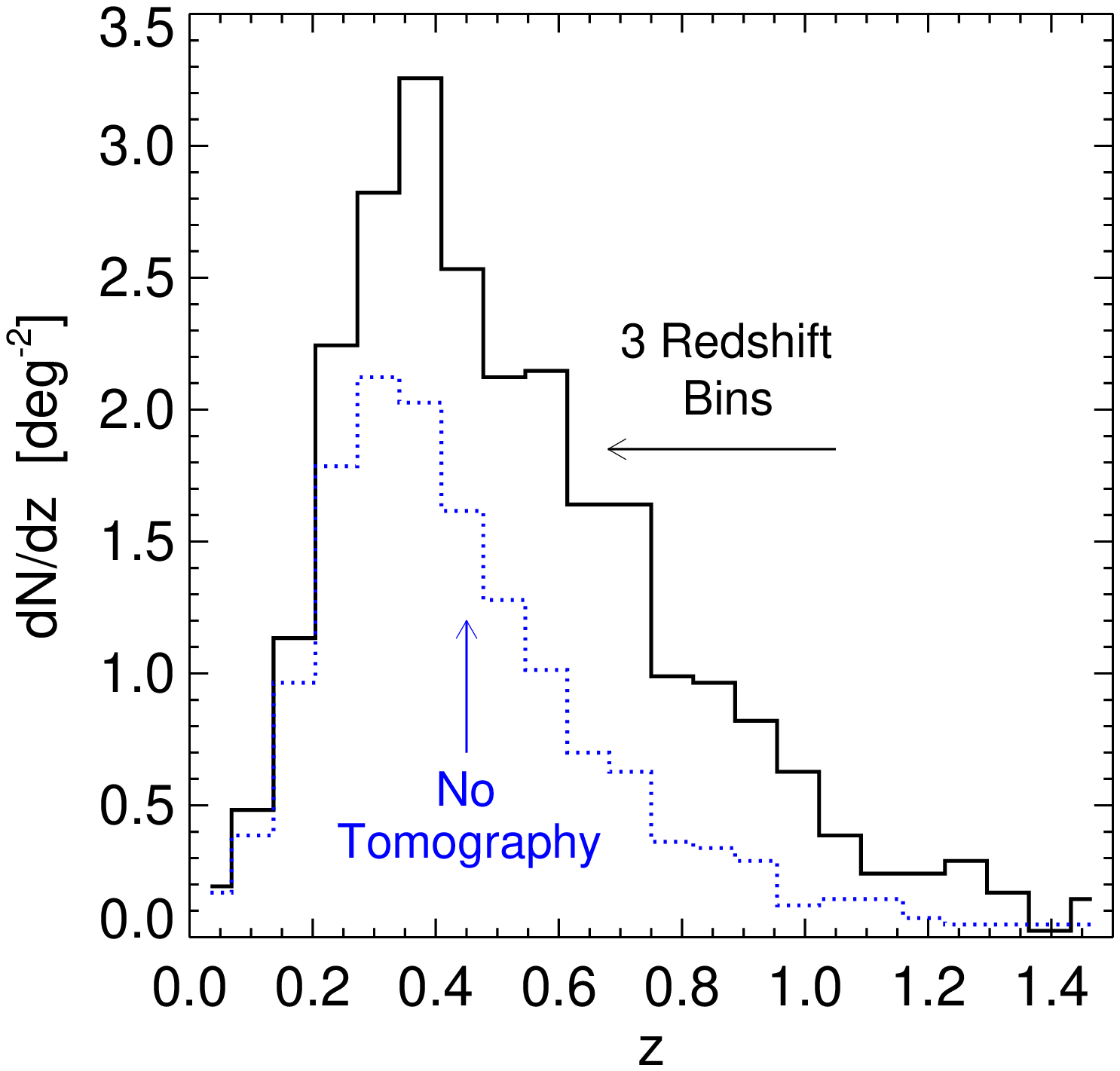,bb=0 0 450
    450,width=0.50\textwidth}\epsfig{file=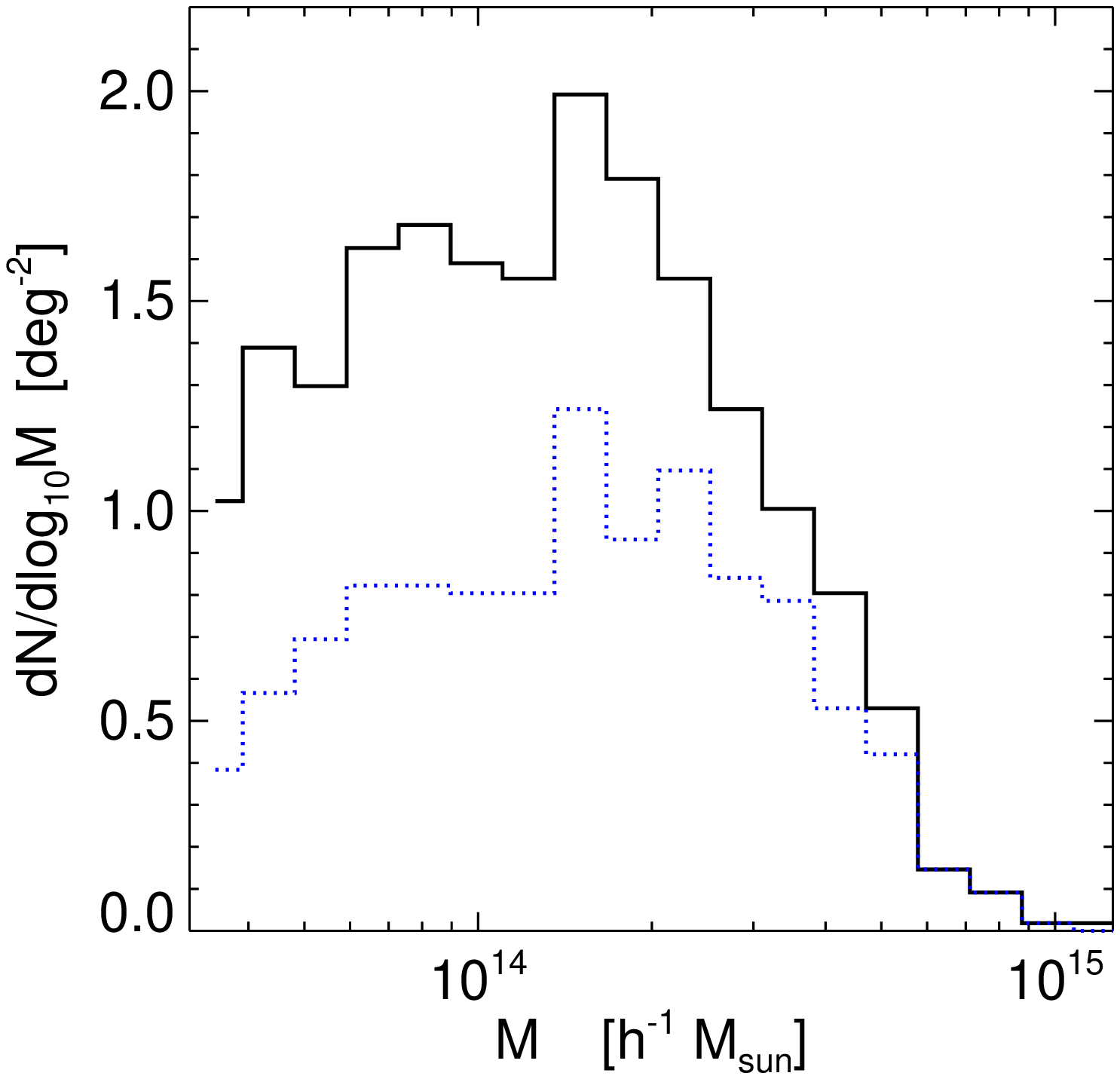,bb=0 0 450
    450,width=0.50\textwidth}}
\caption{Redshift (left) and mass (right) number count distributions
  of clusters detected by the truncated NFW with and without
  tomography.  Dotted (blue) histograms do not use photometric
  redshift information while solid (black) histograms are for the
  source galaxies binned into three coarse redshift bins (TMF3).  The
  cluster samples used to construct these histograms used an
  efficiency cutoff of $e \geq 75\%$ corresponding to a $S/N\gtrsim
  4.5$. Even with just three coarse redshift bins, the TMF increases
  the number of clusters detected by $76\%$ (see Table
  \ref{table:eff_tomo}). Tomography increases the dynamic range of weak
  lensing searches for clusters, detecting more high redshift clusters
  and extending the mass sensitivity down to the scale of large
  groups. }
%For this threshold $70\%$ more clusters 
%  are detected most of which are higher redshift and lower mass clusters.
\label{fig:eff_dist}
  \end{figure*}

\section{Completeness of Shear Selected Samples}

In this section we discuss the completeness of shear selected cluster
samples detected by the filter which performed best in the previous
section. Specifically, we consider the truncated NFW filter with
$\theta_{\rm out}=5.5^{\prime}$ without photometric redshift
information and with source galaxies binned into three coarse redshift
bins (TMF3).  The left panel of Figure \ref{fig:comp_fig} shows the
completeness as a function of redshift for all clusters with
$M>10^{14.3}~\hmsol$. The right panel shows completeness as a
function of cluster mass for clusters in the redshift range $0.2 < z <
0.8$.  Thick lines are for an efficiency cutoff of $e \geq 60\%$
($S/N\gtrsim 3.5$) whereas thin lines are for $e \geq 75\%$ $S/N\gtrsim
4.5$. For the lower efficiency cut of $e \geq 60\%$, the completeness
is $\gtrsim 50\%$ only for the most massive clusters $M\gtrsim
10^{14.3}$ which are near the peak of the lensing efficiency at $z\sim
0.3$. Limiting the cluster search to the higher efficiency threshold
$e \geq 75\%$ further reduces the completeness by $\sim 40\%$.  As a function
of cluster mass the completeness only approaches unity for the most massive
clusters $M \sim 10^{15}~\hmsol$, where we begin to suffer from small 
number statistics. 

Previous studies of shear selected clusters carried out by several
groups \citep{WvWM02,Paddy03,HTY03} come to a similar conclusion.
Namely, that shear selected cluster samples suffer from severe
incompleteness except at the highest masses. Although our results are
broadly consistent with these studies, a direct comparison is
difficult because the noise levels, source redshift distributions,
aperture to match peaks to clusters, and filtering techniques
employed are all different. In particular \citet{WvWM02} and
\citet{HTY03} use a single source redshift plane which implies the
lensing efficiency will be a narrower function of redshift than that
given by the source redshift distributions considered here and in
\citet{Paddy03}. A source redshift distribution results in lower
completeness than a single source plane \citep{Paddy03} because the
broader lensing kernel increases noise from projections of large scale
struture.

\begin{figure*}[t]
  \centering 
  \centerline{\epsfig{file=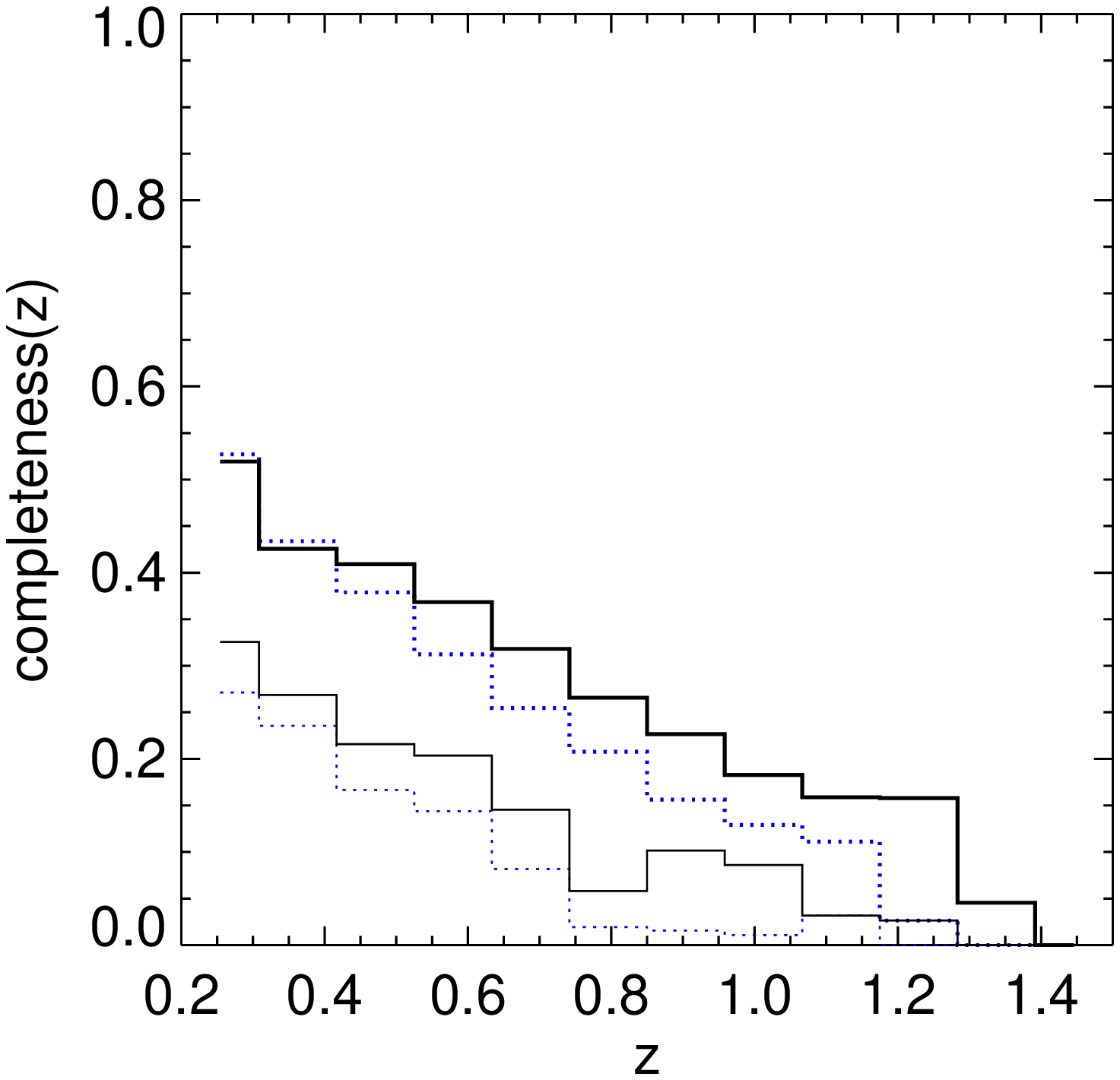,bb=0 0 450
      450,width=0.50\textwidth}\epsfig{file=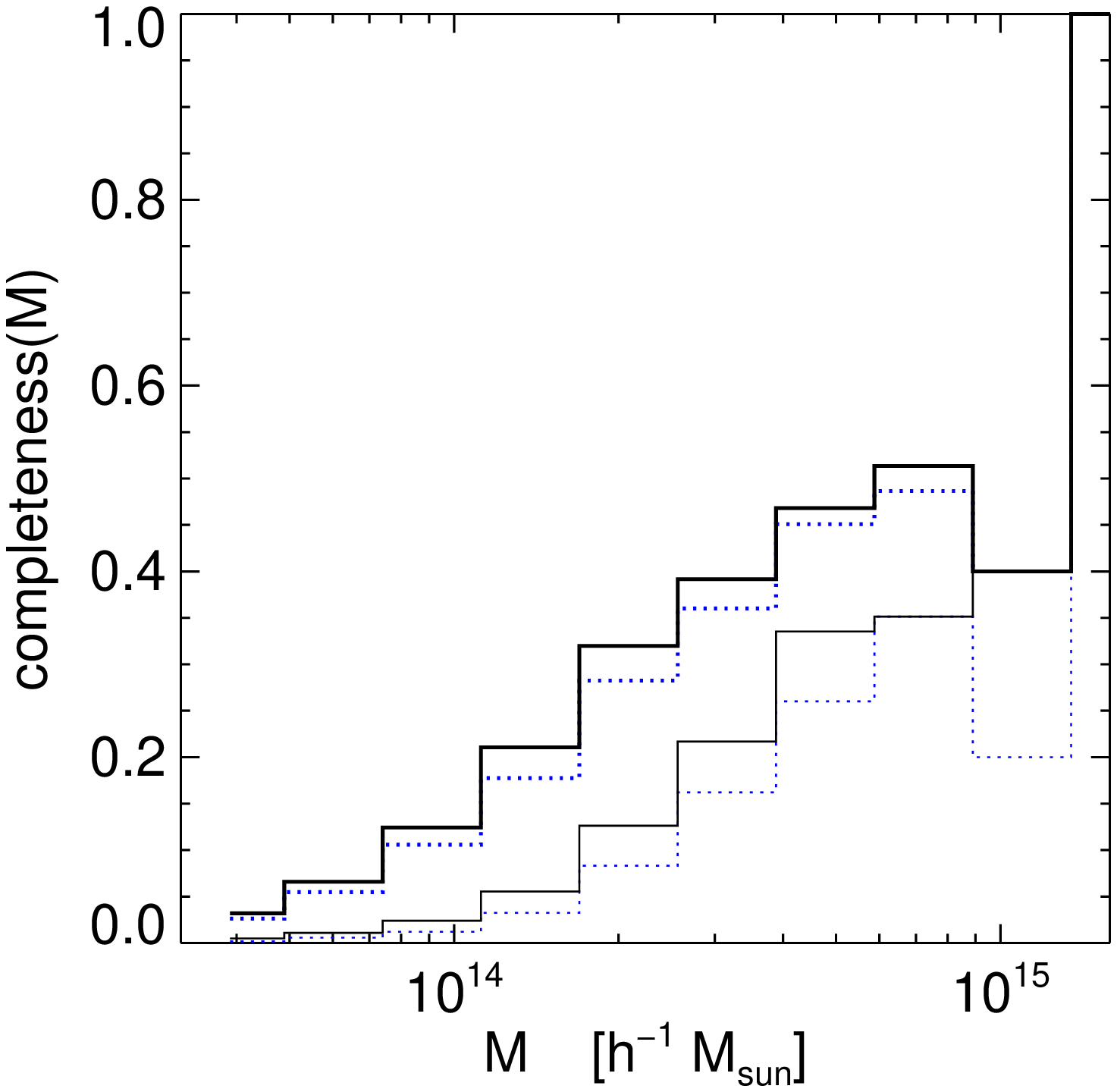,bb=0 0 450
      450,width=0.50\textwidth}}
  \caption{Completeness of shear selected clusters detected with and
    without tomography. The left panel shows completeness as a
    function of cluster redshift for all clusters in the light cone
    above $10^{14.3}~\hmsol$. The right panel shows completeness as
    a function of cluster mass for all clusters in the redshift range
    $0.2 < z < 0.8$ for which the efficiency for lensing is
    appreciable.  Dotted (blue) histograms do not use photometric
    redshift information while solid (black) histograms are for the
    source galaxies binned into three coarse redshift bins (TMF3).
    Thick lines are for an efficiency cutoff of $e \geq 60\%$
    ($S/N\gtrsim 3.5$) whereas thin lines are for $e \geq 75\%$
    ($S/N\gtrsim 4.5$). The left panel is cutoff at $z=0.2$ because
    below this redshift our simulation scheme does not accurately
    reproduce the counts of clusters (see Figure \ref{fig:dndz})}
  %For this threshold $70\%$ more clusters 
  %  are detected most of which are higher redshift and lower mass clusters.
  \label{fig:comp_fig}
\end{figure*}

\section{Tomographic Redshifts}

In this section we characterize the reliability of the tomographic
redshifts determined by maximizing the likelihood in
eqn.~(\ref{eqn:lhood_conv}) (see Figure \ref{fig:pdf}), and study how
reliability depends on detection significance, mass, and redshift. Then we
will vary the filter profile $G(\theta)$ in eqn.~(\ref{eqn:model}) 
and determine the effect on the tomographic redshift errors.

For the rest of this section we focus on a sample of shear selected
clusters obtained with the TMF3 for the truncated NFW profile with
$\theta_{\rm out}=5.5^{\prime}$ applied to noisy data. We showed in
the previous section that this filtering scheme was superior to the
rest for an efficiency cutoff of $e=75\%$ corresponding to $\nu
\gtrsim 4.5$ detections (see Table \ref{table:eff_tomo}). The total
number of clusters in the simulated $608 \ \rm{deg}^2$ above this
cutoff is 1060, sufficient for a statistical study. Although in what
follows we consider both noisy and noiseless data, the cluster sample
will always remain the top 1060 clusters with $e>75\%$ detected in the
noisy maps.  The mass, redshift, and likelihood distribution of the
cluster sample is illustrated by the scatter plots in Figure
\ref{fig:lhood_3d}. The colors and sizes of points reflect the
likelihood, or detection significance.  Note that likelihood of a
cluster is a function of both cluster mass and redshift.

In order to construct the likelihood maps for the TMF3 (Figure
\ref{fig:demo}) we binned the source galaxies both spatially and in
redshift as in eqn.~(\ref{eqn:lhood_bin}), so that the convolutions could
be performed quickly with FFT's.  However, for the purpose of
characterizing the reliability of tomographic redshifts we evaluate
the likelihood \emph{exactly} carrying out the full sums in
eqn.~(\ref{eqn:lhood_conv}).

\begin{figure*}[t!]
\centering
\begin{tabular}{cc}
\begin{minipage}[t]{0.65\textwidth}
\epsfig{file=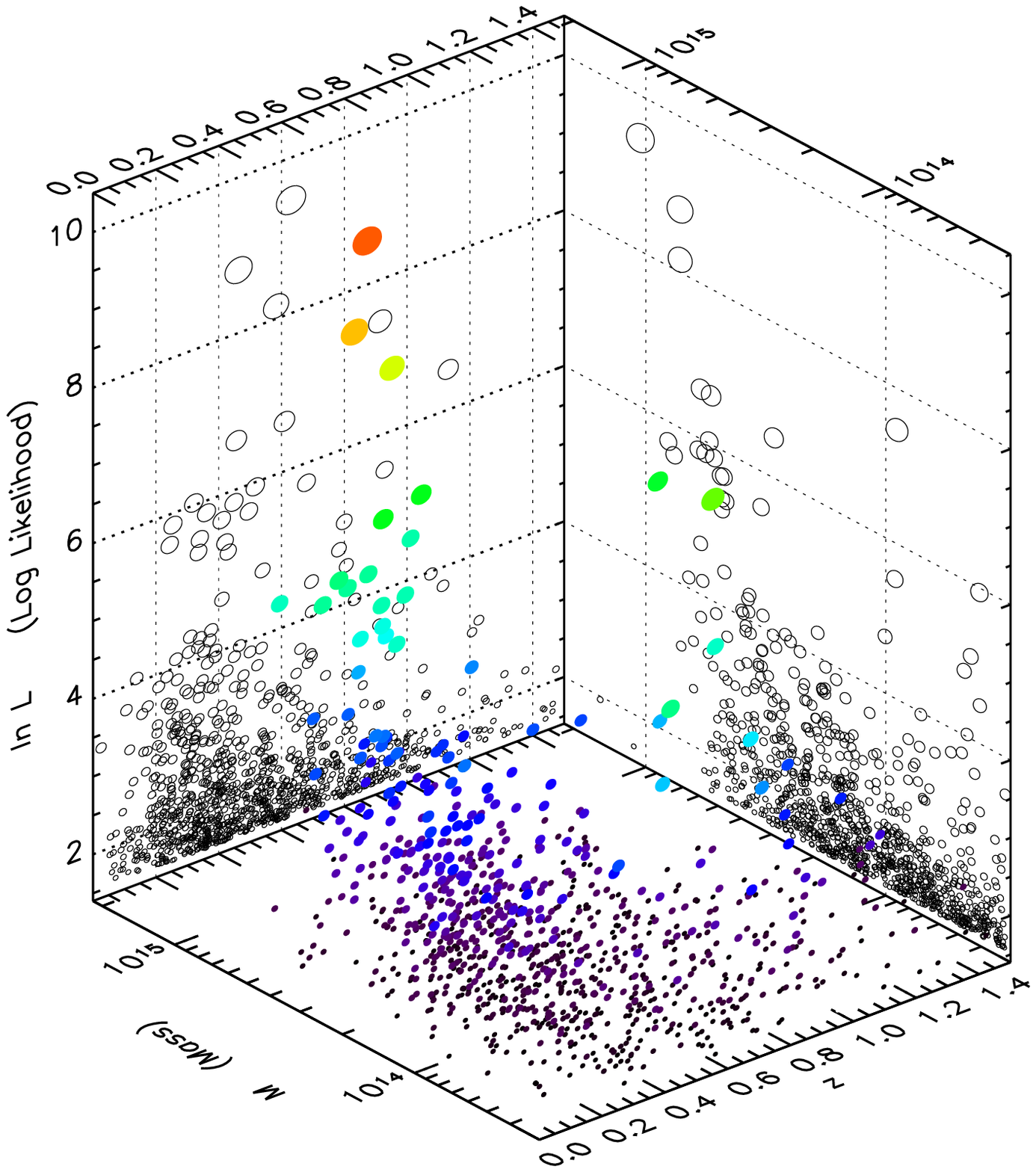,bb=70 40 455 504,width=0.9\textwidth}
\raisebox{2.0cm}{\epsfig{file=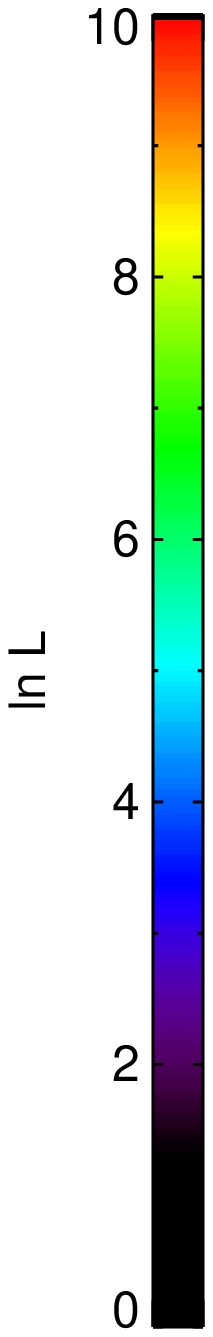,bb=140 100 162 435,clip=,totalheight=7.45cm}}
\end{minipage}\hfill
\hskip -0.3cm
\begin{minipage}[t]{0.35\textwidth}
\raisebox{2.5cm}{\epsfig{file=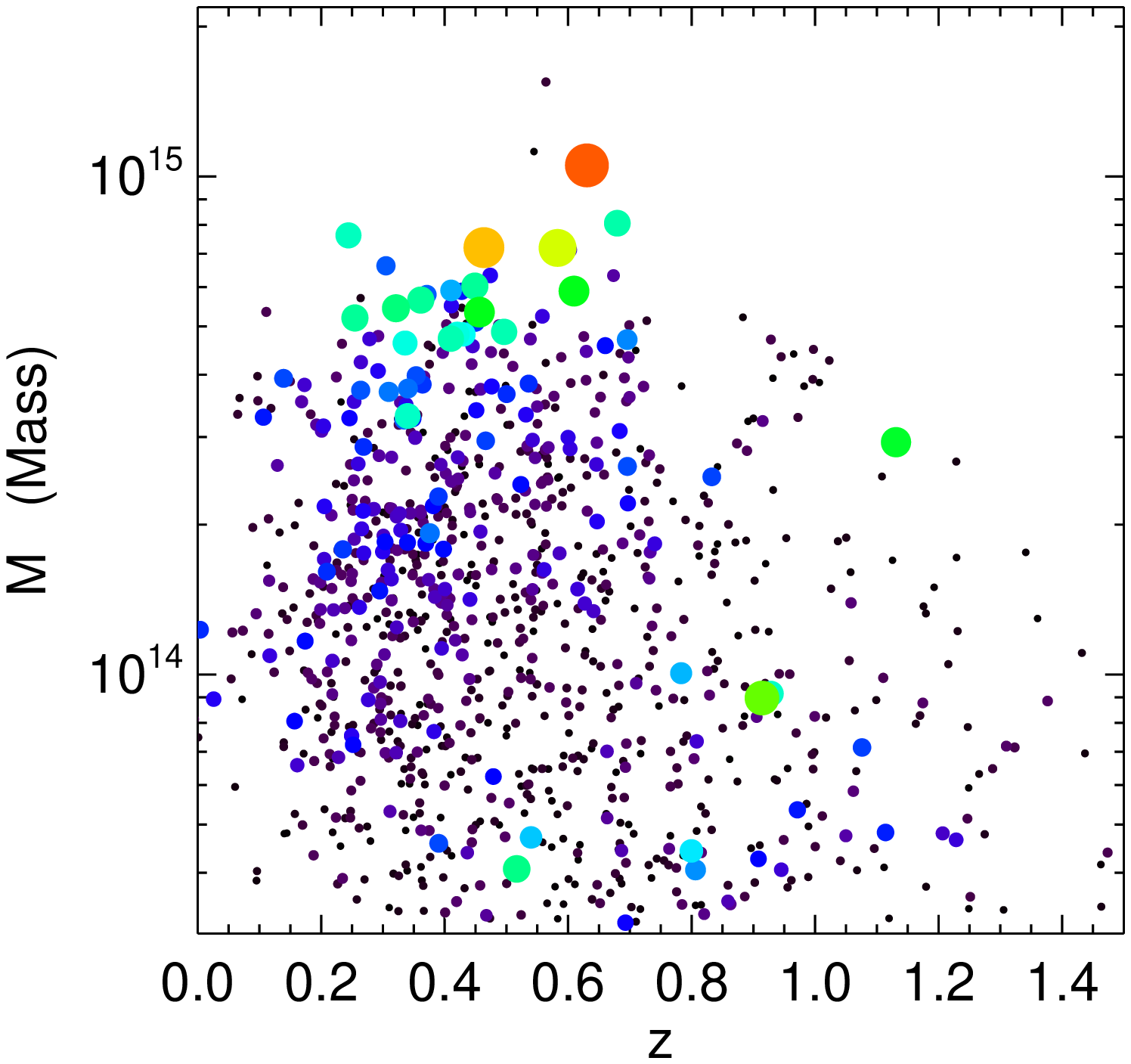,bb=0 0 460 450,clip=,width=1.0\textwidth}}\hfill
\vskip -2.5cm
\caption{\emph{Left:} Scatter plot of likelihood versus cluster mass
  and redshift for 1060 clusters detected with the TMF3 above the
  efficiency cutoff of $75\%$. The colors and sizes of the points
  reflect the detection significance of the cluster. Because of the
  redshift dependence of the lensing kernel, the likelihood is a
  function of cluster redshift in addition to mass.  \emph{Right:} The
  $z-M$ plane of the scatter plot at left.}
\label{fig:lhood_3d}
\end{minipage}
\end{tabular}
\vskip10pt
\end{figure*}

\subsection{Tomographic Redshift Errors}

\begin{figure*}[t]
\centering
\centerline{\epsfig{file=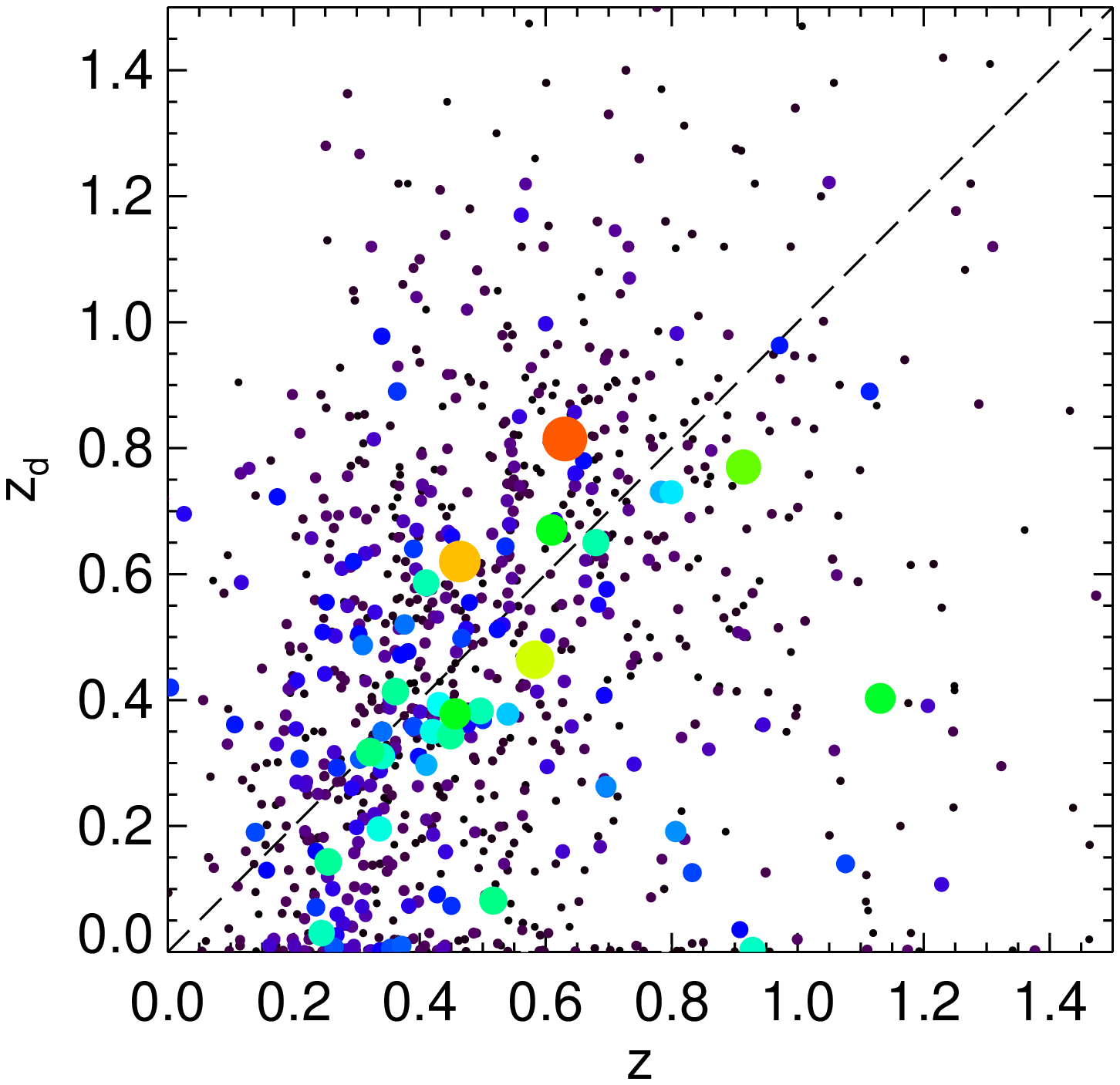,bb=0 20 450 450,width=0.45\textwidth}\epsfig{file=colorbar.eps,bb=70 -8 200 440,clip=,totalheight=7.6cm}\epsfig{file=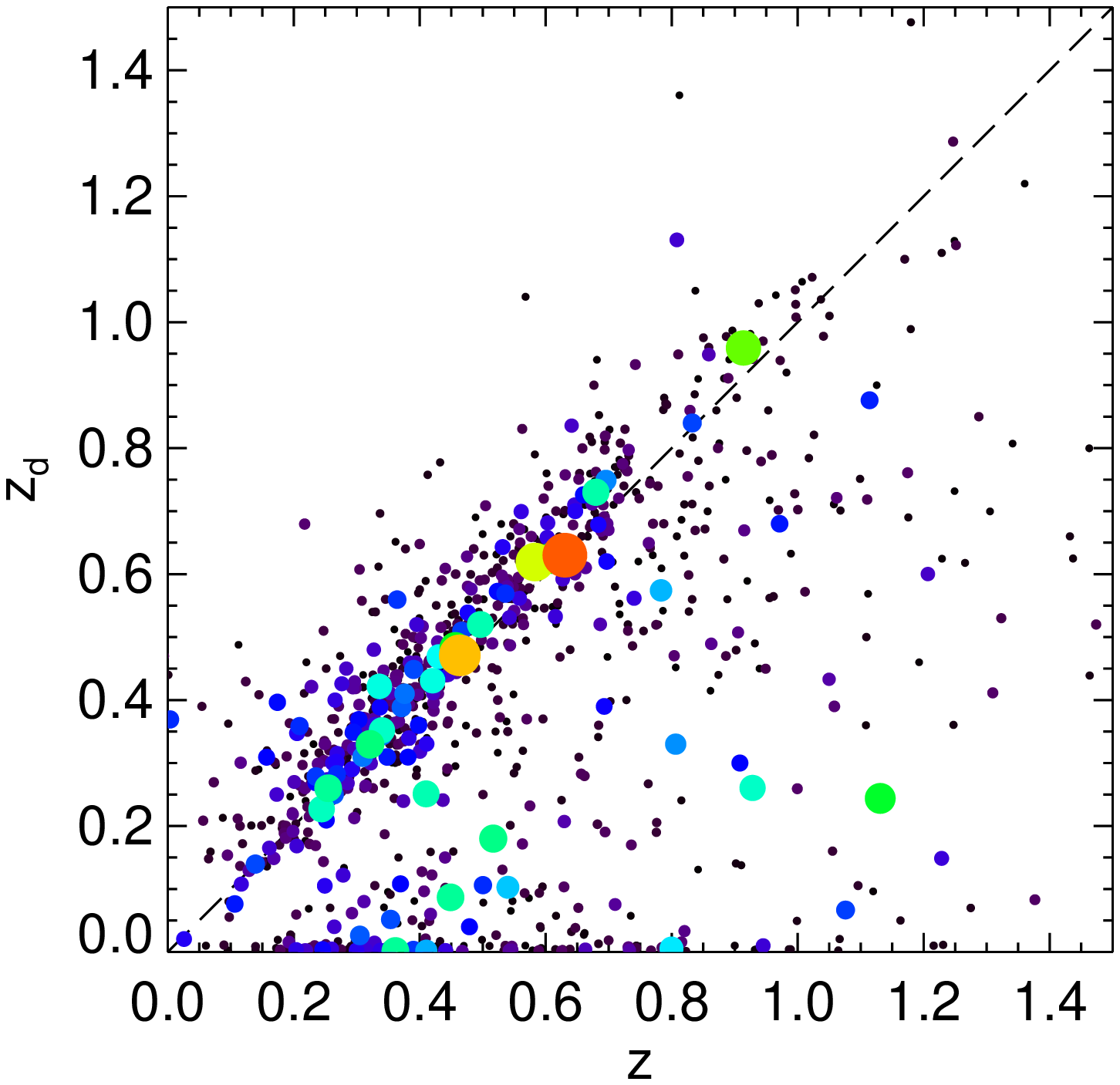,bb=0 20 450 450,width=0.45\textwidth}}
  \caption{ Tomographic redshift versus real redshift for noisy (left)
    and noiseless (right) data. The colors and sizes of the points
    reflect the detection significance of the cluster.  The fraction
    of clusters with tomographic redshift errors $|\Delta z| \leq 0.2$
    is $45\%$ and the root mean square deviation from the real
    redshifts is $\langle(\Delta z)^2\rangle^{1/2}=0.41$ (see Table
    \ref{table:tomo_error}). \label{fig:z_scat}}
\end{figure*}

%The horizontal features in the figure are due to clusters 
%for which the likelihood curve is flat, and are artifacts of our source 
%redshift binning (see fig \ref{fig:pdf} and the discussion below).

In Figure \ref{fig:z_scat} we show scatter plots of tomographic
redshift, $z_d$, versus real redshift where again points are colored
according to their detection significance and higher likelihood points
are larger. The left panel includes noise while the right panel is for
noiseless data.  The filter profile $G(\theta)$ used to determine the
tomographic redshift (see eqn.~\ref{eqn:model}) was the truncated NFW
profile with $\theta_{\rm s}=0.50^{\prime}$ and $\theta_{\rm
  out}=5.5^{\prime}$, which detected more clusters than the
others. For noisy weak lensing data, $45\%$ of clusters have
tomographic redshifts with $|\Delta z| \ \leq 0.2$, where $\Delta
z\equiv z_{\rm real} - z_d$. The rms deviations from the real
redshifts are $\sigma_{\Delta z}\equiv\langle(\Delta
z)^2\rangle^{1/2}=0.41$. For the ideal case of no noise these numbers
change to $67 \%$ within $|\Delta z|\leq 0.2$ and $\sigma_{\Delta z} =
0.34$.  The reliability of the tomographic redshifts for the noiseless
case reflect the intrinsic limitations of this effectively two
dimensional method.  The broad lensing kernel and projections
of large scale structure limit the accuracy to which radial positions
can be determined.

From the colors and sizes of the points in Figure \ref{fig:z_scat} it
is clear that tomographic redshifts are more accurate for clusters
with a higher detection significance. In Figure \ref{fig:dz_Lhood} we
show scatter plots of redshift error $\Delta z$ versus detection
significance, which illustrates that the redshift errors taper down
dramatically for more significant detections. If there were a
candidate `dark clump' corresponding to a very high peak in a
likelihood map, the tomographic redshift might provide the redshift of
the cluster reasonably well, and would be the only means available to
determine a redshift in the absence of a significant overdensity of
galaxies. However, we saw in \S 3.2 that $\sim$ 15 \% of the most
significant peaks were projections, so there would be no guarantee
that the object was an actual cluster, although the tomographic
redshift information could be taken into account when devising a
strategy for follow up.

Scatter plots of the redshift error $\Delta z$ versus mass and
redshift are shown in Figure \ref{fig:dz_logM} and Figure \ref{fig:dz_z}.
As expected, the tomographic redshift errors decrease for more massive
clusters and for clusters closer to the peak of weak lensing kernel
($\sim 0.4$). However, the likelihood of the detection is clearly a
better indicator of the tomographic redshift reliability than mass or
redshift, since the strength of the lensing signal is determined by
both the mass and redshift of a cluster.

\begin{figure*}[t]
  \centering \centerline{\epsfig{file=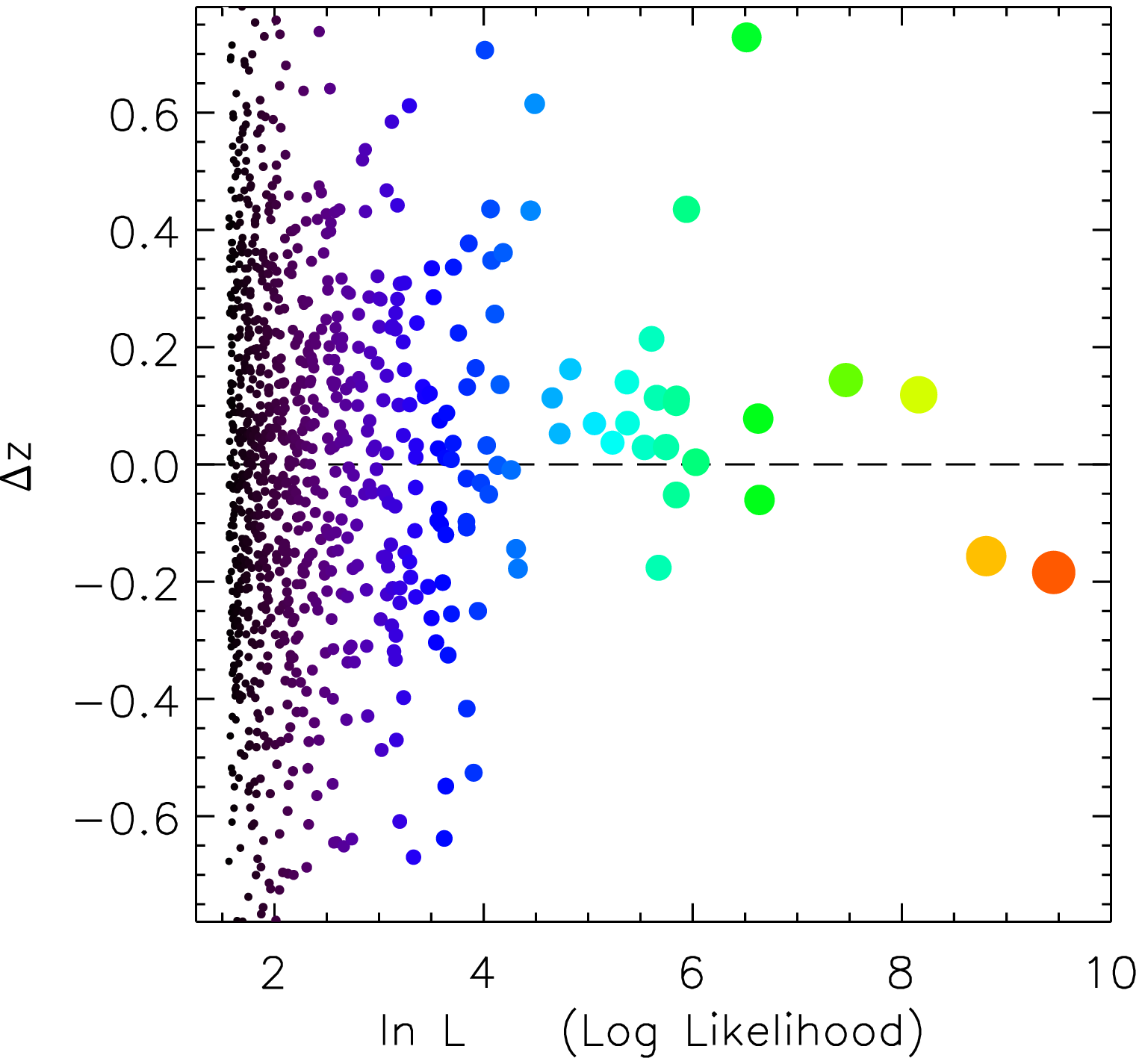,bb=0 20 450
      450,width=0.45\textwidth}\epsfig{file=colorbar.eps,bb=70 -8 200
      440,clip=,totalheight=7.6cm}\epsfig{file=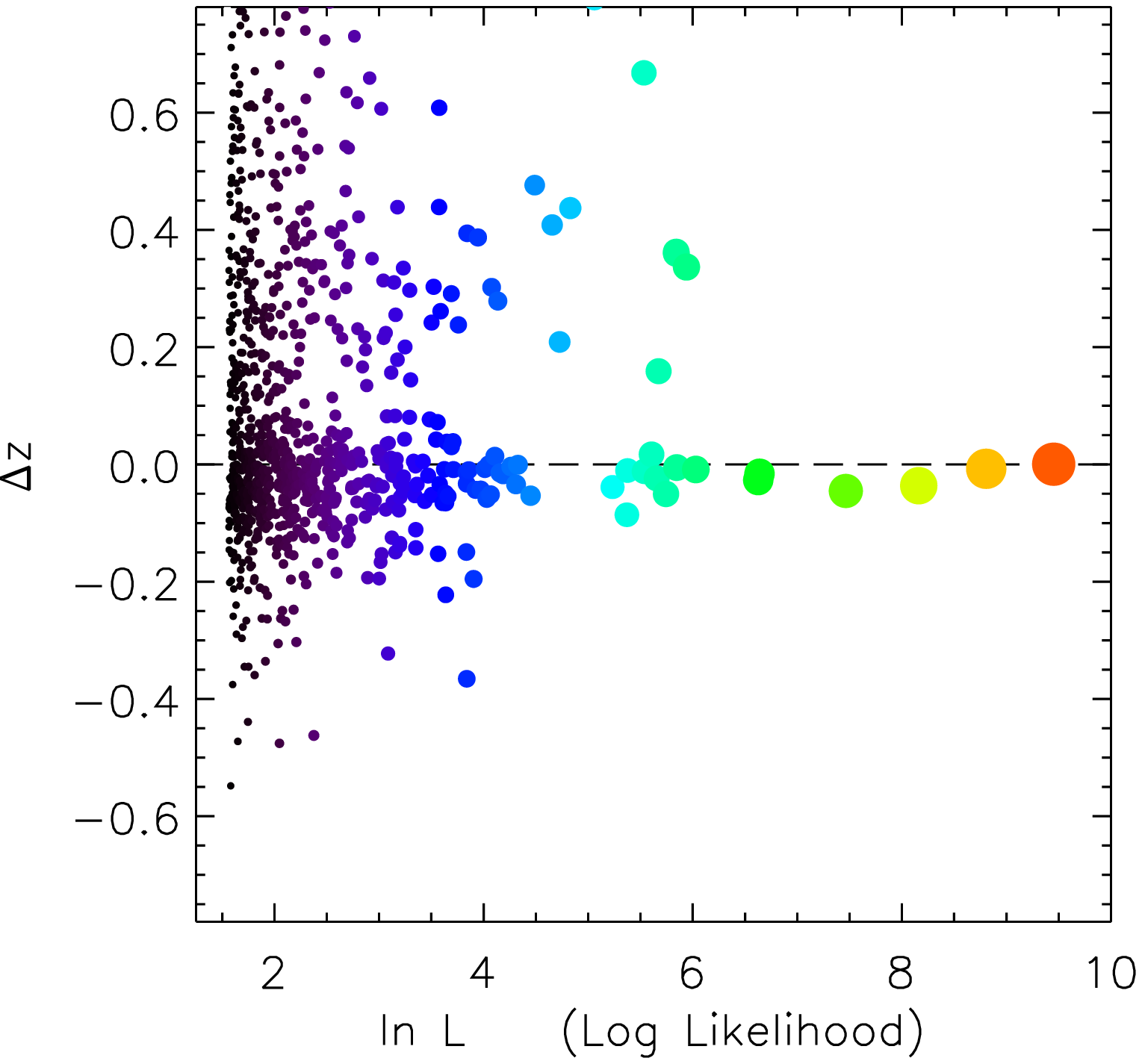,bb=0
      20 450 450,width=0.45\textwidth}}
  \caption{Tomographic redshift errors versus likelihood for noisy
    (left) and noiseless (right) data. The colors and sizes of the
    points reflect the detection significance of the cluster.}
  \label{fig:dz_Lhood}
\end{figure*}

\begin{figure*}[t]
  \centering \centerline{\epsfig{file=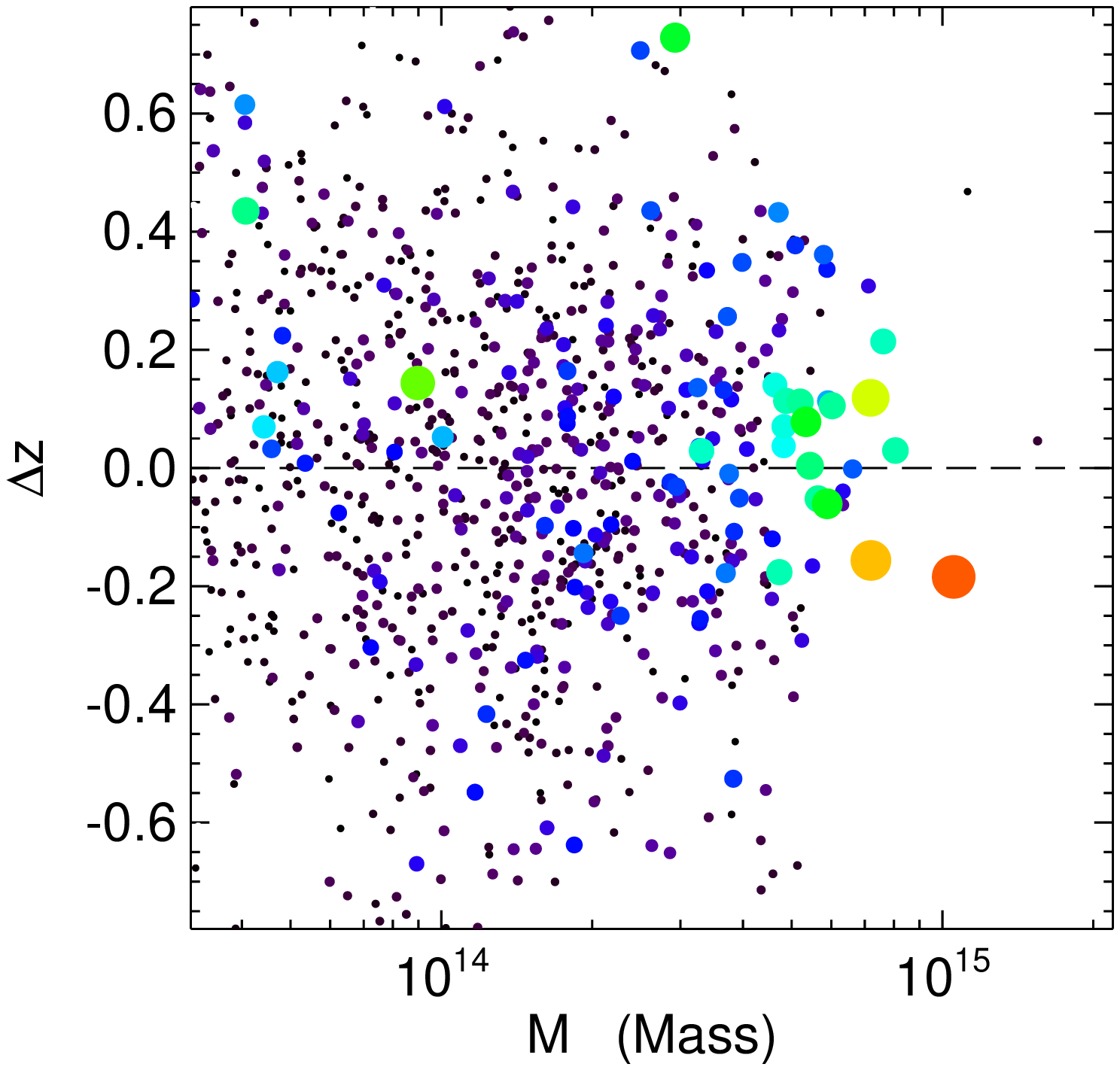,bb=0 20 450
      450,width=0.45\textwidth}\epsfig{file=colorbar.eps,bb=70 -8 200
      440,clip=,totalheight=7.6cm}\epsfig{file=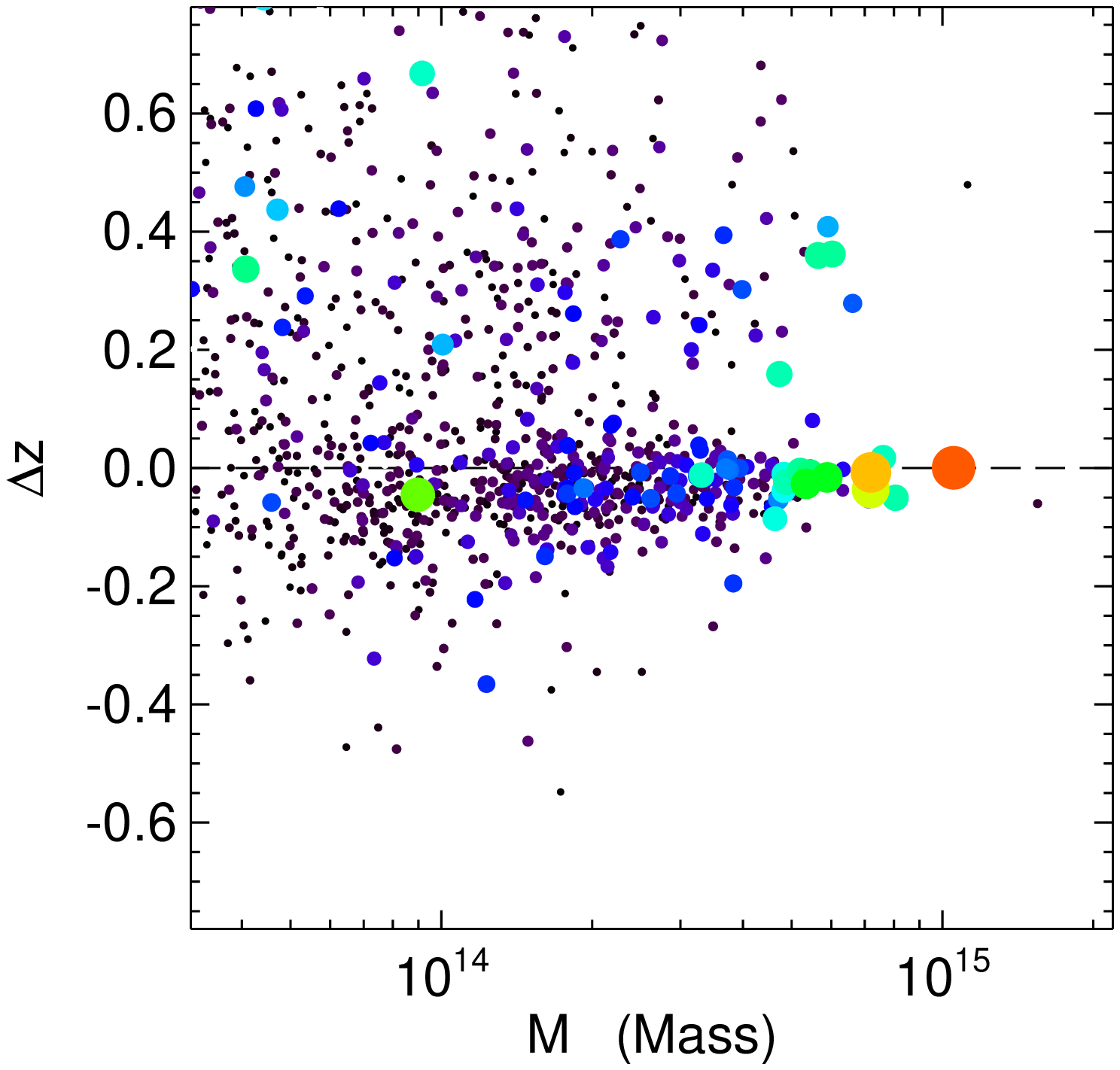,bb=0
      20 450 450,width=0.45\textwidth}}
  \caption{Tomographic redshift error versus cluster mass for noisy
    (left) and noiseless (right) data. The colors and sizes of the
    points reflect the detection significance of the cluster.}
  \label{fig:dz_logM}
\end{figure*}

\begin{figure*}[t]
  \centering \centerline{\epsfig{file=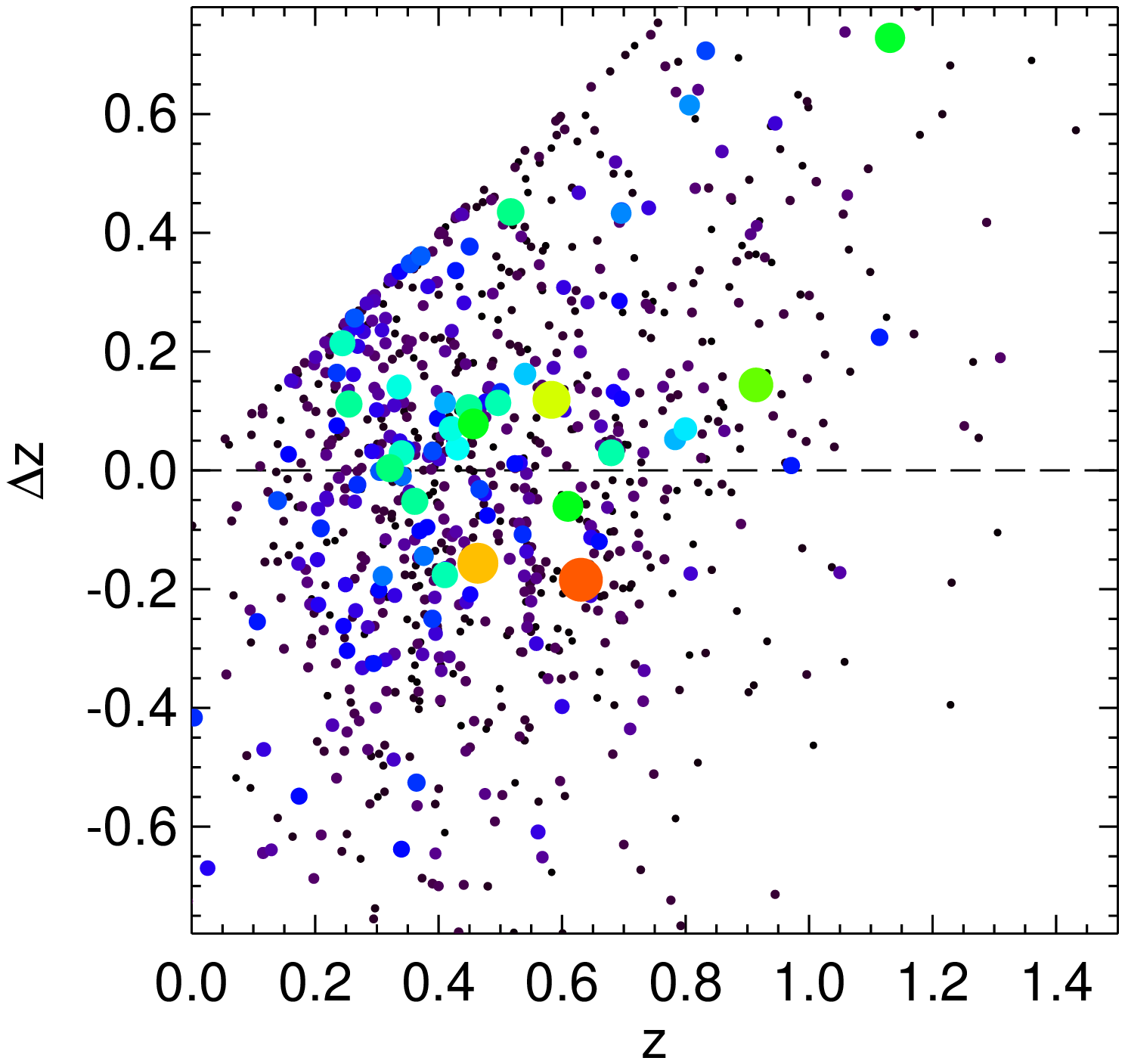,bb=0 20 450
      450,width=0.45\textwidth}\epsfig{file=colorbar.eps,bb=70 -8 200
      440,clip=,totalheight=7.6cm}\epsfig{file=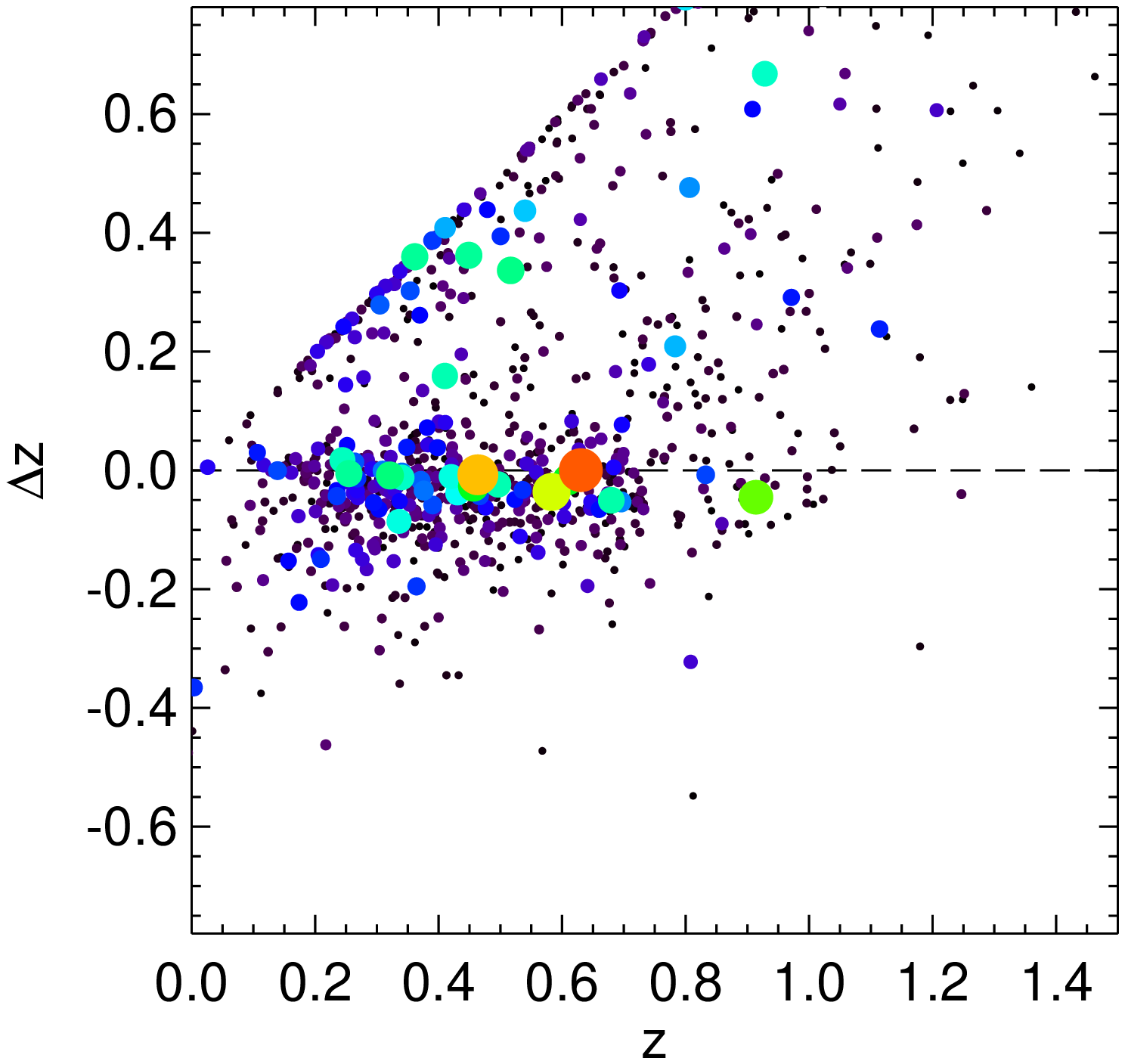,bb=0
      20 450 450,width=0.45\textwidth}}
  \caption{Tomographic redshift error versus cluster redshift for
    noisy (left) and noiseless (right) data. The colors and sizes of
    the points reflect the detection significance of the cluster. The
    linear feature in the upper left corner of the plots are an
    artifact of a slight tendency for the tomography to map low signal
    no noise detections to $z_{\rm d}=0$, hence $\Delta z = z$.
    \label{fig:dz_z}}
\end{figure*}

\subsection{Tests of Tomography}

The truncation scale $\theta_{\rm out}$ in eqn.~(\ref{eqn:G}) that
optimizes the number of clusters detected is not necessarily optimal
for minimizing the error $\Delta_z$ of tomographic redshifts $z_d$ of
the clusters. We applied the TMF to our sample of clusters with
truncation scales $\theta_{\rm out}=3.0^{\prime}, 10.0^{\prime},
15.0^{\prime}$ and found negligible improvement in the tomographic
redshift errors.

Finally, we quantify the degree to which tomographic redshifts depend
on the assumed model cluster profile $G(x)$ in eqn.~(\ref{eqn:model}),
by computing tomographic redshifts for the same sample of clusters but 
using a different shear profile. We consider an isothermal
sphere with a core, where the core radius is set to the same value of
$\theta_{\rm s}=0.50$ used for the NFW scale radius profile and we adopt a
Gaussian truncation scale of $\theta_{\rm out}=5.5$.  An isothermal sphere with 
a core has tangential shear profile 
\beq 
G_{\rm ISO}(x)=\frac{1+\frac{1}{2}x^2}{(1+x)^{3 \slash 2}}, 
\eeq 
\citep{BK87} with $x=\theta\slash\theta_{\rm s}$. This 
profile has an outer power law slope of $\sim \theta^{-1}$, different from the
$\sim \theta^{-2}$ scaling of the NFW profile used with the TMF. Overall, 
the isothermal profile gives tomographic redshift errors similar to the
NFW profile used in this section, although for a particular cluster
the redshifts deduced can be different.  

Table \ref{table:tomo_error} summarizes the results of this section on
the reliability of tomographic redshifts, and its dependence on
truncation scale and filter profile shape.

\begin{table}
\begin{center}
  \caption{Tomography Tests\label{table:tomo_error}}
  \begin{tabular}{lcccc}
     %      \tablevspace{3pt}
    \hline
    \hline
  \hfill\vline   &     \quad\quad\quad  Noisy  &   \quad\quad\quad \vline &    \quad\quad \ \  Noiseless           &   \\
    \hline
    Filter \hfill\vline & $\left|\Delta z\right| \ \leq 0.2$ & $\sigma_{\Delta z}$ \hfill\vline&  $\left|\Delta z\right| \ \leq 0.2$ & $\sigma_{\Delta z}$ \\
    \hline
    NFW $\theta_{\rm out}=3.0$    \hfill\vline & 42\%     & 0.44 \hfill\vline     & 67\%     & 0.36 \\
    NFW $\theta_{\rm out}=5.5$    \hfill\vline & 45\%     & 0.41 \hfill\vline     & 67\%     & 0.34 \\
    NFW $\theta_{\rm out}=10.0$   \hfill\vline & 46\%     & 0.40 \hfill\vline     & 67\%     & 0.35 \\
    NFW $\theta_{\rm out}=15.0$   \hfill\vline & 47\%     & 0.39 \hfill\vline     & 66\%     & 0.36 \\
    Isothermal $\theta_{\rm out}=5.5$ \hfill\vline & 46\% & 0.40 \hfill\vline     & 65\%    & 0.35\\
    \hline
  \end{tabular}
\end{center}
\footnotesize NOTES.--- Summary of tests on accuracy of tomographic
redshifts. The column labeled indicates the filter used, and the left
and right columns indicate the fraction of clusters with $|\Delta z| \leq
0.2$ and the variance of the tomographic redshifts $\sigma_{\Delta z}$ for
noisy and noiseless mock sources.
\vskip -0.5cm
\end{table}

\section{Discussion}

We have described a fast efficient simulation scheme for computing the
statistics of shear selected galaxy clusters from weak lensing. By
tiling the line of sight with 32 unique PM simulations, our algorithm
is unique in that it densely samples the primordial Gaussian
distribution of random phases of large scale structure along the light
cone, providing an accurate representation of the statistics of rare
events. We have confirmed that the two point statistical properties of
the three dimensional density field (Figure \ref{fig:tiling}) and the
abundance of galaxy clusters in the light cone (Figure \ref{fig:dndz})
are accurately reproduced.  Although the dynamic range of our PM
simulations is limited, we argued that high resolution simulations are
not necessary to study shear selected cluster because of the small scale
shot noise limit of weak lensing observations. 

%Further tests of our 
%simulation algorithm will be presented in a future paper \citep{Me04}.

A maximum likelihood tomographic technique was presented which
utilizes the photometric redshifts of source galaxies to determine the
radial position of a galaxy cluster from weak lensing alone. We
introduced a tomographic matched filtering (TMF) shceme, which
optimally incorporates this additional radial information to detect
more clusters.  We applied the TMF to a large ensemble of simulations
and found that it enhances the number of clusters detected with
$S/N\gtrsim 4.5$ by as much as $76 \%$ (see Table \ref{table:eff_tomo}
and \ref{fig:eff_dist}).  As illustrated by Figure \ref{fig:eff_dist},
the TMF increases the dynamic range of weak lensing searches for
clusters, detecting more high redshift clusters and extending the mass
sensitivity down to the scale of large groups.  Furthermore, binning
the sources coarsely using only three redshift bins was sufficient to
reap the benefits of tomography.  Thus the filtering scheme developed
here can be applied to ground based weak lensing observations in as
few as three bands, since two colors and a magnitude provide enough
information to bin sources this crudely.

For the case where the photometric redshifts of sources are known more
precisely, as could be achieved with multicolor ground based data, we
quantified the errors in the tomographic redshifts. For the densities
of sources used in this paper, $45\%$ of clusters detected with signal
to noise ratio $S/N\gtrsim 4.5$ had tomographic redshift errors $|\Delta
z| \leq 0.2$ and the rms deviation from the real redshifts is
$\sigma_{\Delta z}=0.41$, with smaller errors for higher $S/N$ ratios.

The most appealing property of shear selected cluster samples in
comparison to other methods of detecting clusters, is that \emph{the
  expected cluster distributions can be reliably determined for any
  cosmological model using simulations like those described in this
  paper.} Because on the scales of interest only gravity is involved,
comparing theory to observations does not, in principle, depend on
assumptions about the relationship between dark and luminous
matter. However, in practice projection effects and incompleteness are 
likely will complicat the interpretation of shear selected samples. 

Weak lensing searches for clusters are plagued by projection effects.
In particular, we found that the maximum intrinsic efficiency of weak
lensing cluster surveys is $\sim 85\%$.  Phrased in terms of our
criteria for matching peaks with clusters (\S 2.5), of order $\sim
15\%$ of the most significant peaks detected in noiseless weak lensing
maps do not have a collapsed halo with $M >10^{13.5} \ \hmsol $ within
a $3^{\prime}$ aperture.  This intrinsic inefficiency, also noted by
previous investigators \citep{WvWM02,Paddy03,HTY03}, arises because of
the broad weak lensing kernel and confusion from large scale structure
fluctuations.  

The calculated maximum intrinsic efficiency is sensitive to
parameters. Lowering the limiting mass, $10^{13.5} \ \hmsol$, of our
cluster catalog, increasing the size of the aperture, $3^{\prime}$,
used to correlate peaks with halos, or reducing the linking length,
$b=0.2$, of the FOF group finder, would all increase the efficiencies
deduced.  Nevertheless, some of the most significant peaks in the mass
maps do not correspond to actual clusters.  While cluster finding in
numerical simulations involves applying a group finding algorithm to
3-d dark matter particle distributions, observationally determining
whether there is a galaxy cluster at the location of a given peak will
depend on observational criteria like the overdensity of galaxies,
X-ray flux, SZ decrement, or the presence of a brightest cluster
galaxy. \emph{Thus, any criteria employed to distinguish the real
  clusters from the projections will rely on detecting
  baryons}. Because our ability to detect these baryons is a function
of mass and redshift, and because of the scatter between baryonic
observables and the underlying mass of the dark matter halo, the
projection effects in weak lensing searches will degrade the well
defined selection function deduced from `dark matter only'
simulations like those employed in this work.  

%But the dependence on
%this scatter will only be indirect, clearly preferable to the near
%exponential sensitivity to this scatter for cluster surveys that rely
%on a baryonic proxy.

In the context of using cluster counts to constrain cosmological
parameters, the uncertainties in the selection function caused by
projections are exacerbated by the high incompleteness of shear
selected cluster samples.  Even for an efficiency cut of $e \geq
60\%$, the completeness is $\gtrsim 50\%$ only for the most massive
clusters $M\gtrsim 10.0^{14.3}$ which are near the peak of the lensing
kernel at $z\sim 0.3$, and lower everywhere else. The severe
projection effects imply that these missed clusters are likely to
scatter in or out of the detected sample. In light of the fact that
this number of missed clusters is of the same order as the number
detected, an accuracy of $x\%$ in the final cluster sample requires a
theoretical understanding the selection function to better than $\sim
x\%$. This is likely to be very hard. One alternative is to
throw away the number count information and use only the shape of the
redshift distribution of shear-selected clusters to constrain
cosmology since the shape will be much less sensitive to the
precision to which the selection function is known. 

%future optical, SZ, and X-ray cluster surveys will be uncertainties in
%mass-observable relations, and their evolution and scatter.  While we
%have just argued that because of projection effects inherent to shear
%selected samples, the mass-observable scatter will also propagate into
%the weak lensing sample, the dependence on this scatter will only be
%indirect, clearly preferable to the near exponential sensitivity to
%this scatter for cluster surveys that rely on a baryonic proxy.

%Despite the low efficiencies caused by projection effects and the high
%incompleteness, \emph{the appealing property of shear selected cluster
%  samples in comparison to other methods of detecting clusters, is
%  that the expected cluster distributions can be reliably determined
%  for any cosmological model using simulations like those described in
%  this paper.}  Because on the scales of interest only gravity is
%involved, comparing theory to observations does not, in principle,
%depend on assumptions about the relationship between dark and luminous
%matter. However, efforts 

Cross correlation of shear selected cluster samples with overlapping
optical, SZ, or X-ray cluster catalogs is an important cosmological
application (Schirmer \etal 2003,2004; Hughes \etal 2004).  For such a
cross-correlation it is likely to be advantagous to work at lower
efficiency thresholds, considering that the number of clusters per
square degree for an efficiency $e\geq 60\%$ is five times larger than
that with $e\geq 75\%$ (see Table \ref{table:eff_tomo}).  These less
significant peaks will span a larger mass and redshift range
increasing the amount of overlap between the two cluster samples.
Besides addressing probable baryonic biases for baryon selected
clusters, cross correlation could be the most expedient route to
precision cosmology with cluster surveys. In order to recover
parameter sensitivity degraded by the scatter in the mass-observable
relations, \citet{MM03a} advocate calibrating these relations with
extensive follow up mass measurements of a fraction of the clusters in
a `baryon selected' sample.  An alternative to such a follow up
campaign of individual clusters is to conduct a parallel weak lensing
survey, and use the clusters detected from both weak lensing and, say
SZ, to determine the mass-observable relation
\emph{statistically}. Because the mass and redshift distributions of
the shear selected clusters is known for any cosmological model, the
distribution of dual weak lensing-SZ detections can be used to
simultaneously constrain cosmology and the relationship between mass
and SZ decrement.  In addition to the statistical mass-SZ decrement
calibration provided by the distribution of dual detections, stacked
weak lensing mass measurements of the individual SZ clusters can be
used to provide another constraint on this relation. Finally, it is
worth emphasizing that an optical cluster
search \citep{Postman96,Kepner99,GY00,WK02,Koch03} could be conducted
using the same deep imaging data used to measure weak lensing.
Indeed, \citet{Schirm03,Schirm04} have already combined shear
selection with optical color selection, and confirmed several
color-selected optical cluster candidates with weak lensing.  Such
parallel analyses of shear-selected cluster samples and baryonic
cluster samples will provide mass calibration, help break parameter
degeneracies, and eliminate systematics inherent in cluster surveys.

The potential of shear selected clusters to provide precision
constraints on the dark energy, either alone or in conjunction with
another baryonic cluster survey, are topics which merit future
investigation.

%Used in conjunction,  the optical and shear selected
%samples would simultaneously constrain cosmology and the
%(possibly redshift dependent) mass-richness relation.  

%We will explore

\acknowledgments 
~\\

Vijay Narayanan contributed to the early stages of this work and
provided the PM code. We would like to thank Tony Tyson, David
Wittman, and Vera Margoniner for advice on cluster detection and
tomography. We are grateful to Michael Strauss, Peter Schneider, and
Martin White for reading early versions of this manuscript and
providing helpful comments. Thanks to Sheng Wang for pointing out a
notational error in a prevous draft. We acknowledge fruitful
discussions with Gary Bernstein, Paul Bode, Wayne Hu, Robert Lupton,
Jerry Ostriker, Nikhil Padmanabhan, and Uros Seljak. Finally, thanks
to the anonymous referee for constructive comments.  For part of this
study J.~F.~H was supported in part by a generous gift from the Paul
\& Daisy Soros Fellowship for New Americans. The program is not
responsible for the views expressed.  J.~F.~H is currently supported
by NASA through Hubble Fellowship grant \# 01172.01-A awarded by the
Space Telescope Science Institute, which is operated by the
Association of Universities for Research in Astronomy, Inc., for NASA,
under contract NAS 5-26555.

% In addition, Weinberg \& Kamionkowski (2002) claim that a significant number of peaks in weak lensing maps will be due to non-virialized clusters still in the process of gravitational collapse. In principle, one could determine whether a false detection in the simulations is a non-virialized collapsing structure or a projection of cosmic structure by modifying the FOF group finding algorithm.  Observationally, it will be very difficult to make any such distinction.  We conclude that because of the intrinsic inefficiency found here, only statistical statements can be made about any purported population of dark clusters.   

%We demonstrated above, that convolving weak lensing data with a ``matched filter'' maximizes the likelihood, and hence the contrast between signal and background. If one is interested in filtering out a signal 

\end{document}